\documentclass{emulateapj}
\usepackage{natbib}
\usepackage{url}
\def\deg{\ifmmode^\circ\else$^\circ$\fi}

\def\arcs{\ifmmode {''}\else $''$\fi}
\def\arcm{\ifmmode {'}\else $'$\fi}
\def\parcs{\sa=.07em \sb=.03em
    \ifmmode $\rlap{.}$^{\scriptscriptstyle\prime\kern -\sb\prime}$\kern -\sa$
    \else \rlap{.}$^{\scriptscriptstyle\prime\kern -\sb\prime}$\kern -\sa\fi}
\def\parcm{\sa=.08em \sb=.03em
    \ifmmode $\rlap{.}\kern\sa$^{\scriptscriptstyle\prime}$\kern-\sb$
    \else \rlap{.}\kern\sa$^{\scriptscriptstyle\prime}$\kern-\sb\fi}

\def\Msun{M$_{\odot}$}
\def\Myr{\Msun/yr}

\def\spose#1{\hbox to 0pt{#1\hss}}
\def\simlt{\mathrel{\spose{\lower 3pt\hbox{$\mathchar"218$}}
    \raise 2.0pt\hbox{$\mathchar"13C$}}}
\def\simgt{\mathrel{\spose{\lower 3pt\hbox{$\mathchar"218$}}
    \raise 2.0pt\hbox{$\mathchar"13E$}}}
\def\lsim{\rlap{$<$}{\lower 1.0ex\hbox{$\sim$}}}
\def\gsim{\rlap{$>$}{\lower 1.0ex\hbox{$\sim$}}}

\usepackage{amsmath} 
\newcommand{\Angstrom}{\textup{\AA}}

\begin{document}

\title{UVUDF:  Ultraviolet Imaging of the Hubble Ultra Deep Field with
  Wide-field Camera 3\altaffilmark{*}}

\altaffiltext{*}{Based on observations made with the NASA/ESA Hubble
  Space Telescope, obtained at the Space Telescope Science Institute,
  which is operated by the Association of Universities for Research in
  Astronomy, Inc., under NASA contract NAS 5-26555. These observations
  are \#12534.}

\author{Harry I. Teplitz\altaffilmark{1}, 
  Marc Rafelski\altaffilmark{1}, 
  Peter Kurczynski\altaffilmark{2}, 
  Nicholas A. Bond\altaffilmark{3},  
  Norman Grogin\altaffilmark{4},
  Anton M. Koekemoer\altaffilmark{4}, 
  Hakim Atek\altaffilmark{5},
  Thomas M. Brown\altaffilmark{4}, 
  Dan Coe\altaffilmark{4}, 
  James W. Colbert\altaffilmark{1}, 
  Henry C. Ferguson\altaffilmark{4},
  Steven L. Finkelstein\altaffilmark{6}, 
  Jonathan P. Gardner\altaffilmark{3},
  Eric Gawiser\altaffilmark{2}, 
  Mauro Giavalisco\altaffilmark{7},
  Caryl Gronwall\altaffilmark{8,9}, 
  Daniel J. Hanish\altaffilmark{1},
  Kyoung-Soo Lee\altaffilmark{10},  
  Duilia F. de  Mello\altaffilmark{11,3},
  Swara Ravindranath\altaffilmark{12}, 
  Russell E. Ryan\altaffilmark{4},
  Brian D. Siana\altaffilmark{13},
  Claudia Scarlata\altaffilmark{14},
  Emmaris Soto\altaffilmark{11},
  Elysse N. Voyer\altaffilmark{15}, 
  Arthur M. Wolfe\altaffilmark{16}
}

\altaffiltext{1}{Infrared Processing and Analysis Center, MS 100-22,
  Caltech, Pasadena, CA 91125.  hit@ipac.caltech.edu}
\altaffiltext{2}{Department of Physics and Astronomy, Rutgers University, Piscataway, NJ 08854}
\altaffiltext{3}{Laboratory for Observational Cosmology, Astrophysics Science Division, Code 665, Goddard Space Flight Center, Greenbelt MD 20771}
\altaffiltext{4}{Space Telescope Science Institute, 3700 San Martin Drive
Baltimore, MD 21218}
\altaffiltext{5}{
Laboratoire d'Astrophysique, \'{E}cole Polytechnique F\'{e}d\'{e}rale de Lausanne (EPFL), Observatoire, CH-1290 Sauverny, Switzerland}
\altaffiltext{6}{Department of Astronomy, The University of Texas at Austin, Austin, TX 78712}
\altaffiltext{7}{Astronomy Department, University of Massachusetts, Amherst, MA 01003}
\altaffiltext{8}{Department of Astronomy \& Astrophysics, The Pennsylvania State University, University Park, PA, 16802}
\altaffiltext{9}{Institute for Gravitation and the Cosmos, The Pennsylvania State University, University Park, PA 16802}
\altaffiltext{10}{Department of Physics, Purdue University, 525 Northwestern Avenue, West Lafayette}
\altaffiltext{11}{Department of Physics, The Catholic University of America, Washington, DC 20064}
\altaffiltext{12}{Inter-University Centre for Astronomy and Astrophysics, Pune, India}
\altaffiltext{13}{Department of Physics \& Astronomy, University of California, Riverside, CA 92521}
\altaffiltext{14}{Minnesota Institute for Astrophysics, School of Physics and Astronomy, University of Minnesota, Minneapolis, MN 55455}
\altaffiltext{15}{Aix Marseille Universit\'{e}, CNRS, LAM (Laboratoire d'Astrophysique de Marseille) UMR 7326, 13388, Marseille, France}
\altaffiltext{16}{Department of Physics and Center for Astrophysics and Space Sciences, University of California, San Diego,La Jolla, CA 92093-0424, USA}
\begin{abstract}

  We present an overview of a 90-orbit Hubble Space Telescope treasury
  program to obtain near ultraviolet imaging of the Hubble Ultra Deep
  Field using the Wide Field Camera 3 UVIS detector with the F225W,
  F275W, and F336W filters.  This survey is designed to: (i)
  Investigate the episode of peak star formation activity in galaxies
  at $1<z<2.5$; (ii) Probe the evolution of massive galaxies by
  resolving sub-galactic units (clumps); (iii) Examine the escape
  fraction of ionizing radiation from galaxies at $z\sim2-3$; (iv)
  Greatly improve the reliability of photometric redshift estimates;
  and (v)~Measure the star formation rate efficiency of neutral
  atomic-dominated hydrogen gas at $z\sim1-3$. In this overview paper,
  we describe the survey details and data reduction challenges,
  including both the necessity of specialized calibrations and the
  effects of charge transfer inefficiency.  We provide a stark
  demonstration of the effects of charge transfer inefficiency on
  resultant data products, which when uncorrected, result in uncertain
  photometry, elongation of morphology in the readout direction, and
  loss of faint sources far from the readout.  We agree with the STScI
  recommendation that future UVIS observations that require very
  sensitive measurements use the instrument's capability to add
  background light through a ``post-flash.'' Preliminary results on
  number counts of UV-selected galaxies and morphology of galaxies at
  z$\sim$1 are presented. We find that the number density of UV
  dropouts at redshifts 1.7, 2.1, and 2.7 is largely consistent with
  the number predicted by published luminosity functions. We also
  confirm that the image mosaics have sufficient sensitivity and
  resolution to support the analysis of the evolution of star-forming
  clumps, reaching 28-29th magnitude depth at $5\sigma$\ in a
  0\farcs2 radius aperture depending on filter and observing epoch.

\end{abstract}


\section{Introduction}

The great success of the GALEX mission \citep{Thilker:2005}
revolutionized the study of galaxies in the ultraviolet (UV). But it has
left us in the curious position of having extraordinary detail on the
UV emission and structure of the closest galaxies (from GALEX) and quite
distant ones (where the UV redshifts into optical bands), but
having significantly less data in between. The rest-frame 1500 \AA\ continuum
(FUV) is an important tracer of star-formation, because it samples the
output from hot stars directly. The star-formation density of the
Universe peaks in the epoch $1<z<3$, which requires deep
near-ultraviolet (NUV; $\lambda \sim 2000-3500$\ \AA) observations to
measure the redshifted FUV. 

A new generation of {\it Hubble Space Telescope}\ (HST) surveys have
been approved to begin filling this gap through deep, high spatial
resolution imaging.  The Wide Field Camera 3 (WFC3)
UVIS channel provides revolutionary sensitivity in the NUV.  Shortly
after installation, the WFC3 team conducted Early Release Science
observations \citep[ERS;][]{Windhorst:2011}, including a first look,
multi-wavelength extragalactic survey.  The ERS included about 50
square arcminutes of NUV imaging, at 2,2,1 orbit depths in the F225W,
F275W, and F336W filters respectively, reaching 26.9 magnitudes (AB). More
recently, the Cosmic Assembly Near-IR Deep Extragalactic Legacy Survey
\citep[CANDELS;][]{Grogin:2011,Koekemoer:2011} has completed
observations with UVIS.  CANDELS observed the northern field of
the Great Observatories Origins Deep
Survey\citep[GOODS][]{Giavalisco:2004}\ with the F275W filter in the
continuous viewing zone, for a total predicted depth of 27.2 magnitudes (AB;
$5\sigma$\ in a 0\farcs2 radius aperture) over 78 square
arcminutes.

\begin{figure}[t!]
\center{
\includegraphics[scale=0.5, viewport=15 5 500 360,clip]{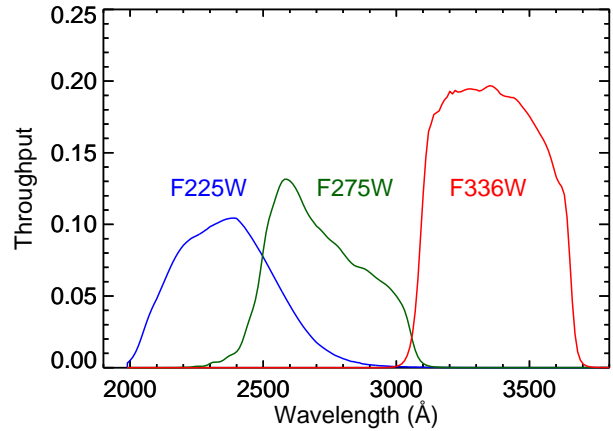}
}
\caption{ \label{fig:filter} Throughput of the WFC3-UVIS filters used
  in the UVUDF: F225W in blue, F275W in green, and F336W in red. These
  throughputs include the quantum efficiency of the CCD.  }
\end{figure}

In this paper, we describe a new program (GO-12534; PI=Teplitz) to
obtain deep, NUV imaging of the Hubble Ultra Deep Field
\citep[UDF;][]{Beckwith:2006}. The UDF provides one of the most
studied fields with a wealth of multi-wavelength data
\citep[][]{Rosati:2002,Pirzkal:2004,Yan:2004, Thompson:2005,
  Labbe:2006, Kajisawa:2006, Bouwens:2006,
  Oesch:2007,Siana:2007,Rafelski:2009, Nonino:2009,
  Voyer:2009,Retzlaff:2010, Grogin:2011, Koekemoer:2011,Bouwens:2011b,
  Elbaz:2011,Teplitz:2011,Koekemoer:2013, Ellis:2013}, enabling the
best return on this new investment of telescope time. This project
obtained deep images of the UDF in the F225W, F275W, and F336W filters
at 30 orbit depth per filter (see Figure \ref{fig:filter}), with the
goal of reaching 28-29th magnitude (AB) depth at $5\sigma$\ in a 0\farcs2
radius aperture.  The program was designed to use $2\times2$\ onboard
binning of the CCD readout to improve sensitivity.  That mode was only
used for the first half of the observations, at which point it became
clear that another strategy is better.  The second half of the
observations were obtained without binning of the CCD readout, but
with the UVIS capability to add internal background light,
``post-flash'', to mitigate the effects of degradation of the charge
transfer efficiency of the detectors.  We will discuss the
implications of these choices for both sensitivity and data
reduction. Combined with previous imaging of the UDF in the
far-ultraviolet \citep[see][]{Siana:2007}, these new observations
(hereafter UVUDF) will complete the pan-chromatic legacy of this deep
field.

We describe the science goals of the project in Section \ref{goals};
survey strategy and observations in Section \ref{obs}; we outline data
reduction and source extraction in Section \ref{data}; we characterize
the data quality and discuss issues related to the charge transfer
efficiency of the CCD in Section \ref{datachar}. In Section
\ref{preview} we describe preliminary analysis of the data and initial
science results, before summarizing in Section
\ref{summary}. Throughout, we assume a $\Lambda$-dominated flat
universe, with $H_0=71$\ km s$^{-1}$\ Mpc$^{-1}$,
$\Omega_{\Lambda}=0.73,$\ and $\Omega_{m}=0.27$.

\section{Science Goals}

\label{goals}

\subsection{Tracing the evolution of star formation}

Observations of UV-bright galaxies trace the evolution of cosmic star
formation and provide key constraints on galaxy formation. The UVUDF
detects galaxies with star formation rates (SFRs) greater than
$\sim0.05~M_{\odot}$/yr at $z\sim2-3$ (in the absence of dust extinction) with the same
UV color selection techniques used at higher redshift.  For example,
the Lyman break galaxy (LBG) selection, whereby high redshift galaxies
are identified by their strong flux decrement at short wavelengths due
to the Lyman break feature, is routinely used in many studies
\citep[e.g.][]{Steidel:1999, Adelberger:2004, Reddy:2008,
  Bouwens:2011b}.  When more photometric information is available,
more complex methods become available (see Section \ref{photoz}).
Measuring the combination of the UV luminosity function and the mass
function of UV-selected galaxies will provide a statistical picture of
the history of star formation in these sources, in redshift slices between
$1<z<2.5$ \citep{Lee:2012}.  UV-selection in this epoch will enable significant
spectroscopic follow-up, with access to vital rest-frame optical
diagnostics of extinction, metallicity, and feedback.  We provide an
initial LBG selection in UVUDF in Section \ref{preview}.

One of the largest sources of systematic error in estimates of the SFR
and the cosmic star formation history is the fact that dust absorbs
and reprocesses approximately half of the starlight in the universe
\citep{Kennicutt:1998b}. The amount of re-radiated light, quantified
by the ratio of integrated IR to UV luminosity,
IRX$\equiv$L$_{IR}$/L$_{UV}$, has been found to be correlated with the
UV spectral index, $\beta$\ (where $f_\lambda \propto \lambda^\beta$),
in local starburst galaxies \citep[e.g.][]{Meurer:1999}.  This
correlation is routinely used to correct UV SFR estimates for dust
attenuation in highly star forming galaxies at all redshifts
\citep[e.g. LBGs and
BzKs;][]{Adelberger:2000,Daddi:2007,Reddy:2010,Reddy:2012a,Kurczynski:2012, Lee:2012b}. 
UV bright galaxies and IR luminous galaxies ($L_{IR} > 10^{11} L_\odot$)
at lower redshifts are found to be broadly consistent with the
starburst IRX-$\beta$ correlation \citep{Overzier:2011}.
Understanding the effects of extinction at high redshift requires
detailed study of normal galaxies 7-10 Gyr into the past (the epoch
probed by the UVUDF), where both the UV slope and the infrared
emission can be measured. 
\citep[e.g.][]{Boissier:2007,Siana:2009,Swinbank:2009,Buat:2010,Bouwens:2012b,Finkelstein:2012a}.

\subsection{The Build-up of Normal Galaxies}

The role (and nature) of feedback, and the relative importance of
merging in galaxy mass growth are still debated issues. Observations
show that ``normal'' galaxies were in place at $z\sim1$, with stellar
population and scaling relations consistent with passive evolution
into the homogeneous population observed in the local Universe
\citep[e.g.][]{Scarlata:2007, Cimatti:2008}. This situation changes
drastically looking back just a few Gyrs. Among the diversity and
complexity of massive galaxy types, two types have been extensively
studied: gas-rich clumpy disks forming stars at rates of 100 \Myr\
that do not have counterparts in the local Universe
\citep[e.g.][]{Daddi:2010a,Elmegreen:2005a,Genzel:2008}, and passive
objects that are observed to be $\sim$30 times denser than any galaxy
today \citep[e.g.][]{Cimatti:2008,vanDokkum:2008}. The former are key
to understanding the role of instability and gas accretion in the
formation of disks and bulges (by migration and merging of the
clumps); the latter are important because we do not yet understand the physics of
quenching of star formation and the role that compactness plays in it.

It is tempting to think of these well-studied populations as different phases in
the formation of local galaxies. Secular evolution of star-forming
sub-structures within gas-rich disks could lead to the formation of
bulges, and the compactness of high$-z$ spheroids would be the result
of the highly dissipative merger of the clumps
\citep{Elmegreen:2008,Dekel:2009b}.  Clumps are predicted to be fueled
by cold ($T< 10^4$\ K) gas streams that efficiently penetrate the hot
medium of the dark matter halo \citep{Keres:2009a}. The UV morphology
of LBGs at $z=3-4$\ are also suggestive of this process
\citep{Ravindranath:2006}. Furthermore, it is still not clear what
mechanism quenches the star formation in the newly formed bulges, what
prevents more gas from cooling and forming stars, and what drives the
size evolution of compact spheroids.

Current HST observations allow us to derive stellar masses, SFR,
surface density of star formation, and the
extinction of individual bright clumps at $z\sim2-3$ by fitting the
spectral energy distribution (SED).  However, without access to the
rest-frame UV, our assessment of star-formation activity becomes
poorer at lower redshifts. At $z\sim2.3$, such structures are found to
have sizes of $\sim 1.8$\ kpc, typical masses of several $10^7
M_{\odot}$, and ages of $\sim 0.3$\ Gyr \citep{Elmegreen:2005b}.

The UVUDF observations are designed to provide the depth and
resolution ($\sim 700$\ pc) to study sub-galactic structures at
$0.5<z<1.5$\ at consistent rest-frame UV wavelengths, offering
continuity with measurements at low and high redshift.  We confirm the
utility of the data for this purpose in Section \ref{preview}. Measurement of
the typical UV sizes and luminosities will constrain stellar-mass and
stellar-population properties using the full SED.  Finally, the data
will enable comparison of the colors of individual sub-galactic units
at different radii for the SF galaxies at $z<2$\ and $z=3$.  A color
gradient would be expected if there is migration of previously formed
structures towards the center to form the bulge.

\subsection{Contribution of galaxies to the ionizing background (below 912 \AA)}

Star--forming galaxies are likely responsible for reionizing the
universe by $z\sim6$, assuming that a high fraction of H{\sc
  i}--ionizing (Lyman continuum; LyC) photons are able to escape into
the IGM.  Recent studies suggest that the escape fraction,
$f_{esc}$\ is higher at high redshift
(\citealt{Shapley:2006,Iwata:2009,Siana:2010,Bridge:2010,Nestor:2013},
but see \citealt{Vanzella:2012a}).  However, it is extremely difficult
to directly measure the LyC at $z>4$ due to the increasing opacity of
the IGM.  Thus, it is important to understand the physical conditions
that allow LyC escape at $2<z<3$\ and to determine if those conditions
are more prevalent during the epoch of reionization.

Ground-based surveys suffer from significant
foreground contamination, and from not knowing from
which part of the source the ionizing emission is escaping.  HST
resolved images of both the ionizing and non-ionizing emission of
galaxies are necessary to confirm the extreme ionizing emissivities
suggested by previous surveys \citep[][]{Iwata:2009,Nestor:2013}.
The UVUDF filters will enable measurement of the LyC escape fraction
at redshifts $z\sim 2.20, 2.45, 3.1$\ in F225W, F275W, F336W,
respectively (see Figure \ref{fig:filter} for the filter throughputs).

\subsection{Improved Photometric Redshifts}

\label{photoz}

Despite intensive spectroscopic surveys that have provided hundreds of
redshifts \citep{Szokoly:2004, LeFevre:2005, Vanzella:2005,
  Vanzella:2006, Vanzella:2008, Vanzella:2009, Popesso:2009,
  Balestra:2010, Kurk:2013}, the majority of sources in the UDF are either too
faint or otherwise inaccessible. Redshifts must therefore be determined
either through color selection or photometric redshifts (photo-z).
However, young star-forming galaxies often lack strong continuum
breaks in the rest-frame optical, making accurate photo-zs nearly
impossible with only optical+NIR data. 

\begin{figure}[b!] \center{
    \includegraphics[scale=0.4]{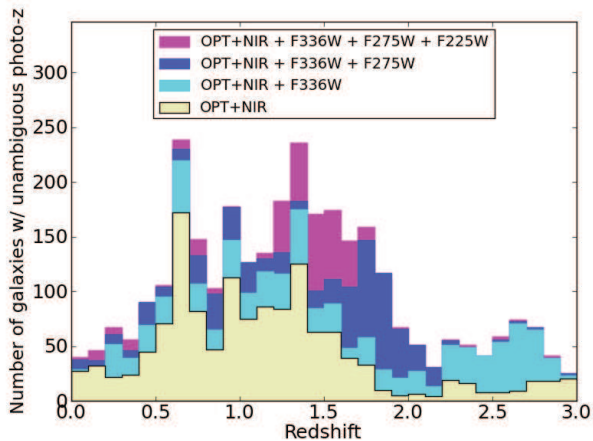}}
  \caption{ \label{fig:photoz} The expected improvement in the number of
unambiguous photometric redshift estimates with the addition of UV filters.
Simulated UV fluxes were added to the catalog of \cite{Coe:2006} assuming
sensitivities of AB=29 in the three UVUDF filters. Photometric redshifts
were then calculated and compared to the results without the UV.  Estimates
determined to have a single, distinct redshift probability peak were
taken as unambiguous.}
\end{figure}

The three UVIS filters target the dominant signature of the galaxies'
SEDs in the redshift range $1.2 \lesssim z \lesssim 2.7$\ -- the Lyman
break. This feature will allow color selection of these galaxies, and
will resolve many of the photo-z degeneracies and thereby improve the
photo-z fits. While photo-z's currently exist for all objects in the
UDF \citep{Coe:2006}, they often have multiple peaks in their
probability distribution functions, $P(z)$, making the true redshift
uncertain. In fact, \citet{Rafelski:2009} found that the introduction
of the ground-based u-band data improved the photo-z's for 50\% of the
$z\sim3$ sample. However, their results suffered from the limited
angular resolution and depth of ground-based u-band data \citep[see
also][]{Nonino:2009}.  The F336W filter significantly improves the
redshifts from $2 \lesssim z \lesssim 3$ and $z\lesssim0.3$, while the
F275W filter improves the redshifts at $1.5 \lesssim z \lesssim 2$ and
$z\lesssim0.2$, and the F225W filter improves them at $1 \lesssim z
\lesssim 1.5$. Figure \ref{fig:photoz}\ shows the expected improvement
in redshift estimation with the addition of UV data with a sensitivity
of AB=29 in each filter. Here we define an unambiguous photometric
redshift as one with 95\% of the probability distribution function ($P(z)$) 
to be within $0.1(1+z)$ with only a single distinct peak in $P(z)$.

\subsection{Star Formation Rate Efficiency of Neutral Atomic-Dominated Hydrogen Gas}

The locally established Kennicutt-Schmidt (KS) relation
\citep{Kennicutt:1998a, Schmidt:1959} relates the gas density and the
SFR per unit area, $\Sigma_{SFR}$ $\propto$ $\Sigma_{gas}^{\beta}$.
While this assumption is reasonable at low redshift for typical
galaxies, it has been shown not to hold for neutral atomic-dominated
hydrogen gas at $z\sim 3$ \citep{Wolfe:2006, Rafelski:2011}.
Nonetheless, cosmological simulations often use the KS relation at all
redshifts, for both atomic and molecular hydrogen gas
\citep[e.g.][]{Keres:2009a}.

Damped Ly${\alpha}$ systems (DLAs; see \citealt{Wolfe:2005} for a
review) are large reservoirs of neutral hydrogen gas. At $z\sim3$, the in situ
SFR of DLAs is found to be less than 5$\%$ of what is expected from
the KS relation \citep{Wolfe:2006}. This means that a lower level of
star formation occurs in DLAs at $z\sim3$ than in modern galaxies.
Another possibility is that in situ star formation may occur at the KS
rate only in DLA gas associated specifically with LBGs. These DLAs are
constrained by measuring the spatially extended low-surface-brightness
(LSB) emission around LBGs. \citet{Rafelski:2011} detect such emission
on scales up to $\sim10$ kpc in a sample of $z\sim3$ LBG's
\citep{Rafelski:2009} in the UDF F606W image (rest-frame FUV). The
emission is measured to $\gtrsim31$ mag arcsec$^{-2}$ and on large
scales by stacking $z\sim3$ LBGs that are isolated, compact, and
symmetric. The resulting SFR around LBGs was found to be $\sim2-10\%$
of what is expected from the local KS relation \citep{Rafelski:2011}.

This result can be used to constrain models of galaxy formation at
$z\sim3$. \citet{Gnedin:2010} conclude that the main reason for the
decreased efficiency of star formation is that the diffuse ISM in high
redshift galaxies contains less dust, and therefore have a lower
metallicity and a lower dust-to-gas ratio, which is needed for
shielding in order to cool the gas and form stars.  This notion
matches the observation that DLA metallicities decrease with redshift
\citep{Rafelski:2012}, and therefore we expect that the efficiency of
star formation may be correlated with redshift.  This effect must be
further understood and taken into account when interpreting
models of galaxy formation and evolution.

The transition from the lower star formation efficiencies at $z\sim3$\
to those on the Hubble sequence at $z\sim0$\ may be observable at
redshifts in between. We plan to find that transition or constrain
when and how it occurs by probing the star formation in the LSB
regions around moderate redshift LBGs.  It is only in the outer
diffuse regions, where the metallicity is lower, that the KS relation
is seen to be evolving.  The NUV coverage of the UDF enables us to
detect this star formation at a range of intervening redshifts by
providing significantly improved photo-z's (Section \ref{photoz}) at
$z\sim 2-3$ in order to identify LBGs to stack in the optical UDF
data, and possibly by stacking the UV data themselves at $z\sim1$, if the
CTE corrected data permit (see Section \ref{ctecorr}).


\section{Observations}

\label{obs}

\begin{figure}[b!] \center{
    \includegraphics[scale=0.33]{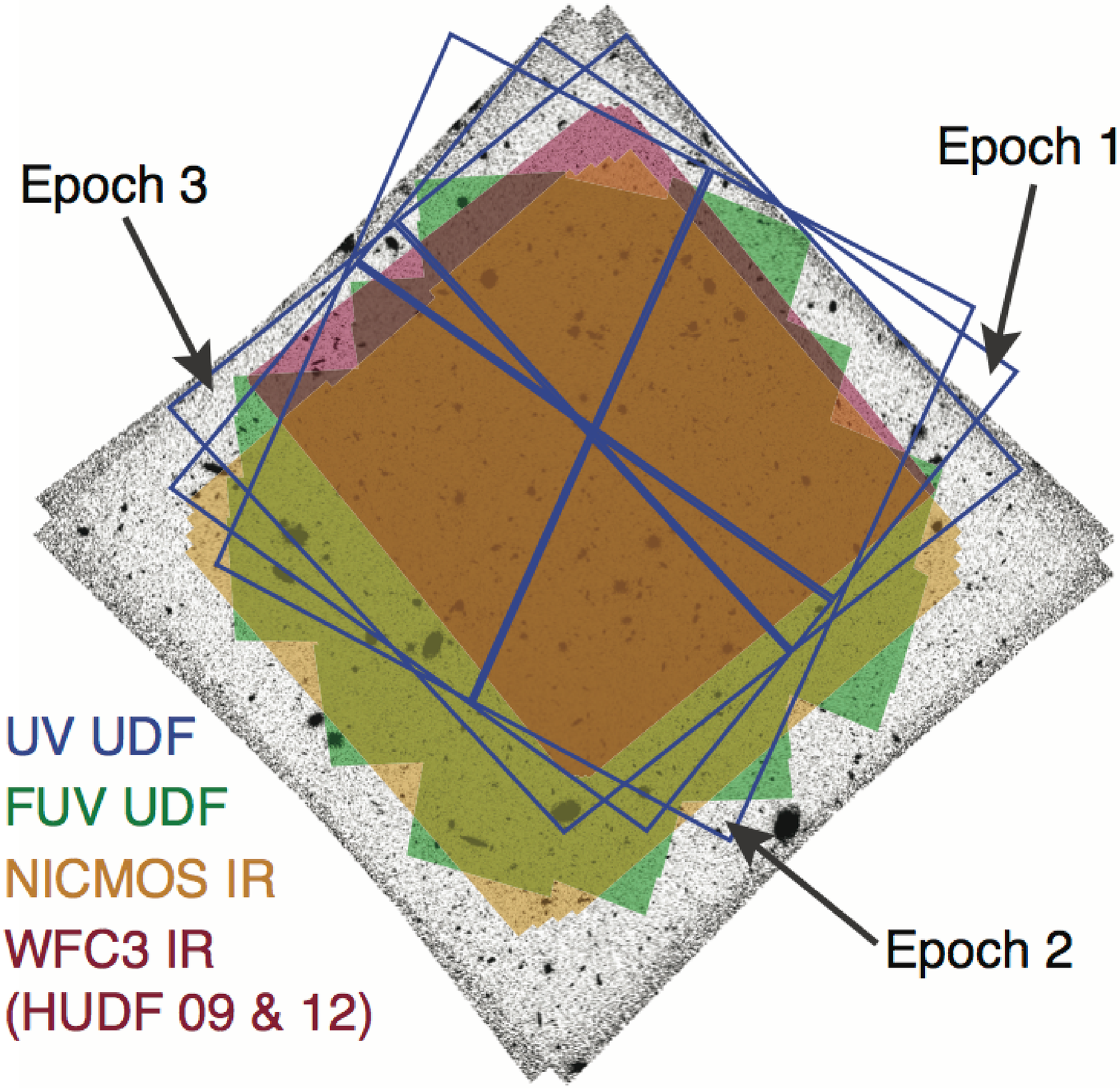}}
  \caption{ \label{fig:uvis_footprint} The footprint of the the UVIS
    pointing for epochs 1, 2, and 3 are shown as blue squares, with
    each epoch individually labeled. The greyscale image is the V-band
    ACS image of the UDF from \citet{Beckwith:2006}, with North up and
    East left. The shaded regions are the footprints of other HST
    imaging of the UDF. The orange represents NICMOS IR
    \citep{Thompson:2005}, the green ACS-SBC FUV \citep{Siana:2007},
    and the red WFC3 near-infrared imaging from HUDF09 and HUDF12
    \citep{Bouwens:2011b, Ellis:2013}. The readout direction is
    perpendicular and away from the blue lines marking the chip gap in
    each epoch, such that the readout is located furthest from the
    chip gap. }
\end{figure}

The UVUDF program was executed in three epochs, due to the heavy
scheduling constraints on HST in Cycle 19 (Fall of 2011 through Fall
of 2012).  Table \ref{tab:epochs} lists the orbit distribution and
position angle of each set of observations.  In each case, a common
pointing center is used: RA: 03 32 38.5471 DEC: -27 46 59.00 (J2000).  Figure
\ref{fig:uvis_footprint} shows the orientation of each epoch compared
to previous UDF programs.

\begin{deluxetable*}{lrrccc}
\tablecaption{UVUDF Observing Epochs \label{tab:epochs}}
\tablehead{
\colhead{Epoch} &
\colhead{Observing Window} &
\colhead{ORIENT\tablenotemark{1}} &
\colhead{Orbits per UVIS filter} &
\colhead{Orbits per } &
\colhead{CCD Readout} \\
\colhead{} & 
\colhead{} &
\colhead{} &
\colhead{F225W:F275W:F336W} &
\colhead{ACS filter} &
\colhead{Mode\tablenotemark{2}} 
}
\startdata

Epoch 1 & March 2$-$March 11 2012 & 96.0 & 6:6:6 & 4:3:11\tablenotemark{a} & binning\\
Epoch 2 & May 28$-$June 04 2012 & 197.25 & 8:8:10\tablenotemark{b} & 20:2:2:2\tablenotemark{c} & binning  \\
Epoch 3 &August 3$-$September 7 2012 & 264.0 & 16:16:14 & 46\tablenotemark{d} & post-flash \\ 
Total   & March - September 2012  & \nodata & 30:30:30 & \nodata & \nodata

\enddata
\tablecomments{List of orbit distribution and position angle for each set of observations.}
\tablenotetext{1}{The ORIENT keyword is defined in Section
  \ref{sec:par}}
\tablenotetext{2}{See discussion in Section \ref{cteintro}}
\tablenotetext{a}{Parallel orbits per filter in order F435W:F606W:F814W.}
\tablenotetext{b}{Due to two failed visits, F336W has 10 orbits per filter, while F275W and F225W have 8. The failed 
visits were re-observed in Epoch 3. }
\tablenotetext{c}{Parallel orbits per filter in order F435W:F606W:F775W:F850L.}
\tablenotetext{d}{Parallel orbits in F435W.}

\end{deluxetable*}

The UVIS focal plane consists of two CCDs, each with $4146\times
2051$\ pixels.  The plate scale is 0\farcs0396/pixel at the central
reference pixel.  After accounting for the overscan regions, the
usable area of each CCD is $4096\times 2051$\ pixels.  There is a
physical gap between the CCDs that corresponds to about 30 pixels
(1\farcs2).

WFC3/UVIS has a field of view of $162^{\prime \prime}\times
162^{\prime \prime}$, larger than the WFC3/IR channel ($136^{\prime
  \prime}\times 123^{\prime \prime}$) but smaller than the optical
field of the Advanced Camera for Surveys' Wide Field Camera
\citep[ACS/WFC; $202^{\prime \prime}\times 202^{\prime
  \prime}$;][]{Ford:2002}. The UVUDF observations are well matched to
the WFC3/IR pointings from two observing programs, as shown in Figure
\ref{fig:uvis_footprint}.  The first program (GO-11563,
PI=Illingworth) was excecuted in 2009
\citep[HUDF09;][]{Oesch:2010a,Oesch:2010b,Bouwens:2011b}. The second
program (GO-12498, PI=Ellis) was executed after UVUDF at the same pointing
as the HUDF09 \citep{Ellis:2013}.  The footprint of previous UV
imaging of the UDF taken with the ACS Solar Blind Channel (SBC)
\citep{Siana:2007} and IR imaging taken with NICMOS
\citep{Thompson:2005} is also shown in the Figure
\ref{fig:uvis_footprint}.

Observations were obtained in visits of 2 orbit duration in order to
maximize schedulability. Each visit used a single UVIS filter. These
visits were linked in groups of 3 in the scheduling instructions to
guarantee that all three filters were obtained at the same
orientation. During each 2-orbit visit, four exposures were taken.
Typically this schedule allowed about 1300 seconds of integration per
exposure. In total, we obtained $\sim82,000$ seconds of integration
per filter in the full overlap region (see Table~\ref{tab:sens}).
Half the data were taken with binning of the CCD readout, while the other
half were taken without binning, but with the use of the post-flash
capability (see Section~\ref{cteintro}). The unbinned Epoch 3
exposures were dithered with the standard WFC3-UVIS-DITHER-BOX, which
is a 4 point dither pattern with a point spacing of 0\farcs173. The
binned Epoch 1 and 2 exposures are dithered in an analogous way, but with
doubled spacing of 0\farcs346.

\begin{deluxetable*}{lcrccclccccccc}
\tablecaption{UVUDF Sensitivities \label{tab:sens}}
\tablehead{
\colhead{Filter} &
\colhead{Zero Point\tablenotemark{a}} &
\colhead{Epoch} &
\colhead{Exposure Time} &
\colhead{$5\sigma$ 0\farcs2 ETC\tablenotemark{b}} &
\colhead{$5\sigma$ 0\farcs2 RMS} &
\colhead{50\% completeness} \\
\colhead{} &
\colhead{(mag)} &
\colhead{} &
\colhead{(s)} &
\colhead{(mag)} &
\colhead{(mag)} &
\colhead{(mag)}
}
\startdata

F225W & 24.0403 & 1\&2 & 39278 &28.3& 28.3& 28.6  \\
F275W & 24.1305 & 1\&2 & 39106 &28.5& 28.4& 28.6  \\
F336W & 24.6682 & 1\&2 & 45150 &29.0& 28.7& 28.9 \\
F225W & 24.0403 & 3 & 44072 &27.8& 27.9& 27.7  \\
F275W & 24.1305 & 3 & 41978 &27.7& 27.9& 27.7  \\
F336W & 24.6682 & 3 & 37646 &28.3& 28.3& 28.2 
\enddata
\tablecomments{UVUDF filters, zeropoints, and sensitivities.}
\tablenotetext{a}{Zeropoint information is available at \url{http://www.stsci.edu/hst/wfc3/phot\_zp\_lbn}.}
\tablenotetext{b}{Exposure Time Calculator (ETC) modified to work with binned and post-flashed data.}

\end{deluxetable*}

An exception to the observing plan occurred in two visits (``1N'' and
``2H'' in the HST schedule), resulting in the loss of
both visits in Epoch 2, one for F275W and one for F225W.  These visits
were rescheduled during Epoch~3 (as visits ``5N'' and ``6H''), and executed as
planned at that time.

The area of full overlap between dithered exposures, and thus full
sensitivity, is 6.2 arcmin$^2$, or 86\%\ of the area of the UVIS
detector. The full NUV UVIS overlap region and all of Epoch 3 are
completely covered by the deep ACS optical data. The footprint of the
UVIS pointing is overlaid on the ACS F606W image of the UDF in 
Figure~\ref{fig:uvis_footprint}.  The full WFC3/IR pointings (HUDF09 and
HUDF12) are covered by the NUV UVIS data.

\subsection{Charge Transfer Inefficiency}
\label{cteintro}

Over time, radiation causes permanent damage to the CCD lattice,
decreasing the charge transfer efficiency (CTE) during readout. The
CTE degradation is a serious problem for low background imaging of
faint sources, resulting in decreased sensitivity and uncertain
calibration for extended sources. The effect is worse for objects that
are far from the CCD readout, that is for objects close to the gap
between the two detectors in the case of UVIS. The degradation of the
UVIS CCDs has been {\it faster} than in the early years of ACS,
already causing significant signal loss of $\sim20$\% in moderately
bright sources (those with $\sim1000$ e$^{-}$/read), and $\sim 50$\% for somewhat
fainter sources \citep[those with $\sim300$ e$^{-}$/read;][]{Noeske:2012}. This faster
degradation is believed to be due to the extreme solar activity
minimum, and resulting cosmic ray maximum, during the initial flight years of UVIS. The resulting loss of
data quality can be partially mitigated by post-processing. The effect
is worse for very faint sources, which can be completely lost to
``traps'' before readout \citep{MacKenty:2012,Anderson:2012c} and
cannot be recovered later.  In the literature, CTE degradation is
often referred to and measured as charge transfer inefficiency (CTI =
1-CTE) \citep[e.g.][]{Massey:2010b}, and we use this terminology
interchangeably below.

After Epoch 2 of the UVUDF had already been obtained, the Space
Telescope Science Institute (STScI) 
released a new report on mitigating CTI \citep{MacKenty:2012}.  The
strong recommendation is to use the ``post-flash'' capability of the
instrument to illuminate the detector and add background light to the
observation.  This additional background will fill the traps and
ensure that faint objects are not lost, as well as significantly improve the 
accuracy of pixel-based CTE corrections.  This benefit comes at the
cost of decreased sensitivity, however, due to the noise introduced by
the added background.

Considering that many of the science goals of the UVUDF rely on
measuring (or setting limits on) the faintest sources, and require
accurate photometry, we chose to follow the recommendation for
post-flash.  In Epoch 3, we applied a post-flash to bring the average
background (the sum of post-flash, sky and dark current) up to about
13 electrons per pixel.  In practice, this meant adding 11e$^{-}$ in
F225W and F275W, and 8e$^{-}$ in F336W.  The spatial distribution of
post-flash light is not uniform \citep{MacKenty:2012,Anderson:2012c},
so target levels were set to ensure both a reasonable average and a
sufficient background in the less illuminated regions.

\subsection{Binning the CCD Readout}

Without post-flash, the UVIS detectors are read-noise limited in the
F225W and F275W filters, even in long exposures such as those needed
for the UVUDF.  The noise from the readout and from the sky background
is about equal in F336W.  As a consequence, there is the potential for
tremendous sensitivity gain by binning the CCD pixels $2\times2$\
during readout.  In principle, $2\times2$ binning results in a gain of
a factor of 2 in signal to noise (S/N) ratio, or 0.75 magnitudes.  One
concern in the decision to bin the CCD readout is the loss of spatial
resolution.  However, the large number of repeated observations allow
for excellent sub-pixel image reconstruction.

Once the post-flash capability became available to mitigate the effects of CTI,
the benefit of binning the CCD readout was greatly reduced.  At that
point, the signal-to-noise gain would be under 20\%, while still
reducing the spatial resolution.  As a result of these considerations,
we chose to take the second half of the UVUDF data, that is Epoch 3,
without binning the CCD readout.

\vspace{5mm}

\subsection{Parallel ACS Observations}
\label{sec:par}

\begin{figure}[b!]
\center{
\includegraphics[scale=0.26]{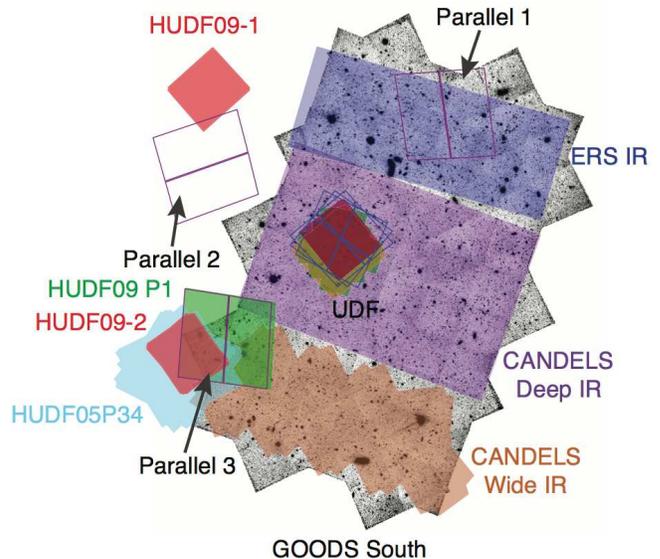}}
\caption{ \label{fig:acs_par_footprint} The footprint of the the ACS
  parallel pointings for Epochs 1, 2, and 3 are shown as purple
  squares.  The greyscale image is the V-band ACS map of GOODS-South
  from \cite{Giavalisco:2004}, with North up and East left.  The blue 
  squares and nearby shaded
  regions indicate the footprint of the UVUDF UVIS pointings and
  complementary data from Figure \ref{fig:uvis_footprint}.  The blue
  shaded region indicates the footprint of the ERS imaging
  \citep{Windhorst:2011}, the purple and brown indicate the footprint
  for CANDELS Deep and Wide respectively \citep{Grogin:2011}, and the
  shaded red regions indicate the footprint of the
  near-infrared imaging from the HUDF09 \citep{Bouwens:2011b}. The
  green shaded region indicates the footprint of the HUDF09 parallel
  1, and the cyan shaded regions represents the HUDF05 parallel P34
  \citep{Oesch:2007}. }
\end{figure}

Coordinated parallel exposures were obtained with the ACS/WFC3 during
the primary WFC3/UVIS observations. Figure
\ref{fig:acs_par_footprint}\ shows the location of the parallel fields
in comparison to other data in GOODS-South.  Table \ref{tab:epochs} gives
the specification for each parallel field, with position angle
specified by the HST ORIENT keyword, which is the position angle of
the U3 axis, where U3$=-$V3. The V3 angle is an angle based on the
reference frame of the telescope, where V3 is perpendicular to the
solar-array rotation axis. This angle describes the angle of rotation
of the WFC3 UVIS exposure on the sky, and the position and rotation of
the parallels.

The Epoch 1 parallel exposures fall within the ERS field. The Epoch 2
parallel exposures fall outside of the main CANDELS and GOODS
footprint, but still within the field observed by the GEMS program
\citep{Rix:2004}, and the ground-based U- and R-bands
\citep{Nonino:2009}.  Scheduling constraints did not permit a more
favorable orientation.  The Epoch 3 orientation was chosen
specifically to place the parallel field at the position of the
one of the parallels to the HUDF09 (the HUDF09-2 parallel field). 
The distribution of exposures per filter varies with the position of the
parallel data (see Table \ref{tab:epochs}).

The parallels of Epoch 1, which fall within the ERS, consist of 18
orbits.  Given the depth of existing data in that field, we chose to
obtain images with the three standard optical filters F435W, F606W, F814W.
Four orbit depth was obtained in the B-band (F435W), to more than
double the previous imaging exposure time.
The V (F606W) and wide I-band (F814W) exposure times were chosen so
that when combined with previous imaging, the ratio would be $\sim$1:2,
following the strategy for parallel imaging in CANDELS \citep[][their Section 6]{Grogin:2011}.

For Epochs 2 and 3, we obtained very deep B-band imaging.  There is a
growing recognition that HST's UV and blue optical capabilities are a
unique resource which should be used now to prepare for later years
when space-based observing will be focused on the near-infrared, with
missions such as the James Webb Space Telescope \citep{Gardner:2006},
the Wide Field Infrared Survey Telescope \citep{Dressler:2012a}, and
Euclid \citep{Laureijs:2012}.  With 20 and 46 orbits in Epochs 2 and 3
respectively, we obtained deep-field quality images.  At the position of the
Epoch 3 parallel field, the HUDF09-2 field has already been
observed for 10 orbits in the B-band \citep{Bouwens:2011b}, enabling a combined deep pointing of 
56 orbits of observation.  For comparison,
the original ACS/WFC3 B-band images of the UDF were also obtained in 56
orbits.  We note, however, that the detector performance was better at
that time (see Sections \ref{cteintro} and \ref{cte}).  In Epoch 2, we
also obtained shallow imaging in the V (F606W), $i$\ (F775W), and $z$\
(F850LP) filters, to augment the shallower imaging available from
GEMS.  The failed visits described in section \ref{obs} shifted 4
orbits from planned B-band exposures in Epoch 2 to Epoch 3.


\section{Data Reduction}
\label{data}
The UVUDF data set consists of four exposures and 2 orbits per visit,
with visits divided into three observing epochs as described above.
In total, there are 15 visits (30 orbits) for each of the three
filters.

In this section, we describe the data reduction process needed to
produce science quality images from the UVUDF observations.  We plan
to release fully reduced images and catalogs at a later time, but not
in combination with this paper (see Section \ref{summary}).
Nonetheless, it is important to document the many issues with the data
from this HST Treasury program, and for the reader to understand the
process that led to the images used for the analysis in the later
sections of the paper.  The same lessons learned here will be relevant
to planning of future UVIS observations.

Binned and unbinned data (and data with and without post-flash) must be
processed differently, and they require
different calibration files. The software pipeline that we use for
UVUDF data begins with the standard
Pyraf/STSDAS\footnote{Further documentation for all the PyRAF/STSDAS
  data reduction software is provided at
  \url{http://stsdas.stsci.edu/}} {\sc calwf3}\ modules, though
calibration files needed to be constructed with some care 
as described in this section. The processing of ACS parallel data
closely follows the procedures used by CANDELS \citep{Koekemoer:2011},
and is not described here.

\subsection{Calibration Pipeline}
\label{calpipe}

Calibration exposures (darks, biases, flat fields) for UVIS are
obtained by STScI as part of the standard calibration observations.
In most cases, these calibrations are taken without binning the CCD readout,
though a few binned calibration observations have been obtained.  The
CCD detectors are periodically heated in order to mitigate hot pixels
that develop over time, called annealing. Specifically, $\sim500$ new
hot pixels appear per day, while the annealing process removes
$\gtrsim70$\% of hot pixels \citep{Borders:2009}. The number of
permanent hot pixels that can not be fixed by anneals is growing by
0.05-1\% per month (WFC3 instrument handbook). In order to minimize
the number of hot pixels at any given time, the detector is annealed
once per month.

New calibration files are needed for the calibration pipeline: new
biases, darks, and flats for the binned data, and new darks for the
unbinned data. Only the bias files used data that were taken with
onboard binning. In the other cases, unbinned calibration data are the
basis of creating new files, with after-the-fact binning applied where
necessary. We validated this latter procedure using the limited set of
onboard-binned calibrations that are available. We use a combination
of custom scripts and standard STSDAS routines to make these
calibration files. The steps involved to construct each type of
calibration file are described below.

The standard calibration pipeline begins with an overall bias correction,
calculated by fitting the overscan region in a master bias frame, and
removing the electronic zero point bias level. Next, a bias reference
frame is subtracted from the full image to correct for pixel-to-pixel
bias structure. For the binned Epoch 1 and 2 data, this reference file
is created using the STScI software wfc3\_reference.py
\citep{Martel:2008} to average ten onboard-binned bias frame exposures
(Baggett, CAL-12798). For the unbinned Epoch 3 data, we use the
standard unbinned bias frames provided by STScI.

The next calibration step is the subtraction of a dark reference file
to correct dark current structure and to mitigate hot pixels that
can cause significant artifacts in the images.  STScI releases new
darks every 4 days that are based on the average of $\sim10-20$ dark
exposures with integration times of $\sim900$s each. This is necessary due to the large number
of new hot pixels per day, and the drastic change in hot pixels after
each anneal. However, binned darks are not obtained on a regular
basis. Therefore, unbinned darks are binned after the fact using
custom IDL scripts. We validate this approach by measuring the dark
current in one set of binned dark exposures obtained for this purpose
(Baggett, CAL-12798).

We find the standard processing of the dark calibration is
insufficient for the UVUDF data. The STScI-processed darks were
created with two choices that are not optimized for this case. First,
the process uses an unaggressive definition of a hot pixel as a
$\sim10\sigma$ deviation. The choice results in warm-to-hot pixels not
being masked in the UVUDF images, which add significant artifacts to
the highly sensitive mosaics. This effect is augmented by CTI causing
many hot pixels and their CTI trails to fall below this threshold designed
for data without CTI issues.
Second, the standard processing uses the
median value of the average darks (with hot pixels masked) as the
value of all pixels in the dark frame. This median-value dark with hot
pixels is the calibration file available from STScI. It is not
suitable for UVUDF data, because there is a low-level gradient present
in the dark that is not subtracted. This gradient is typically small
compared to the sky background in the optical, with a peak-to-valley
deviation of $\sim3$ e$^-$/pix/hr. However, in the low-background NUV
images, it is the dominant structure. While this background can be
corrected in the background subtraction phase of the pipeline, it is
more accurate to model this background and subtract it before dividing
by the flats.

We therefore reprocess the darks, starting with the raw dark
observations, using a procedure based on the one provided to us by
STScI (J. Biretta, private communication) and the wfc3\_reference.py
code \citep{Martel:2008}. We make two significant modifications to the
STScI procedure \citep{Borders:2009} to fix the issues identified
above. First, we use an iterative $\sim3\sigma$ cutoff for defining a
hot pixel, applied to cosmic-ray cleaned darks made from the average
of a minimum of 10 exposures. This change significantly increases the
number of hot pixels masked ($\sim7\%$ of the image), but decreases
the extra systematic noise. Secondly, we fit a $7^{th}$\ order
polynomial to the remaining non-flagged pixels in the image of
each UVIS CCD. Then, we create a final dark frame by combining this
polynomial fit, as the background value, and the hot pixels
superimposed and flagged in the data quality array.  In the case of
the binned data, the new darks are binned after the fact.

The next calibration step is the application of the flat field
reference files. For the binned data, flats are binned after
the fact from the unbinned calibration data.  For the unbinned data, the flats
provided by STScI are applied. The final calibration step is
populating the photometry keywords in the FITS header using the current filter throughput
curves and detector sensitivity information using {\sc calwf3}.

The last processing step is the background subtraction of the individual
calibrated images. The unbinned data have an artificial background
introduced by the post-flash process. We subtract the post-flash
reference files provided by STScI from the unbinned data. These reference files
are generated by STScI from stacks of post-flashed exposures, and then
scaled to the flash count rate when applied to the data. However, both
these post-flash-subtracted images, and the binned images, have a
residual nonuniform background.  We therefore fit a
background to each individual image via a custom inverse distance
code. This code masks large fractions of each image for cosmic rays,
sources, hot pixels, and bad pixels. It then interpolates the
background value at any given pixel based on an inverse-distance
weighting within a subgrid region. These backgrounds are then
subtracted from all science images. The final products of the
calibration pipeline are basic calibrated background-subtracted
images, together with data quality maps, that can be used as input to the
mosaicking pipeline.

Image registration and mosaicking are performed following the
procedures used for CANDELS.  We refer the interested reader to
Koekemoer et al. (2011).  UV mosaics are registered to the ACS B-band
image \citep{Beckwith:2006}. 


\subsection{Object Detection and Photometry}
\label{objdet}

We use the Source Extractor software version 2.5
\citep[SExtractor;][]{Bertin:1996} for object detection and
photometry. SExtractor is used in dual image mode, where objects are
detected in the deeper F435W (B-band) mosaic \citep{Beckwith:2006},
and the photometry is measured in a combined Epoch 1 and 2 mosaic and
Epoch 3 mosaic for each filter. In this way, colors of sources are
measured using the same isophotal apertures, and fluxes are measured
for all B-band detected objects regardless of any flux decrement in
the NUV mosaics due to the Lyman Break. Edge regions and the central
chip gaps of the mosaics are excluded, and are set to the sky level
with the same noise properties as the mosaics such that SExtractor
does not find spurious sources along the edges or in the central chip
gap.

The detection parameters for the B-band mosaic are tuned such that no
sources are detected in the negative image. This is accomplished by
setting the minimum area of adjoining pixels to 9 pixels, and a
1$\sigma$ detection threshold.  A Gaussian filter is applied on the
mosaics, with a full width at half maximum (FWHM) of 3 pixels for
object detection.  SExtractor is provided an RMS weight map for both
the detection and analysis image. The gain parameters are set to the
exposure time, such that SExtractor calculates the uncertainties
properly. All source photometry has the the local background
subtracted by SExtractor, using a local annulus that is 24 pixels wide
(with the inner radius depending on source size).  Zero points of
24.0403, 24.1305, and 24.6682 are applied for the F225W, F275W, and
F336W mosaics respectively (see Table \ref{tab:sens}). 
We note that since the B-band is significantly more sensitive than the
NUV images, the resulting catalog contains B-band objects too faint to
be measured in the UV, and thus cuts on the catalog are used as needed
for each scientific purpose.

The photometry of objects is measured with SExtractor using both
isophotal and \citet{Kron:1980} elliptical apertures. Isophotal
apertures are used whenever measuring the color of a source, such as
in the color-color selections used in section \ref{lymanbreak}. For
this purpose, we also run SExtractor on the F606W (V-band) mosaic
\citep{Beckwith:2006} in dual image mode, still using the B-band 
as the detection image. This procedure results in aperture-matched photometry,
although it is not corrected for variation in the point-spread
function (PSF). Since the PSFs of the NUV and the
optical B- and V-bands are quite similar, this correction will be small
for these bands. For this overview paper, these color measurements are
sufficient. Uncertainties will be dominated by CTI effects (see
section \ref{cte}).  Kron elliptical apertures are used to measure the
total magnitude of each source via SExtractor's MAG\_AUTO
parameter. These magnitudes measure the total flux from a source, and
are used whenever a total magnitude is needed, such as in the number
counts of LBGs (see section \ref{lymanbreak}).

\section{Data Characterization}

\label{datachar}

\subsection{CTI effects}

\label{cte}

Radiation damage sustained by the CCD degrades its ability to transfer
electrons from one pixel to the next, trapping electrons (in part
temporarily) during readout, while other electrons are moved to the
next pixel. This results in trails of electrons in the direction of
the CCD readout, with regions of the CCD furthest from the readout
affected most severely \citep[e.g., ][and the references therein]{Rhodes:2010, Massey:2010a}. 
The three different orientations of the three
UVUDF epochs enables the measurement of CTI effects in the
data. Specifically, Epoch 1 and 2 are at an angle of 101.25 degrees
relative to each other, resulting in some galaxies located close to
the readout in one epoch, and far from the readout in the other (see
Figure \ref{fig:uvis_footprint}). This configuration allows the
characterization of the effect of CTI on the photometry and
morphology, as well as an estimate of the number of faint galaxies
that are completely lost.

\subsubsection{Corrections for CTI}
\label{ctecorr}
There are currently two methodologies to correct the photometry for
CTI losses. The best method is a pixel-based CTE correction of the raw
data based on empirical modeling of hot pixels in dark exposures
\citep{Anderson:2010, Massey:2010a}. Such a correction not only
corrects the photometry, but also restores the morphology of sources
(see section \ref{morph}). A preliminary version of software to implement
such a correction for unbinned WFC3/UVIS data was released in March
2013\footnote{For more information about the pixel-based CTE
  correction for WFC3/UVIS, see
  \url{http://www.stsci.edu/hst/wfc3/tools/cte\_tools}}, but
significant improvements and verification will be needed before the
correction is stable enough to warrant the public release of corrected
high level science products for the UVUDF. There are three major
issues that need to be overcome for the software to fully support the
UVUDF data: 1)~The code only works for unbinned data, and half the
UVUDF data are binned. 2)~The current algorithm over predicts the CTE
correction for low background faint sources \citep{Anderson:2013a},
and the binned half of the data have very low backgrounds. 3)~Read
noise mitigation in the algorithm results in under-correction for
faint sources \citep{Anderson:2013a}. The WFC3 team at STScI is aware
of the latter two limitations and is working on improvements. In
addition, while the post-flash Epoch 3 data can have the CTE algorithm
applied in a straightforward manner, post-flashed CTE corrected darks
are required to match the hot pixels.

The second method to mitigate the effect of CTI is to apply a
correction to the measured flux densities of sources, based on their
location on the detector, the observation date, and their flux in
electrons \citep[e.g.][Bendregal et al. 2013]{Cawley:2001, Riess:2004,
  Rhodes:2007,Noeske:2012}. However, the current WFC3 UVIS
implementation of this catalog-based calibration \citep{Noeske:2012} has many
limitations. First, it can only be applied for a small number of
quantized background levels, including virtually no background,
$\sim3$e$^-$/pix, and 20-30e$^-$/pix. Thus, it is only applicable
to the UVUDF Epoch 1 and 2 data for F275W and F225W, and these
corrections have slightly higher backgrounds than the UVUDF data. The
F336W data and all the Epoch 3 data have backgrounds that are not
similar to any of the standard calibrations. The poor sampling of
background levels in the calibrations makes interpolating between them
unreliable. Second, the calibration was measured for relatively bright
point sources, and the correction is uncertain at the faint end, which
encompasses the majority of UVUDF sources. Third, it does not take into
account other nearby sources which can fill charge traps and thereby
shield the sources. Lastly, it does not take into account the
morphology of sources (i.e. size and shape), and therefore does not
account for effects such as self-shielding that accompany non-point
sources, as electrons from the part of the source closer to the
readout will shield the other part from charge traps.

Keeping these several limitations in mind, we apply the
\citet{Noeske:2012} correction to the F225W and F275W mosaics of
Epochs 1 and 2 separately. This correction enables us to refine our investigation
of the effects of CTI (e.g. Sections \ref{phot} and \ref{loss}).
However, we can not apply the calibration to the {\it combined} Epoch 1 and
2 mosaic nor the F336W mosaics, so the science investigations in
Section \ref{lymanbreak} do not include the correction. Those
investigations use the combined Epoch 1 and 2 mosaic, which partially
mitigates the CTI effect because objects far from the readout in one
epoch are averaged with their counterparts closer to the readout in
the other epoch. Future work using this data will apply the
pixel-based CTE correction (when it is stable) to obtain more reliable
photometry.

\subsubsection{CTI effects on photometry}
\label{phot}

In order to characterize the CTI in the UVUDF data, a new catalog was
created, with a method differing from that described in section
\ref{objdet}. For each single epoch mosaic in each filter, SExtractor
was run in dual image mode, with the combined Epoch 1 and 2 mosaic as
the detection image. The detection threshold was set such that we do
not detect sources in the negative image. Objects near the edges or
near the chip gaps for any of the three epochs were excluded, and
objects were required to be covered by all three epochs of
observation. The catalog was trimmed to only include sources with S/N
ratios greater than 5$\sigma$ in all three single epoch mosaics.
Galaxies in the NUV images are often clumpy, which results in single
galaxies appearing as multiple clumps in the images. Regardless of the
deblending parameters used with SExtractor, these galaxies are
detected as separate sources. This is not an issue for the CTI
measurements described below, and the main catalog is not strongly
affected by this, since the B-band is used as the detection image in
that case.

The effects of CTI are worst in exposures with low background
\citep{MacKenty:2012}, thus the measured UVUDF CTI effects are
described for F275W, which has a lower background than F336W, yet
sources are brighter than in F225W, enabling us to measure more
sources.  Specifically, the unbinned equivalent average backgrounds
are $\sim5.8$ e$^-$/pix/hr, $\sim6.2$ e$^-$/pix/hr, and $\sim12.2$
e$^-$/pix/hr for F225W, F275W, and F336W exposures, corresponding to
$\sim2.4$ e$^-$/pix, $\sim2.5$ e$^-$/pix, and $\sim5.1$ e$^-$/pix in
the half orbit exposures used. The backgrounds in F225W and F275W are
consistent with the expected value due to dark current. The CTI
effects are present in all three bands, but expected to be
at a lower level in the higher-background F336W mosaics. The basic
effect of CTI on the photometry is that the objects lose a fraction of
their flux proportional to their distance from the readout, as
electrons encounter more charge traps the further they travel.

\begin{figure}[b!]
\center{
\includegraphics[scale=0.5, viewport=15 5 500 360,clip]{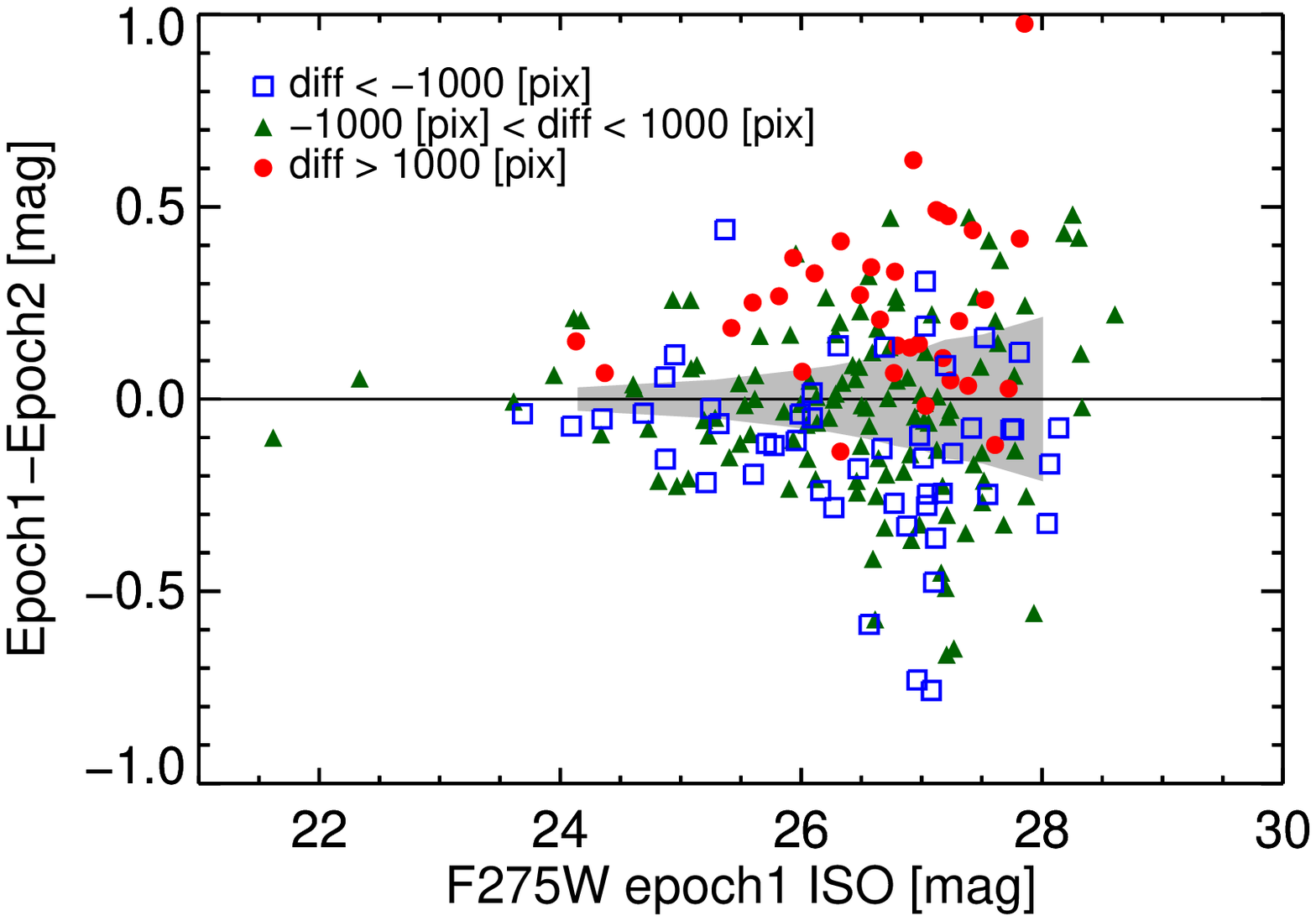}
\includegraphics[scale=0.5, viewport=15 5 500 360,clip]{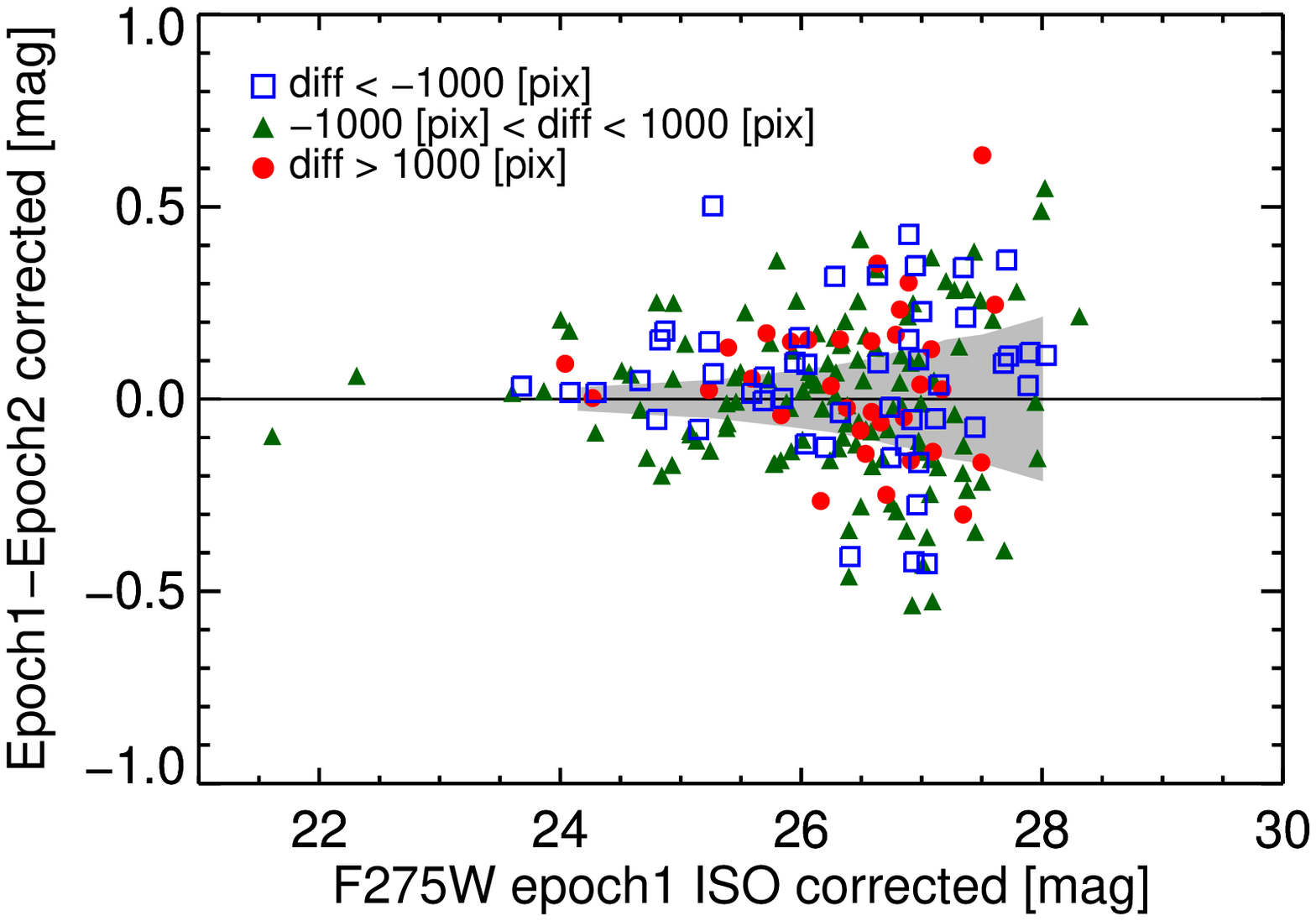}
}
\caption{ \label{fig:magepoch12} Photometry comparison of sources in
  Epochs 1 and 2 F275W mosaics illustrating a larger photometric
  scatter than expected from the uncertainties, likely due to CTI
  effects.  For objects far from the readout, the actual 1$\sigma$ dispersion 
  is larger by more than a factor of 2 than expected. Both panels are the same except that the
  bottom panel includes a catalog based CTE correction for both Epochs
  1 and 2 assuming point source morphology (see Section
  \ref{ctecorr}).  The difference in isophotal magnitudes between
  Epochs 1 and 2 should be zero, with a scatter that increases with
  increasing magnitude.  The black line is the zero difference line,
  and the expected 1$\sigma$ dispersion is shown as the gray shaded
  region from uncertainties as measured by Source Extractor. 
  The colors denote the difference in source distance to readout between
  the epochs.  The blue open squares are sources close to the
  readout in Epoch 1 and far from the readout in Epoch~2, the green
  filled triangles are sources an intermediate distance from the readouts, and the
  red filled circles are sources far from the readout in Epoch 1 and close to the
  readout in Epoch 2. }
\end{figure}

The uncorrected photometry of Epochs 1 and 2 are compared in the top
panel of Figure \ref{fig:magepoch12}, which plots the difference in
isophotal magnitude of Epoch 1 and 2 as a function of the Epoch 1
isophotal magnitude. The scatter is much larger than the expected
1$\sigma$ dispersion (shown as the gray shaded region) likely due to
the effects of CTI. For objects far from the readout, the actual 1$\sigma$ dispersion 
is larger by more than a factor of 2 than expected. 
The photometric scatter is characterized as a
function of the difference in distance to the readout between the
epochs, as measured on the drizzled images. When the difference
is a large negative number, the sources are close to the readout in
Epoch~1 and far from the readout in Epoch~2 (open blue squares). When the
difference is a large positive number, the sources are close to the
readout in Epoch~2 and far from the readout in Epoch~1 (red filled circles).
If CTI is the cause of the large scatter, then the expected behavior
is for the blue squares to be primarily below the zero line, and the
red circles to be primarily above the zero line. This behavior is
indeed what is observed, confirming that CTI is the most likely cause
of the large observed scatter.

The CTE-corrected photometry of Epochs 1 and 2 are compared in the
bottom panel of Figure \ref{fig:magepoch12}, which plots the same
quantities as the top panel, with the addition of a catalog-based CTE
correction (see Section \ref{ctecorr}). The CTE correction reduces the
scatter observed in the top panel, and it removes the systematic
offset of the red circles furthest from the readout. However, the
scatter remains larger than expected, possibly due in part to the
limitations of the catalog-based CTE corrections described in Section
\ref{ctecorr}. On the other hand, the scatter could result from
imperfect image registration, or CTI effects on source morphology
causing inappropriate apertures to be used in the photometric
measurements (see Section \ref{morph}). The image registration is
unlikely to be the cause, because the Epoch 1 and 2 have relative
astrometric accuracy of
better than 0\farcs05.
It is possible that the CTI effects on source morphology is the cause, though the use
of the combined Epoch 1 and 2 mosaic as the detection image 
somewhat reduces this effect (but see Section \ref{morph}).

\begin{figure}[t!]
\center{
\includegraphics[scale=0.5, viewport=15 5 500 360,clip]{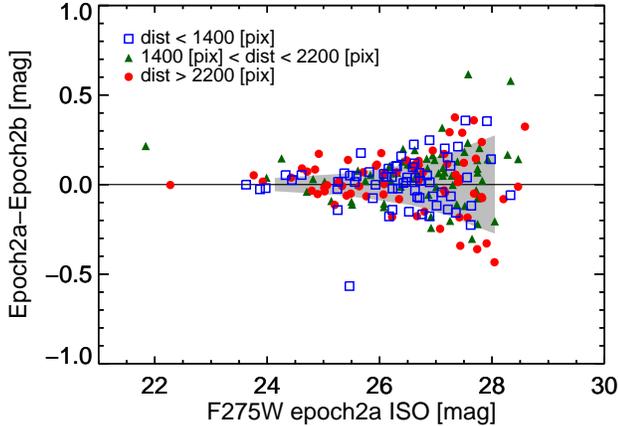}}
\caption{ \label{fig:magepoch2ab} Photometry comparison of Epochs 2a
  and 2b F275W with a photometric scatter mostly consistent with that
  expected from the uncertainties. No correction for CTI is applied,
  because the correction would be a function of distance to the
  readout and therefore the same in both halves of the epoch 2 data.
  The expected 1$\sigma$ dispersion
  is shown as the gray shaded region.
 The blue open squares are sources close to the readout ($<700$ pixels away), the green filled triangles are
  sources an intermediate distance from the readout ($>700$ pixels and
  $<2200$ pixels away), and the red filled circles are sources far from the readout
  ($>2200$ pixels away). Regardless of distance to the readout, source
  magnitudes are mostly consistent with the 1$\sigma$ scatter (gray
  shaded region). }
\end{figure}

We test the hypothesis that something other than CTI is the cause of
the scatter by making a comparison that is mostly insensitive to the
distance to the readout.  We compared two subsets of the Epoch 2 data,
each consisting of half the exposures (2a and 2b). Figure
\ref{fig:magepoch2ab} plots the difference in isophotal magnitude
between the two halves of the Epoch 2 data as a function of the Epoch
2a isophotal magnitude. The points are color coded by distance to the
readout, as no difference in readout distance exists.  Regardless of
the distance to the readout, magnitude differences are consistent
with random scatter, although with a slightly larger magnitude than
expected from the measurement uncertainties (gray shaded 1$\sigma$
dispersion). This minor remaining difference is most likely due to a slight
underestimation of the uncertainties by SExtractor, possibly caused by
SExtractor not including the uncertainty in local sky subtraction. It
has been noted several times in the literature that SExtractor
underestimates the true uncertainties
\citep{Feldmeier:2002,Labbe:2003,Gawiser:2006,Becker:2007,Coe:2013}.

\begin{figure}[t!]
\center{
\includegraphics[scale=0.5, viewport=15 5 500 360,clip]{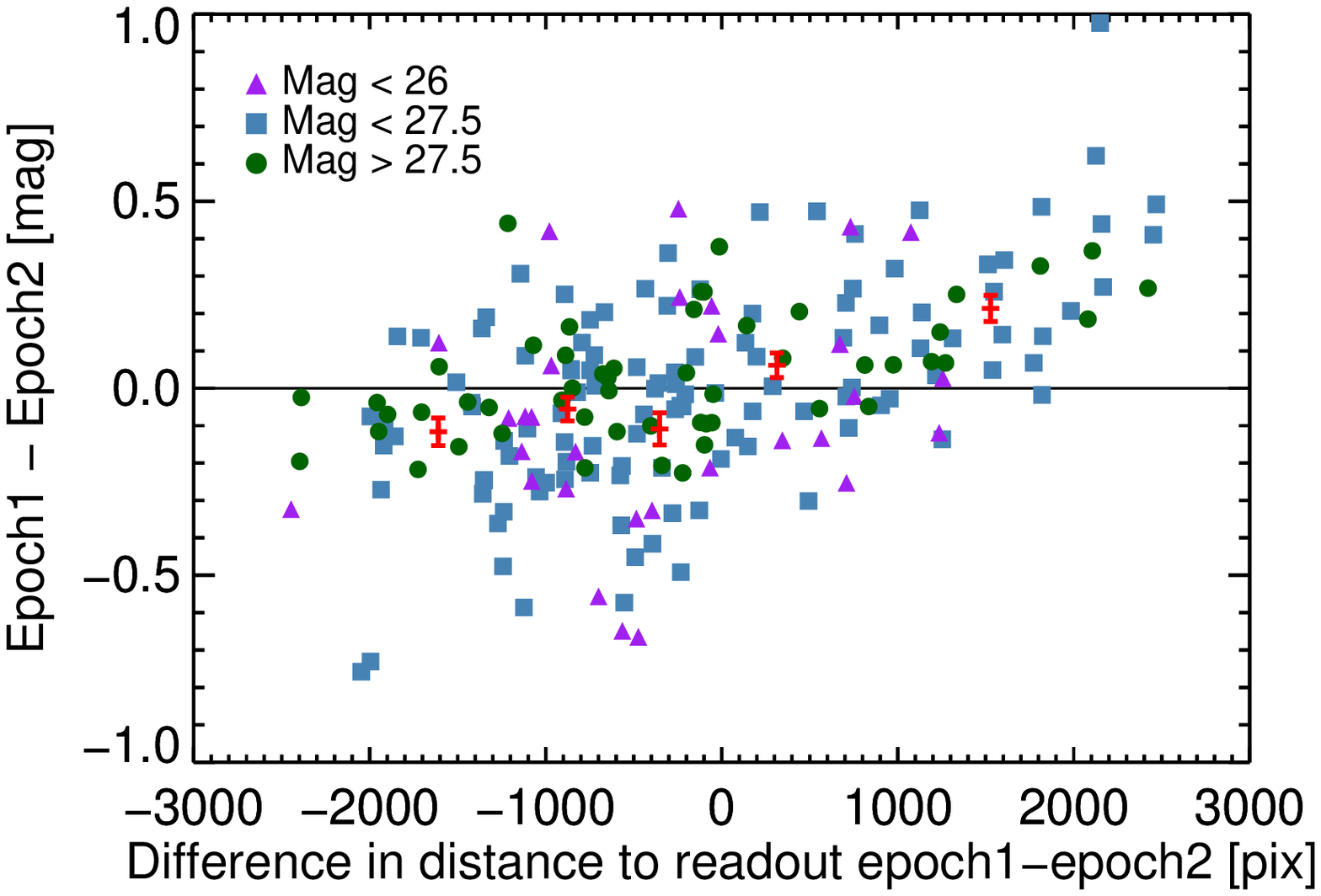}
\includegraphics[scale=0.5, viewport=15 5 500 360,clip]{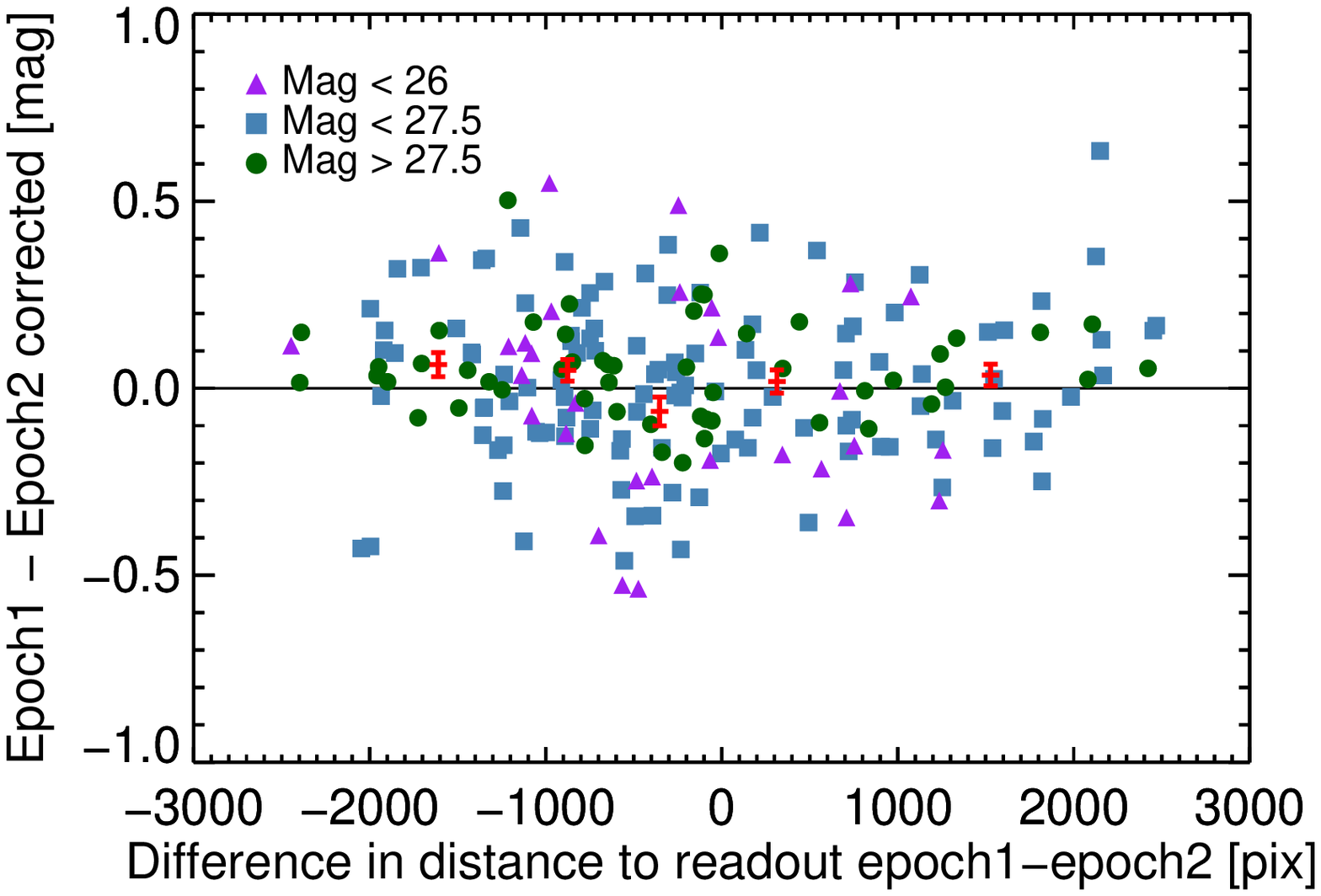}
}
\caption{ \label{fig:readepoch12} Photometry comparison of sources in
  Epochs 1 and 2 F275W mosaics as a function of the difference in
  distance to readout.  Sources falling to the left in this figure are
  close to the readout in Epoch 1, and far from the readout in Epoch
  2, while sources falling to the right in this figure are close to
  the readout in Epoch 2, and far from the readout in Epoch 1.
  Sources are color coded by their Epoch 1 magnitudes. The black line
  is the zero difference line. The red points with error bars are average binned
  values, with equal numbers of sources in each bin. Observed
  photometry is consistent with CTI effects, with the difference in
  magnitudes being negative in the top side of the figure, and
  positive in the right side of the figure. The slope of the effect is
  removed (bottom panel) when applying the catalog based CTE correction (see Section
  \ref{ctecorr}).  }
\end{figure}

Another method to visualize the CTI effects is to plot the magnitude
difference in Epochs 1 and 2 versus the difference in distance to the
readout (top panel, Figure \ref{fig:readepoch12}). Sources falling to
the left in this figure are close to the readout in Epoch 1 and far
from the readout in Epoch 2, while sources falling to the right in
this figure are close to the readout in Epoch 2 and far from the
readout in Epoch 1.  Sources for which electrons travel larger
distances to the readout lose more flux, so CTI effects would cause
the difference in magnitude to be negative in the left half of the
figure and positive in the right side of the figure. The sources used
in the figure are color coded by magnitude, with purple triangles
representing the brightest, and green circles representing the
faintest. The purple triangles have a smaller scatter, consistent with
the fact that bright sources are less severely affected by CTI than
faint sources \citep{Massey:2010b}. The red points with error bars in
Figure \ref{fig:readepoch12}, which show the average values in bins of
equal numbers per bin, emphasize the trend. The uncertainties are the
standard deviation of the points in each bin divided by the square
root of the number of points per bin. The bottom panel of Figure
\ref{fig:readepoch12} is the same as the top, with the addition of a
catalog-based CTE correction (see Section \ref{ctecorr}). The CTE
correction somewhat reduces the scatter observed in the top panel,
and removes the slope observed in the data.

Given that the increased photometric scatter is correlated with the
readout direction, and that there is significantly less scatter when
comparing the subsets of Epoch 2 data, 
we conclude that CTI is the dominant cause of the large scatter in
photometry observed in Figures \ref{fig:magepoch12} and
\ref{fig:magepoch2ab}. It is possible that other calibration
issues contribute as well, but they would require effects that are also
dependent on source position on the detector.

\subsubsection{CTI effects on morphology}
\label{morph}

CTI affects the shape of galaxy images as well as their
photometry. \citet{Rhodes:2010} investigated the effects of CTI on
galaxy morphology using simulations and found that small
galaxies are more affected by CTI than large ones. They also found
that small bright galaxies are slightly less affected by CTI than
small faint ones, but this dependence is not observed for
large galaxies. The net effect of CTI on image morphology is a
smearing out of the flux in the readout direction. Thus CTI results in
circular objects appearing elongated in the readout direction.

\begin{figure}[h!]
\center{
\includegraphics[scale=0.47]{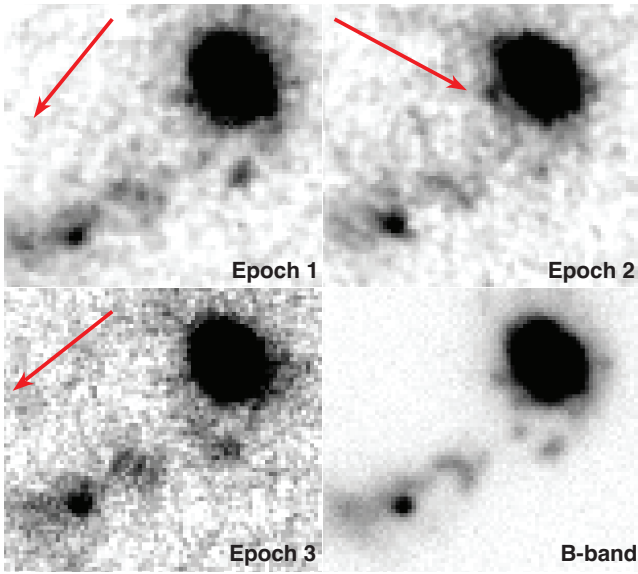}}
\caption{ \label{fig:ctemorph} Example of a galaxy affected by CTI in
  the F275W images. 
  The top left panel is from Epoch 1, the top right panel is from Epoch 2,
   the bottom left panel is from Epoch 3, and the bottom right panel is the B-band image. 
   The red arrows correspond to the readout direction, and the galaxy is elongated in the readout
  direction in each case. The elongation is worst in Epoch 2, as it is
  furthest from the readout in that Epoch.  The elongation is reduced
  in the post-flashed Epoch 3 compared to Epoch 1.  }
\end{figure}

This elongation effect is observed in the UVUDF data, as shown in the example
in Figure \ref{fig:ctemorph}. This galaxy is located
about two thirds the length of the detector away from the readout in Epoch
1, and almost as far as possible from the readout in
Epoch 2. In this example, both the bright galaxy and the nearby smaller
structures are elongated in the readout direction, as marked by the red
lines. The near 90 degree separation of Epochs 1 and 2 shows the
magnitude of the elongation in each direction. We note that this
elongation also affects the astrometry,
limiting the precision of the alignment between
epochs and between UVIS and ACS images.

\begin{figure}[h!]
\center{
\includegraphics[scale=0.5, viewport=15 5 500 360,clip]{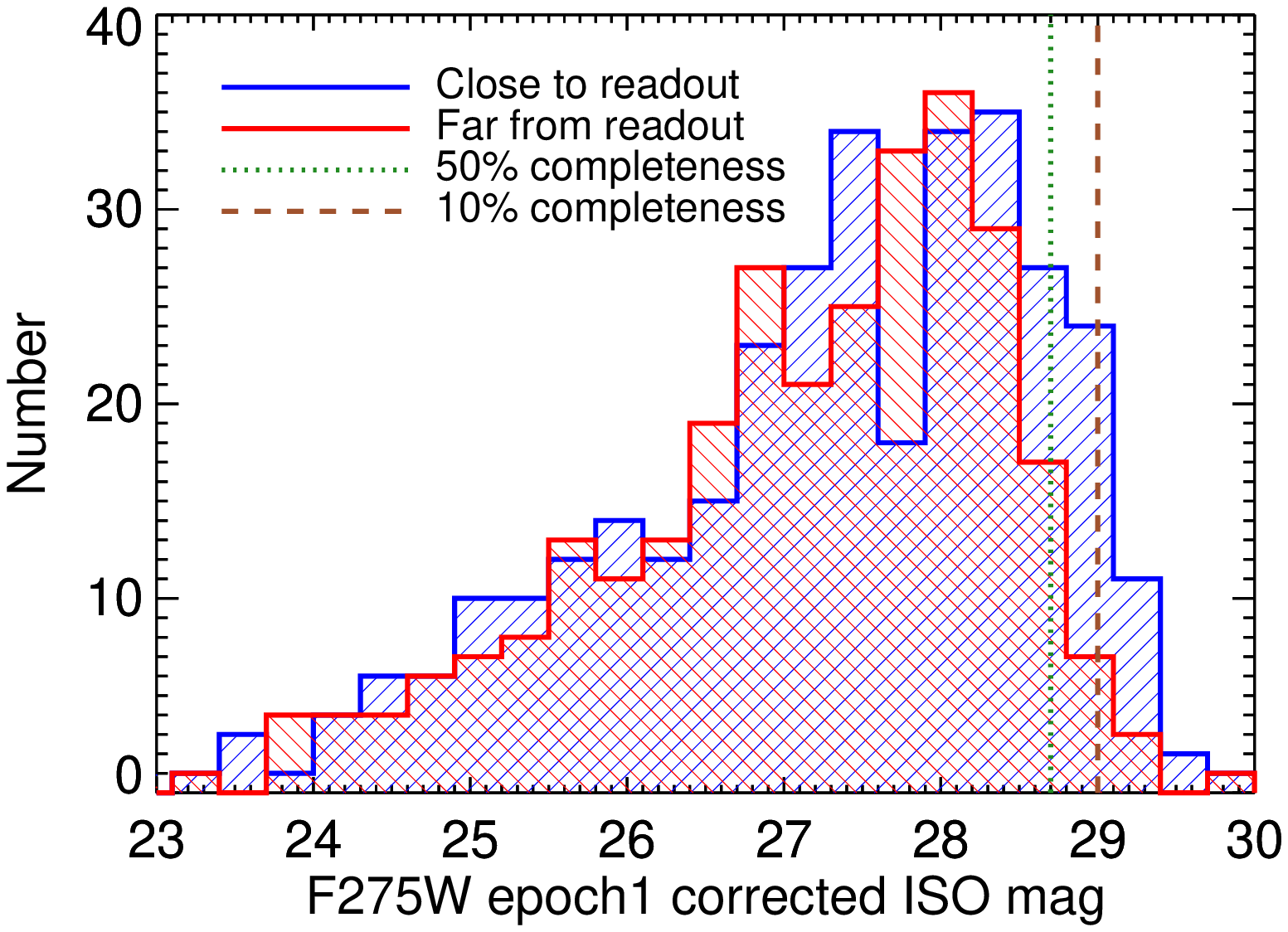}
\includegraphics[scale=0.5, viewport=15 5 500 360,clip]{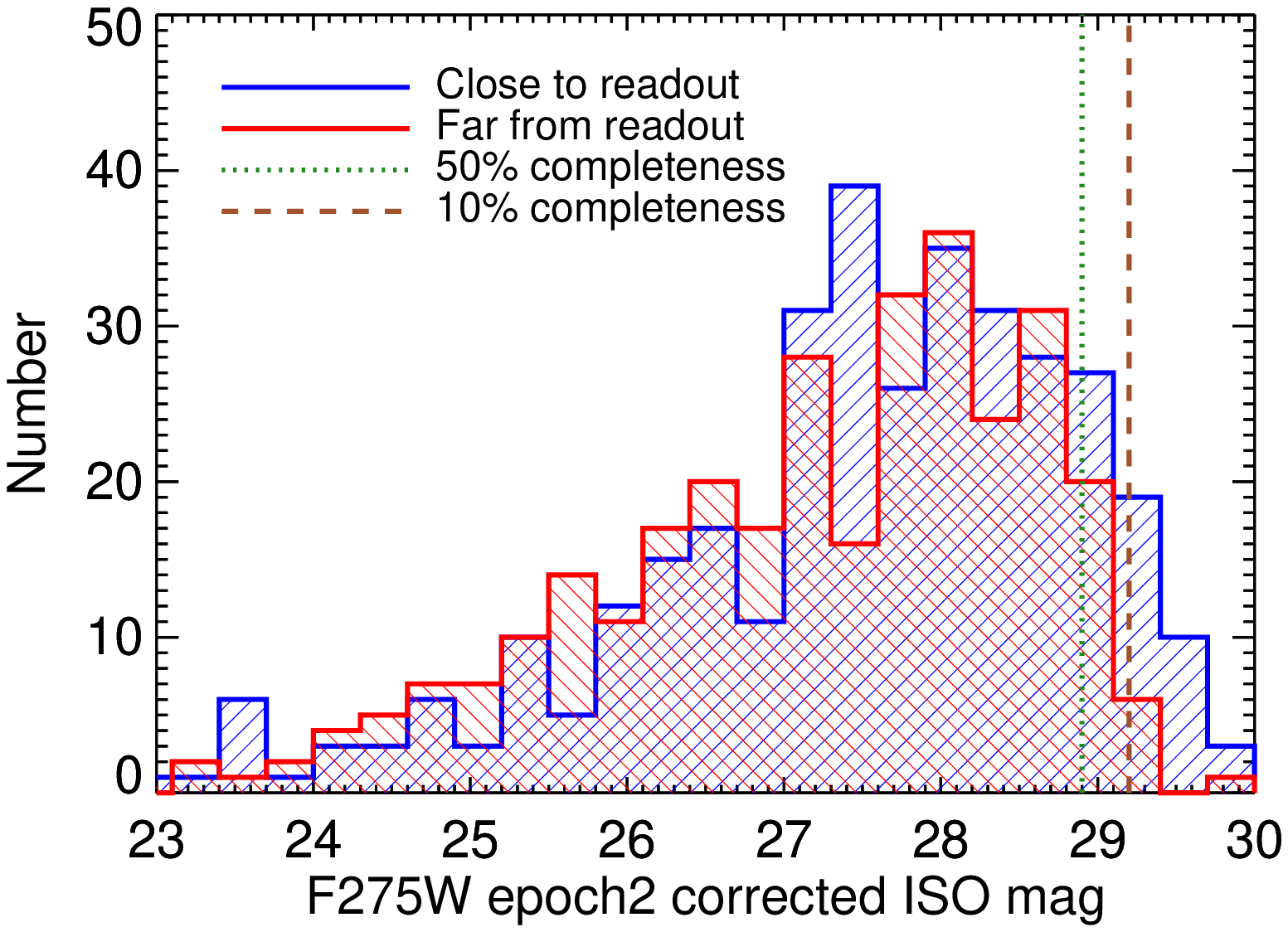}
\includegraphics[scale=0.5, viewport=15 5 500 360,clip]{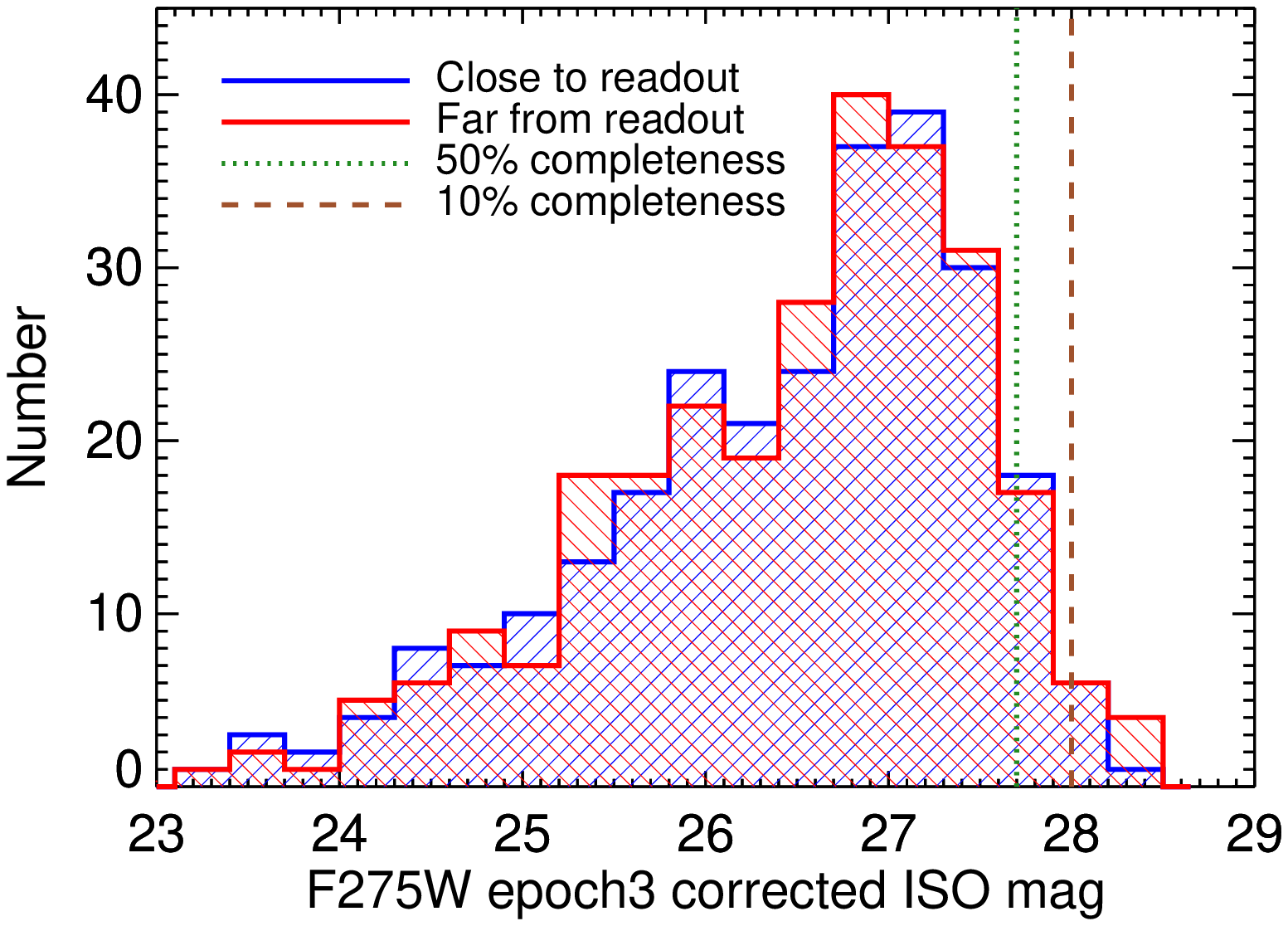}}
\caption{ \label{fig:ctemaghist} Histograms of sources detected in the
  F275W Epoch 1, 2, and 3 mosaics based on their isophotal CTE
  corrected magnitudes. Sources are split into two groups based on
  distance to the readout, with sources in the halves of the chips
  close to the readout shown in blue, and the sources in the other
  halves far from the readout shown in red. The 50\% and 10\%
  completeness levels (see section \ref{sensitivity}) are plotted in
  green and brown respectively.  Sources close to the readout appear to
  have a tail beyond the 10\% completeness, while sources far from the
  readout drop more steeply. This suggests that we are losing sources
  far from the readout that are not lost close to the readout.  }
\end{figure}

\subsubsection{CTI effects resulting in source loss}
\label{loss}
Another effect of CTI on the images is the possibility of losing faint
sources completely. Studies of warm pixels in long dark exposures show
that the number of warm pixels decrease drastically further away from
the readout, and the effect is worse for fainter warm pixels
\citep{MacKenty:2012}. That study is a worst case scenario, because
warm pixels are not shielded by other nearby pixels as is the case for
pixels associated with faint astronomical sources. Nonetheless,
post-flash calibration observations of Omega Centuri confirm that
faint sources in low backgrounds can disappear completely due to CTI
\citep{Anderson:2012c}. The sensitivity limit of observations is thus
set by the exposure time of each individual exposure rather than the
average of a stack. This depth varies as a function of distance to the
readout, morphology, and position of other sources on the detector.

A simple test of source losses is a comparison of the number counts of
detected objects as a function of magnitude for sources close and far from the readout.
We start with the B-band selected catalog described in Section
\ref{objdet}, and consider sources down to $10\sigma$ detections in the B-band
and 3$\sigma$ detections in F275W. Except for Epoch 3, we apply the
catalog based CTE correction (see Sections \ref{ctecorr} and
\ref{phot}) to reduce the effects of CTI on the photometry. The
prediction is that some sources far from the readout in Epochs 1 and 2
will be lost completely, while source losses should be greatly reduced
in the post-flashed Epoch 3 data.

A histogram of detected sources based on their isophotal CTE-corrected
magnitudes is shown in Figure \ref{fig:ctemaghist} for all three
epochs. For Epochs 1 and 2, more faint sources are found close to the
readout (blue) than are found far from the readout (red), suggesting
that some sources far from the readout have been lost.  There is no
significant difference in the number of B-band sources in the same
sample areas.

The sources lost due to CTI are very faint, and the number counts at
these faint magnitudes are suppressed by the incompleteness due to
lack of sensitivity (Section \ref{sensitivity}). It is therefore
difficult to distinguish sources lost due to CTI from sources that
would not be detected because of insufficient sensitivity. Most of the
losses from CTI are at magnitudes close to the 10\% completeness limit
for the Epoch 1 and 2 F275W mosaics. Keeping this limitation in mind,
as well as the small number statistics at the faintest magnitudes
where incompleteness is very high, we estimate the number of lost
sources by comparing the number counts close to and far from the
readout. For sources fainter than the 50\% completeness limit of
AB$\sim$28.3 mag (see section \ref{sensitivity}), we find that at
least $\sim30$ sources are lost in each Epoch 1 and Epoch 2 (out of
$\sim$600 source positions that are common between the epochs), while
no sources are lost in Epoch 3 (out of $\sim500$). The total number of
lost galaxies is likely larger than those found above, because sources
in the middle of the CCDs may also be lost. These sources are not
close to the readout in either Epochs 1 and 2 and fall below the
sensitivity limit of Epoch 3, making them difficult to identify. We
expect that the number of these sources per area is smaller than the
number found far from the readout, suggesting the total number of lost
sources is likely within a factor of two of those observed to be lost.
Our best estimate is a loss of $\lesssim100$ sources out of $\sim$600.
The small number of losses suggests that the results presented in
Section \ref{lymanbreak} are not strongly biased due to CTI.

\begin{figure}[b!]
\center{
\vspace{3mm}
\includegraphics[scale=0.47]{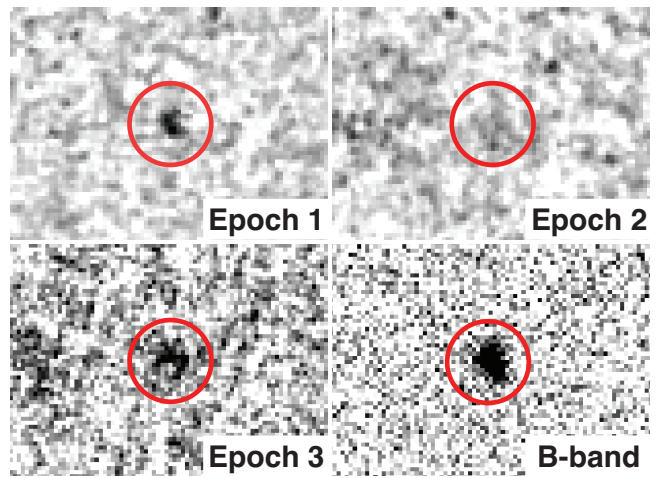}}
\caption{ \label{fig:ctelost} Example of a galaxy lost due to CTI in
  the F275W Epoch 2 mosaic. The left panel is a cutout of the F275W
  Epoch 1 mosaic, the middle panel is from the Epoch 2 mosaic, and the
  right panel is from the Epoch 3 mosaic. The galaxy is present in the
  shallower Epoch 1 data which is close to the readout, and is not
  detected in the slightly more sensitive Epoch 2 data which is far from the readout. 
  This galaxy is observed in the optical ACS images, and is object 4188 in the
  catalog by \citet{Coe:2006}. It has an F275W isophotal magnitude of
  $28.6\pm0.1$, and a F435W total magnitude of $27.9\pm0.04$
  \citep{Coe:2006}.  The galaxy is detected at $8\sigma$ in Epoch 1,
  and should have been observed at a higher significance in Epoch 2
  were it not for CTI. The galaxy is also detected in the shallower
  post-flashed Epoch 3.  }
\end{figure}

Another empirical test of source losses due to CTI is to compare
individual sources that are detected close to the readout in one epoch
but whose position is far from the readout in another epoch.  The
Epoch 2 mosaic is slightly more sensitive than the Epoch 1 mosaic (8
orbits F275W in Epoch 2 compared to 6 orbits for Epoch 1), so sources
that are detected in Epoch 1 close to the readout, but not detected in
Epoch 2 far from the readout demonstrate the effect. In searching for
such sources, we also required them to be detected in the
significantly more sensitive B-band image \citep{Beckwith:2006}. There
exist a few such sources, and an example is shown in Figure
\ref{fig:ctelost}. The left panel is a cutout of the F275W Epoch 1
mosaic and the middle panel is from Epoch 2, and the right panel is
from Epoch 3. This source is observed in the optical ACS images, and
is object 4188 in the catalog by \citet{Coe:2006}. It has an F275W
isophotal magnitude of $28.6\pm0.1$, and a F435W total magnitude of
$27.9\pm0.06$ \citep{Coe:2006}. This source is detected at $8\sigma$
in Epoch 1, and should have been observed at least at that S/N in the
Epoch 2 data.  

The potential loss of faint objects is one of the primary reasons that
we decided to use the post-flash option in Epoch 3. The other
motivations for the post-flash include reducing other CTI effects and
significantly improving pixel-based CTE corrections by taking data
with higher backgrounds \citep{MacKenty:2012}. 

\begin{figure}[h!]
\center{
\includegraphics[scale=0.47]{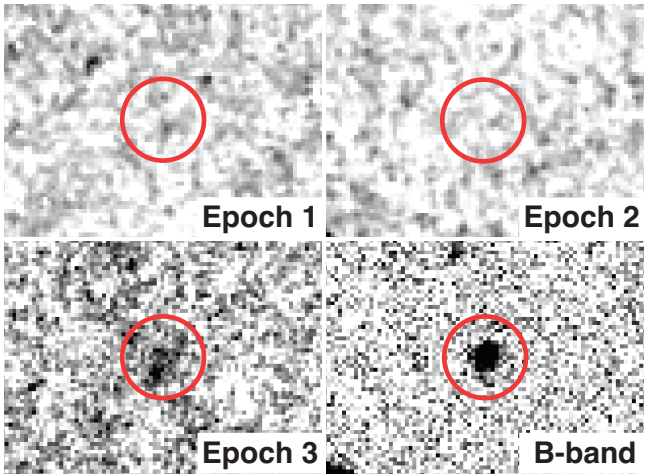}}
\caption{ \label{fig:ctelost2} Example of a galaxy lost due to CTI in
  the F275W Epoch 1 and 2 mosaics, but preserved in Epoch 3 due to
  post-flash. The left panel is a cutout of the F275W Epoch 1 mosaic,
  the middle panel is from the Epoch 2 mosaic, and the right panel is
  from the Epoch 3 mosaic. The galaxy is present in the shallower
  Epoch 3 data, and is not detected in the more sensitive Epoch 1 and
  2 data. The galaxy is approximately in the middle of the chip in all three epochs. 
  This galaxy is observed in the optical ACS images, and is
  object 8020 in the catalog by \citet{Coe:2006}.  It has an F275W
  isophotal magnitude of $27.8\pm0.1$, and a F435W total magnitude of
  $27.9\pm0.04$ \citep{Coe:2006}. The galaxy is detected at $9\sigma$
  in Epoch 3, and would have been easily detected in both Epochs 1 and
  2 were it not for CTI.  }
\end{figure}

The evidence that no sources have been lost in Epoch 3 is encouraging,
though the sensitivity limit is necessarily worse. While the F275W
exposure time in Epoch 3 is about double that of Epochs 1 and 2
individually, Epoch 3 is significantly less sensitive (see Section
\ref{sensitivity}). In fact, most of the sources that appear to be
lost in Epochs 1 and 2 due to CTI would not have been detected in the
Epoch 3 mosaic in the first place. Thus there are very few examples of
sources that were lost in either of the Epochs without post-flash but are present in
Epoch 3. One such example is shown in Figure \ref{fig:ctelost2}. This galaxy is
observed in the optical ACS images, and is object 8020 in the catalog
by \citet{Coe:2006}. It has an F275W isophotal magnitude of
$27.8\pm0.1$, and a F435W total magnitude of $27.9\pm0.04$
\citep{Coe:2006}. The source is detected at $9\sigma$ in Epoch 3, and
would have been easily detected in both Epochs 1 and 2 were it not for
CTI.

It is difficult to measure precisely how many sources may have been
lost in Epoch 1 and 2 due to the effects of CTI.  We can estimate
the magnitude of the problem by referring to the comparison presented
in Figure \ref{fig:ctemaghist}.  Significantly more faint sources are
detected at positions on the CCD close to the readout than far away
from it in Epochs 1 and 2, which lack the additional post-flash
background.  If objects were evenly distributed on the detector, which
they may not be, the histograms would suggest that $\sim 5$\% of the
total sources may have been lost to the effects of CTI, and as many as
$\sim 30$\% of sources fainter than the 50\% completeness limit.

The CTI effects create a dichotomy between the first two UVUDF epochs
and the third epoch. The combined Epoch 1 and 2 mosaic is more
sensitive than the Epoch 3 data, but suffers more from CTI, and some
objects may be lost completely. Once pixel-based CTE corrections
are applied, the Epoch 3 data will be the best characterized NUV
mosaic available. We agree with the STScI recommendation that future
WFC3 UVIS observations that require very sensitive measurements use
the post-flash.

\subsection{Sensitivity}
\label{sensitivity}

We use two common methods to characterize the sensitivity of the UVUDF
data. First, we measure the sky fluctuations of the images. Secondly,
we measure the 50\% completeness limit, measured by recovery of
simulated sources placed in the science mosaics. The completeness test
takes into account both the sky surface brightness and the spatial
resolution of the mosaics, yielding a good sense of the usable depth
of an image \citep{Chen:2002,Sawicki:2005, Rafelski:2009,
  Windhorst:2011}.

The sky noise of each image is measured via the pixel-to-pixel rms
fluctuations. These fluctuations are measured in 51$\times51$\ pixel boxes at
1000 semi-random locations, such that the boxes are entirely on the
image, do not fall on a detected object, and the boxes do not overlap
other boxes.  This technique is designed to be less sensitive to any
residual gradient in the image than simply using the rms of all
unmasked pixels. The rms in each mosaic is the iterative sigma clipped
mean of the rms in each box, which is determined with an iterative
sigma clipped standard deviation. This rms is then multiplied by the
noise correlation ratio to account for the correlated noise from
drizzling the mosaics. The approximate correlation ratio of the UVUDF
data is $\sim2.5$ and $\sim1.5$ for the binned and unbinned data, respectively,
based on equation 9 from \citet{Fruchter:2002}. These rms values
corrected for correlated noise match the expected values from the rms
images. The resulting $5\sigma$ rms magnitudes (assuming 0\farcs2 radius
aperture) for the
mosaics are tabulated in Table \ref{tab:sens}. These values are within
$0.1-0.2$ mags of the 0\farcs2 aperture, $5\sigma$ magnitudes predicted by the STScI
exposure time calculator modified for binning or post-flash (see Table
\ref{tab:sens}).

\begin{figure}[t!]
\center{
\includegraphics[scale=0.5, viewport=15 5 500 360,clip]{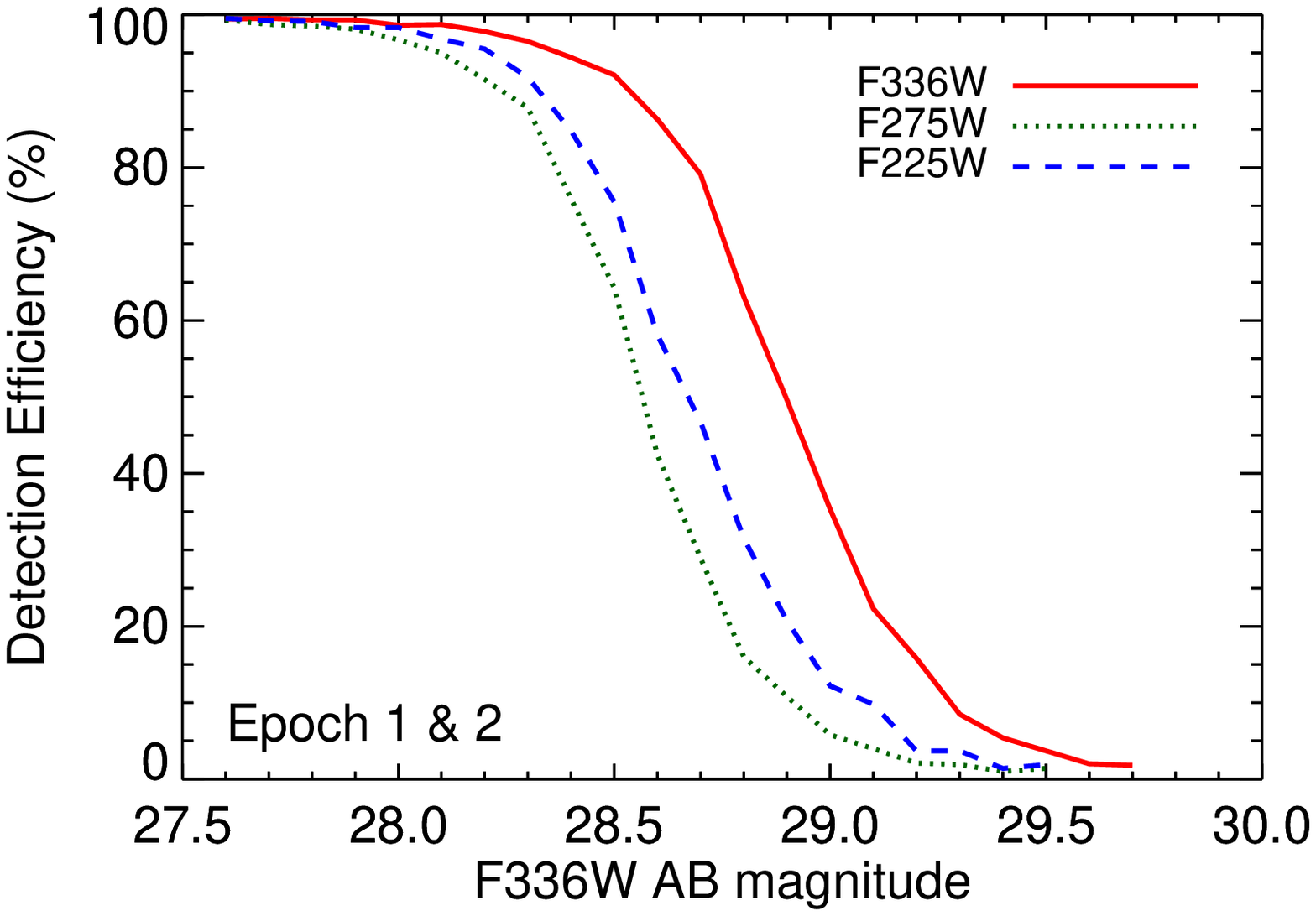}
\includegraphics[scale=0.5, viewport=15 5 500 360,clip]{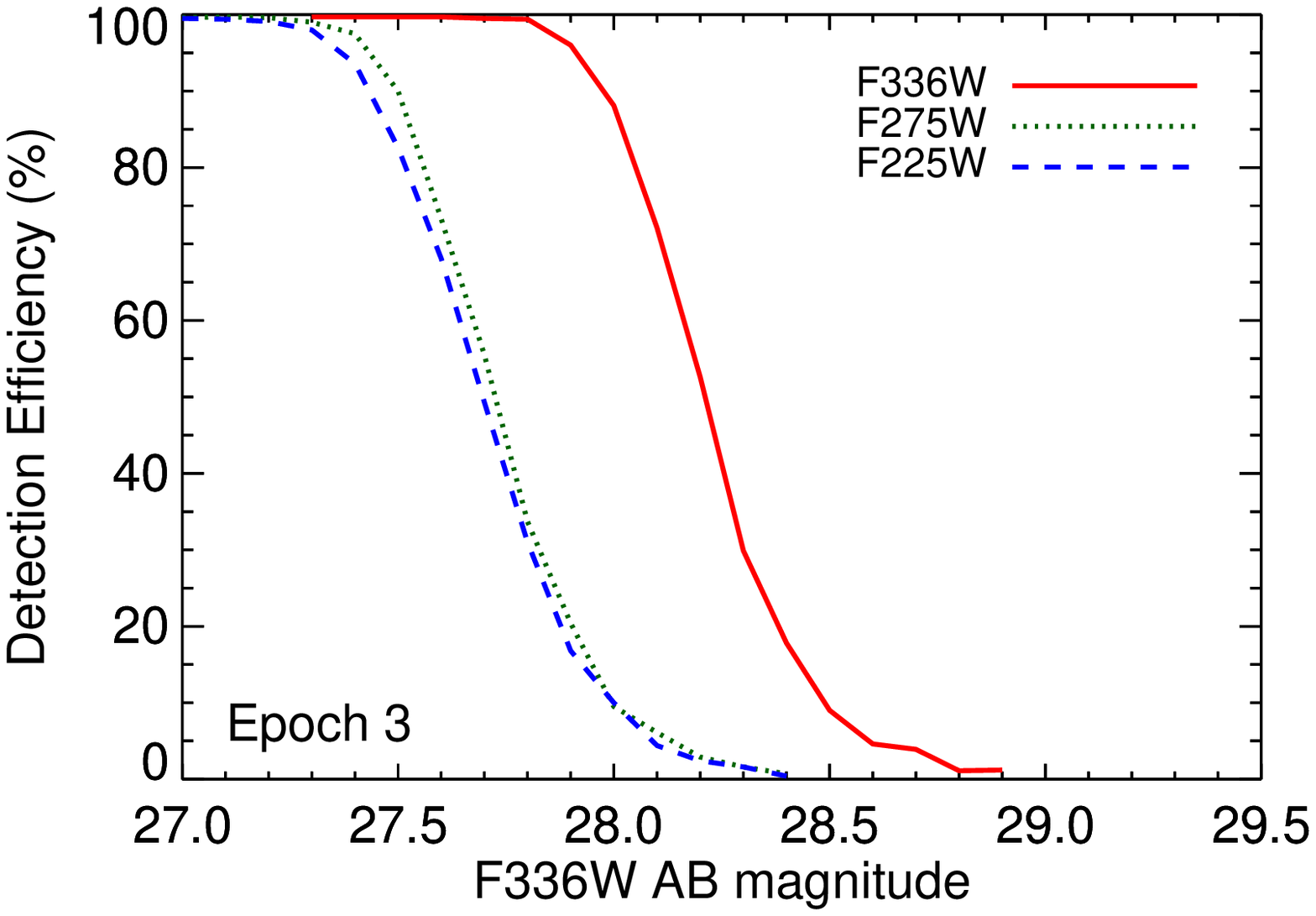}}
\caption{ \label{fig:det} Detection efficiency of the Epochs 1 \& 2 (top) and Epoch 3 (bottom)
  mosaics as a function of total magnitude. These are the recovery
  percentages of simulated point sources in the images. The limiting
  magnitude is defined as the magnitude at which 50\% of the sources
  are recovered. The limiting magnitudes for Epoch 1\&2 are 28.6, 28.6, and 28.9 mag, and for Epoch 3 are 27.7, 27.7, and 28.2 mag,
  for F225W, F275W, and F336W (see Table \ref{tab:sens}). This
  completeness test does not account for source losses due to CTI. }
  \vspace{5mm}
\end{figure}

We performed a standard completeness test to confirm the noise
characteristics of the data by planting and recovering simulated
objects.  This test does not take into account the loss of sources at
the faint end due to CTI, and so the results of the test are an upper
limit on the completeness. Specifically, the 50\% completeness
magnitude limit due to noise is measured by planting Gaussian PSFs
for a range of magnitudes in the mosaics at semi-random locations, and
counting the fraction of sources that are recovered with SExtractor.
The PSF FWHMs are matched to those measured in the data for each
filter. Unresolved sources are selected from the published catalogs of
stars in the UDF \citep{Pirzkal:2005}. However, there are only a small
number of identified sources bright enough in the NUV to be used for
PSF determination. Three sources are used for F336W, and two sources
are used for F225W and F275W. The sources are each registered to their
subpixel centers, normalized by the peak value, and coadded with a
mean. The resulting PSFs are worse than measured by
\citet{Windhorst:2011} in the ERS, because half the data are binned
and CTI affects the source morphology.  For the combined Epoch 1 and 2
mosaic, we measure PSF FWHMs of 0\farcs133,0\farcs133, and
0\farcs127 for F225W, F275W, and F336W.  For the Epoch 3 mosaic, we
measure PSF FWHM's of 0\farcs134, and 0\farcs121 for F275W and
F336W, and use the F275W PSF for the F225W PSF as it is not well
determined.  The locations of the planted sources are constrained such
that they do not fall off the edges, fall on a real detected object,
or fall on any previously planted source. The detection efficiency as
a function of magnitude is shown in Figure \ref{fig:det}, and the 50\% 
completeness magnitudes are tabulated in Table \ref{tab:sens}.


\section{Initial Results}
\label{preview}

In this section, we briefly present initial results, representative of
those that will be possible with the UVUDF.  We describe the color
selection of galaxies using the Lyman break technique, and we
demonstrate the utility of the deep NUV images for morphological
analysis.  In both cases, these results are presented based on the
combined Epoch 1 and 2 mosaics, without the application of the
pixel-based CTE correction.  We anticipate that future papers will
improve the analysis once that correction is stable and can be
confidently applied.

\subsection{Lyman Break Galaxies}
\label{lymanbreak}

The selection of high redshift galaxies by the identification of the
strong Lyman break feature in their SED using broad-band photometry
has been extremely successful
\citep[e.g.][]{Steidel:1996a, Steidel:1996b, Steidel:1999,Steidel:2003, Adelberger:2004,
  Bouwens:2004, Bouwens:2006, Bouwens:2010, Bouwens:2011b,
  Rafelski:2009, Reddy:2008, Reddy:2009, Reddy:2012b}.  Although less
precise than a full SED-fit such as those used in photometric redshift
estimates, the Lyman break identification is a standard in the
literature. Here, we have taken a first look at selecting LBGs in the
UVUDF at redshifts where the Lyman break falls in the NUV filters:
1.7, 2.1, and 2.7 in F225W, F275W, and F336W, respectively. We
directly compare these initial results with published results from the
ERS \citep{Windhorst:2011}, which used the same filters in shallower
data (AB=26.9) over a larger area (about 50 square arcminutes).

We implement the dropout criteria used on ERS data by
\citet{Hathi:2010} (hereafter H10) and \citet{Oesch:2010eb} (hereafter
O10). These consist of both color-color criteria as well as S/N
criteria for candidates to be considered dropouts. 

Faint stars were removed from the sample by position matching sources
in the UVUDF catalog with the published catalog of unresolved sources
in the UDF \citep{Pirzkal:2005}. 25 sources are found to match this
catalog (0.1$''$ matching radius); 22 of these sources are identified
as stars according to the criteria of \citet{Pirzkal:2005}. There are
7, 10, 18 stars detected at S/N = 3 threshold in F225W, F275W, F336W,
respectively.

\begin{deluxetable*}{ccccccccccccc}[h]
\tablecolumns{13}
\tablecaption{Summary of Dropout Galaxies in ERS and UVUDF.} 
\tablehead{
\multicolumn{2}{c}{Dropout} &
\multicolumn{3}{c}{ERS} &
\colhead{ } &
\multicolumn{3}{c}{UVUDF (ERS Depth)} &
\colhead{ } &
\multicolumn{3}{c}{UVUDF (Full Depth)} \\
\colhead{} &
\colhead{} &
\colhead{{\it Predicted}} &
\multicolumn{2}{c}{{\it Observed}} &
\colhead{ } & 
\colhead{{\it Predicted}} &
\multicolumn{2}{c}{{\it Observed}} &
\colhead{ } & 
\colhead{{\it Predicted}} &
\multicolumn{2}{c}{{\it Observed}} \\
\colhead{} &
\colhead{} &
\colhead{} &
\colhead{} &
\colhead{Surface} &
\colhead{} &
\colhead{} &
\colhead{} &
\colhead{Surface} &
\colhead{} &
\colhead{} &
\colhead{} &
\colhead{Surface} \\
\colhead{Filter} &
\colhead{Method} &
\colhead{Number} &
\colhead{Number} &
\colhead{Density} &
\colhead{} &
\colhead{Number} &
\colhead{Number} &
\colhead{Density} &
\colhead{} &
\colhead{Number} &
\colhead{Number} &
\colhead{Density} \\
\colhead{(1)}&
\colhead{(2)} & 
\colhead{(3)} & 
\colhead{(4)} & 
\colhead{(5)} &
\colhead{} &
\colhead{(6)} &
\colhead{(7)} &
\colhead{(8)} & 
\colhead{} & 
\colhead{(9)} & 
\colhead{(10)} & 
\colhead{(11)}
}
\startdata
F336W	&	H10	   &	  394   &   256 & 5.1    &&      49  & $37\pm6.6$  & $6.0\pm1.1$         &&      185 & $211\pm14.5$  & $34\pm2.3$     \\
$z\sim2.7$	&	O10	   &      448 &  403 & 8.6     &&    56 & $67\pm8.2$  & $10\pm1.2$         &&       224 & $304\pm17.4$  & $49\pm2.8$   \\ 
		&		   &		 &  	       &           &&	          &        &               &&               &         &        \\
F275W     &       H10	   &	  228 &     151 & 3.0    &&     28   & $22\pm5.2$  & $3.5\pm0.8$        &&        86  & $88\pm9.4$    & $14\pm1.5$   \\
$z\sim2.1$	&      O10	   &	   102 &      99  & 2.1    &&    13   & $10\pm3.7$   & $1.6\pm0.6$       &&       125 & $146\pm12.1$  & $24\pm2.0$   \\
		&		   &		 &  	       &           &&	          &        &             &&               &          &        \\
F225W	&  	H10	   &	    62 &       66  & 1.3    &&     8  & $4\pm2.5$     & $0.6\pm0.4$      &&          36 & $25\pm5.5$    & $4.0\pm0.9$    \\
$z\sim1.7$		&	O10 	   &	   99 &      60  & 1.3    &&     12  & $9\pm3.5$     & $1.5\pm0.6$      &&          111 & $61\pm7.8$   & $9.8\pm1.3$
\enddata
\tablecomments{Column (1) indicates the dropout filter and redshift
  bin.  Column (2) indicates the reference to the dropout method and
  luminosity function used to identify and predict source counts:  H10
  \citep{Hathi:2010}; O10 \citep{Oesch:2010eb}.   Columns (3-11)
  compare predicted and observed source counts for each dropout type.
  ERS refers to the Early Release Science data \citep{Windhorst:2011}.
  UVUDF(ERS) refers to UVUDF data analyzed to a comparable depth as
  ERS, UVUDF(Full) refers to UVUDF data analyzed to its full depth.
  Dropout sky density values are in units of arcmin$^{-2}$.
  Uncertainties on the observed number and density of sources are
  Poissonian. Errors to predicted source counts are discussed in the text.}
\label{TABLE:DropoutLargeSummary}
\end{deluxetable*}

\subsubsection{F336W, F275W, F225W Dropouts}

For reference, the
H10 criteria for dropout galaxies are given below. F336W dropouts
require observed magnitudes and signal-to-noise ratios, S/N, satisfy
each of the following 

\begin{equation} 
\begin{cases}
    F336W - F435W > 0.8 \\
    F435W \leq 26.5 \\
    F435W - F606W < 1.2 \\
    F435W - F606W > -0.2 \\
    F336W - F435W > 0.35 + [1.3 \times (F435W - F606W)] \\
    S/N(F435W) > 3 \\
    S/N(F336W) < 3 \\
    S/N(F275W) < 1 \\
    S/N(F225W) < 1 
\end{cases} 
\end{equation}

Similarly, F275W dropouts are identified by the criteria:

\begin{equation}
\begin{cases}
F275W - F336W > 1.0 \\
F336W \leq 26.5 \\
F336W - F435W < 1.2 \\
F336W - F435W > -0.2 \\
F275W - F336W > 0.35 + [1.3 \times (F336W - F435W)] \\
S/N(F336W) > 3 \\
S/N(F275W) < 3 \\
S/N(F225W) < 1 \\
\end{cases}
\end{equation}


F225W dropouts require all of the following criteria:

\begin{equation}
\begin{cases}
F225W - F275W > 1.3 \\
F275W \leq 26.5 \\
F275W - F336W < 1.2 \\
F275W-F336W > -0.2 \\
F225W - F275W > 0.35 + [1.3 \times (F275W - F336W)] \\
F336W - F435W > -0.5 \\
S/N(F275W) > 3 \\
S/N(F225W) < 3 \\
\end{cases}
\end{equation}

\begin{figure*}[b!]
\begin{center}
\includegraphics[scale=0.5]{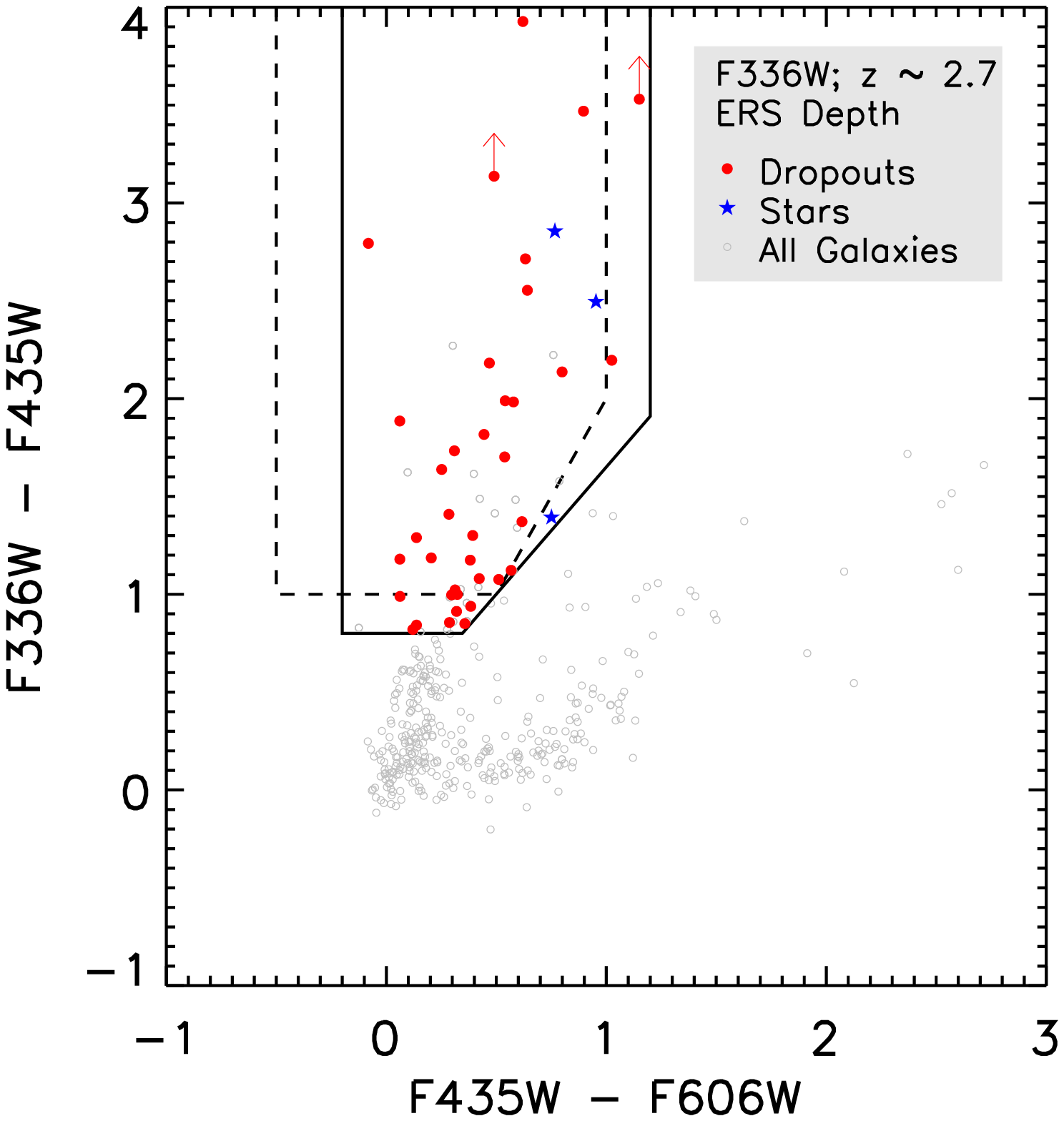}
\includegraphics[scale=0.5]{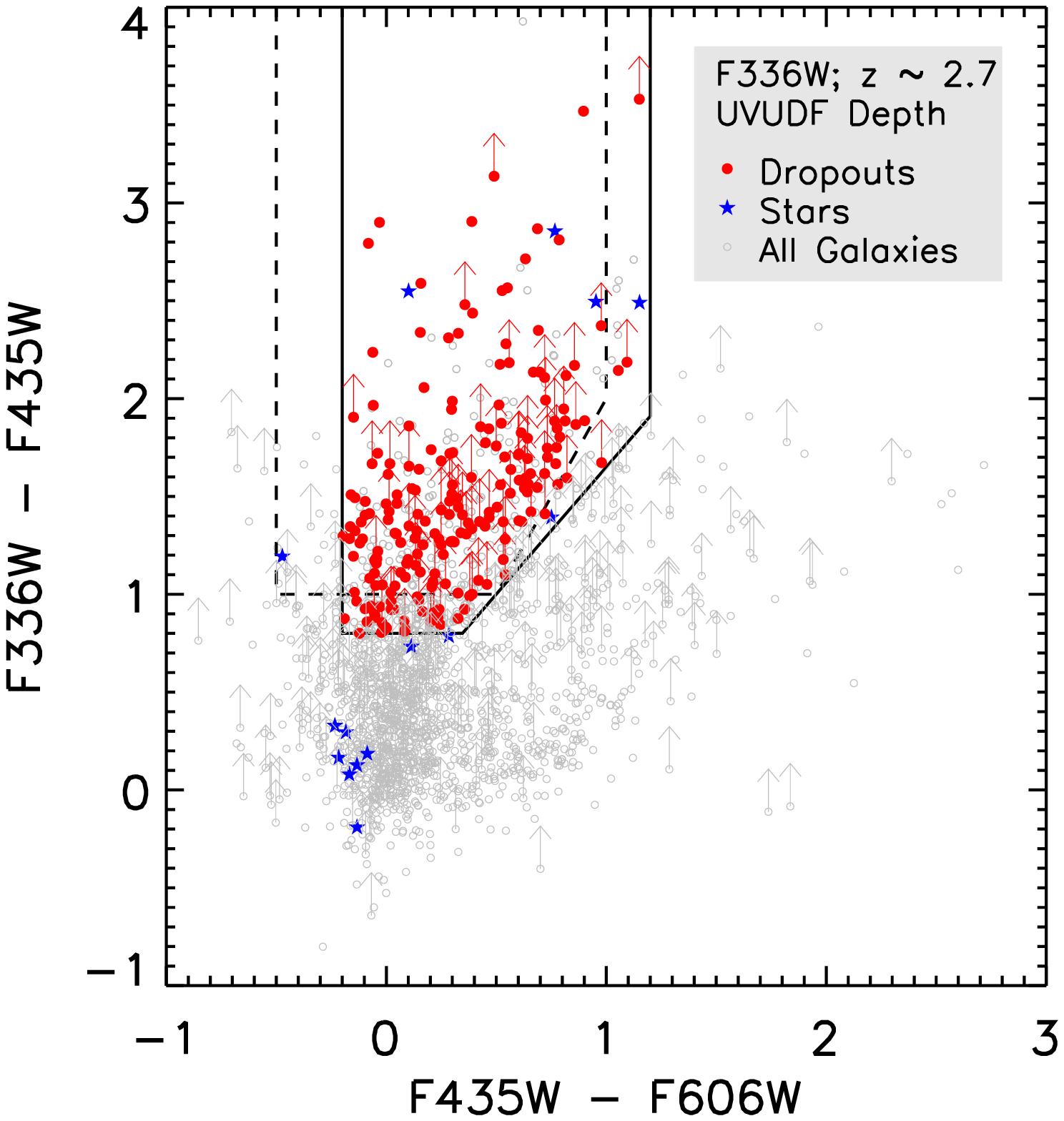}

\end{center} \caption{(left panel) Color-color diagram illustrating
  F336W dropout candidates in UVUDF to ERS depth using the method of
  Hathi et al 2010 (H10). Colors are computed from subtracting
  magnitudes in the appropriate HST wavebands. Sources are illustrated
  to ERS depth (F435W = 26.5 AB) and have S/N degraded to match ERS
  observations (as discussed in the text). Red circles are dropout
  galaxies according to criteria of H10. Stars are indicated as blue
  asterisks, and upper arrows indicate non-detections in the dropout
  band replaced with a 1$\sigma$ upper limit. Gray points indicate all
  sources in the UVUDF catalog to this depth limit; gray points in the
  color selection region fail to meet the S/N criteria of bona fide
  dropout galaxies. Solid lines indicate the color-selection region of
  H10; dashed lines indicate the color-selection region of O10. (right
  panel) Color-color diagram illustrating F336W dropout candidates to
  the full depth of UVUDF. Colors and symbols are the same as in the
  left panel.}
\label{FIGURE:F336WDropoutDiagrams}
\end{figure*}

\begin{figure*}[h]
\begin{center}

\includegraphics[scale=0.5]{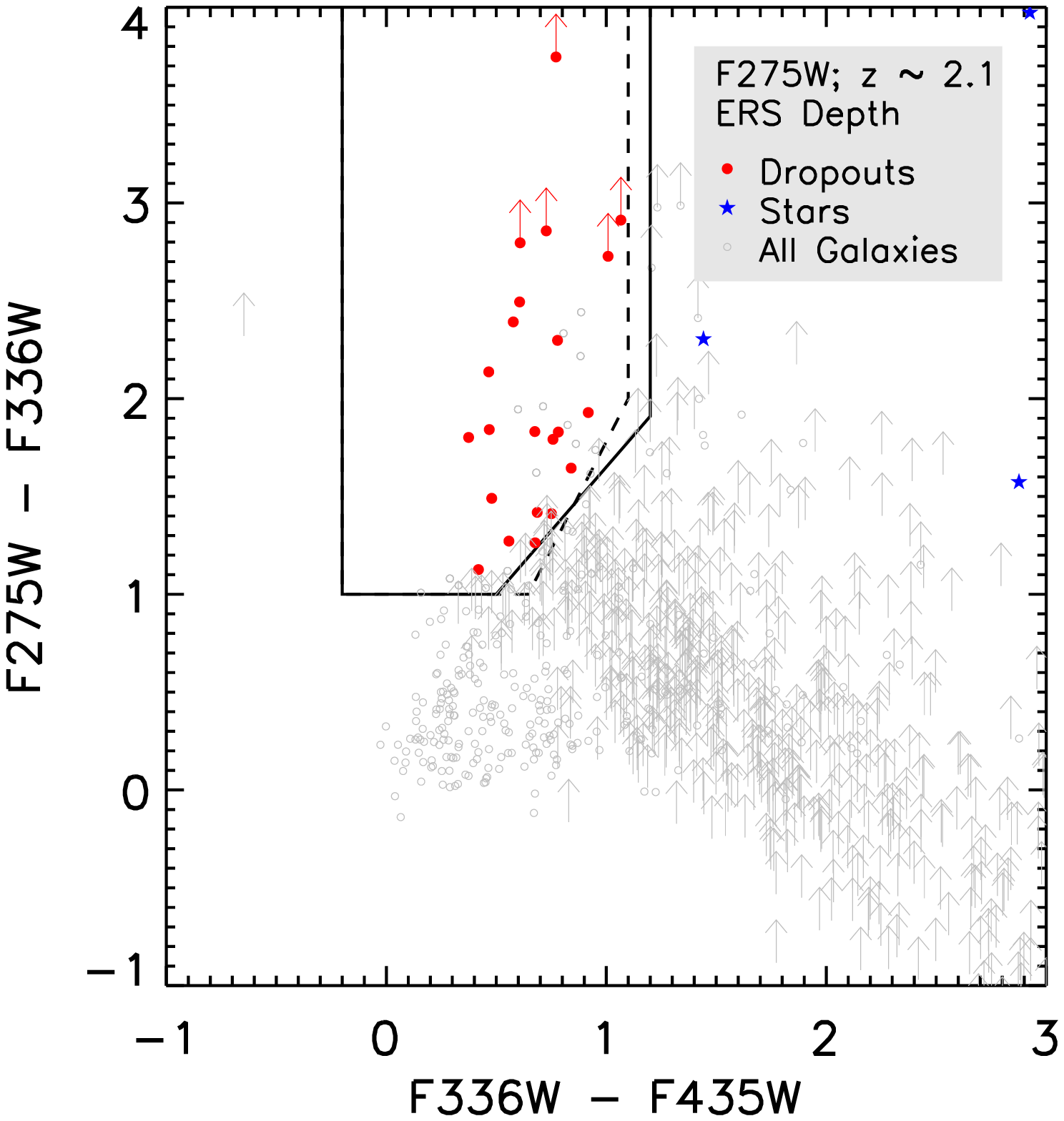}
\includegraphics[scale=0.5]{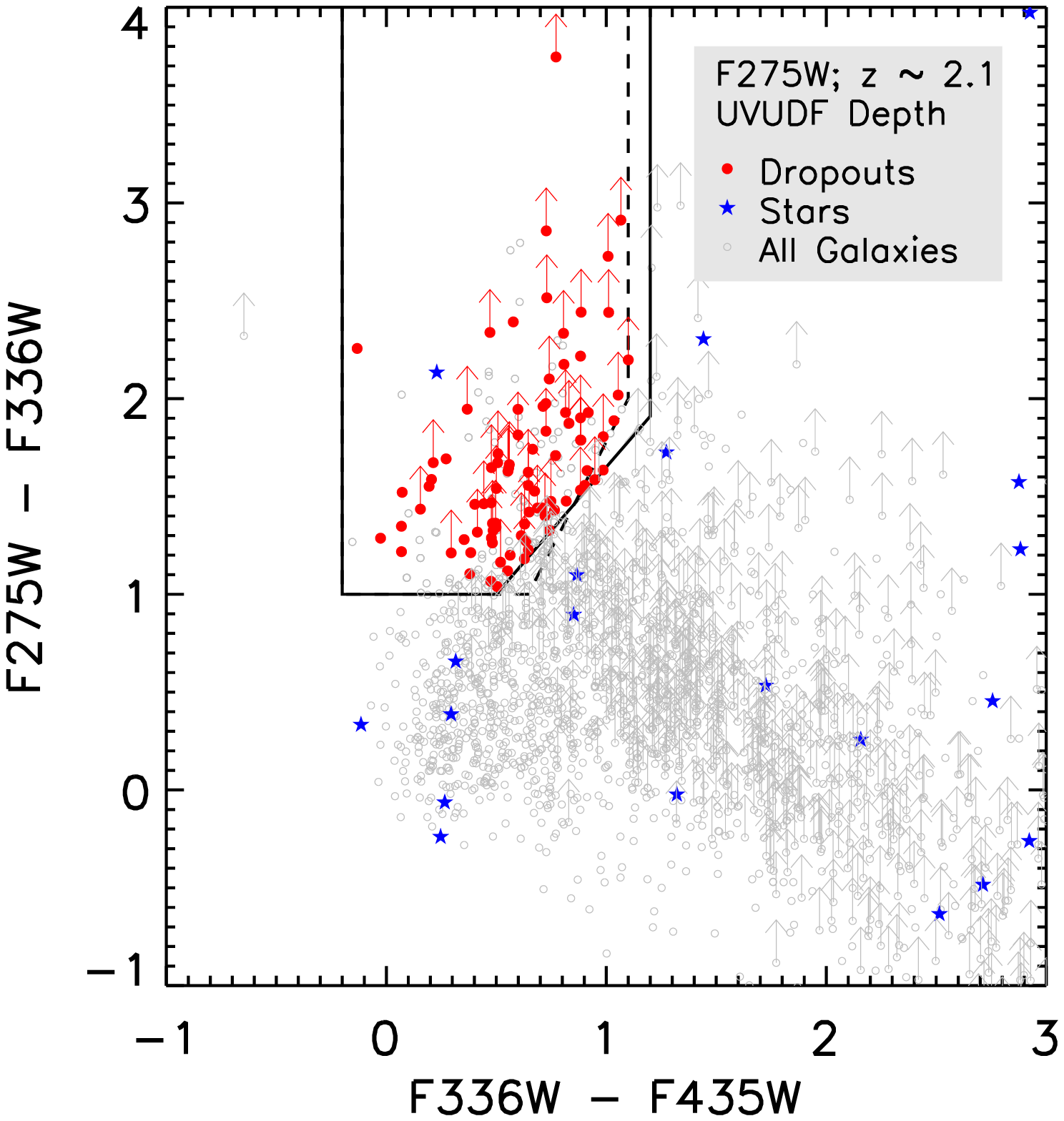}

\end{center} \caption{(left panel) Color-color diagram illustrating
  F275W dropout candidates in UVUDF to ERS depth using the method of
  Hathi et al 2010 (H10). Colors are computed from subtracting
  magnitudes in the appropriate HST wavebands. Sources are illustrated
  to ERS depth (F336W = 26.5 AB) and have S/N degraded to match ERS
  observations (as discussed in the text). Colors and symbols are the
  same as in Figure \ref{FIGURE:F336WDropoutDiagrams}. (right panel)
  Color-color diagram illustrating F275W dropout candidates to the
  full depth of UVUDF.}

\label{FIGURE:F275WDropoutDiagrams}
\end{figure*}

\begin{figure}[h!]
\begin{center}
\vspace{-5mm}

\includegraphics[scale=0.5]{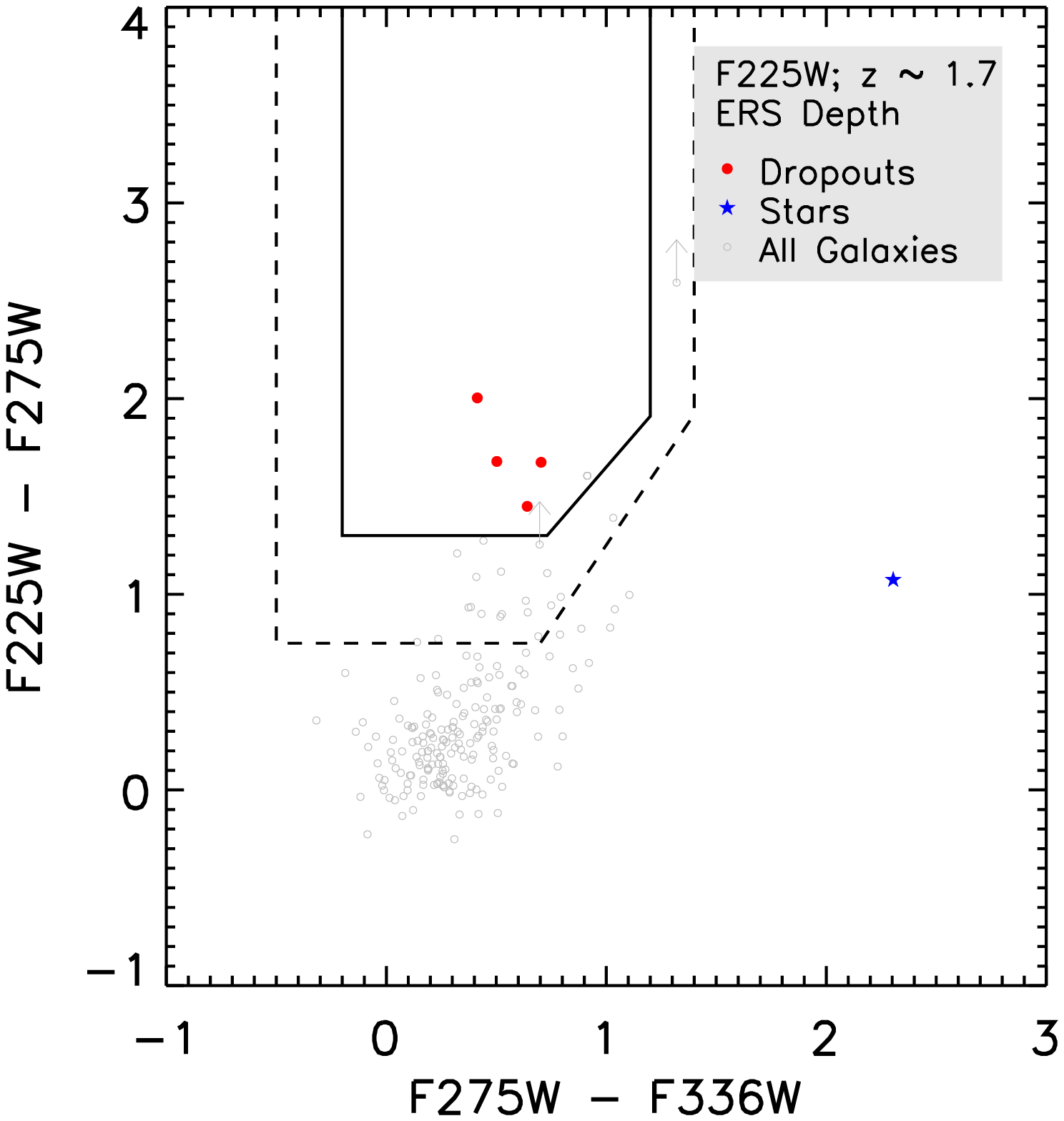}
\includegraphics[scale=0.5]{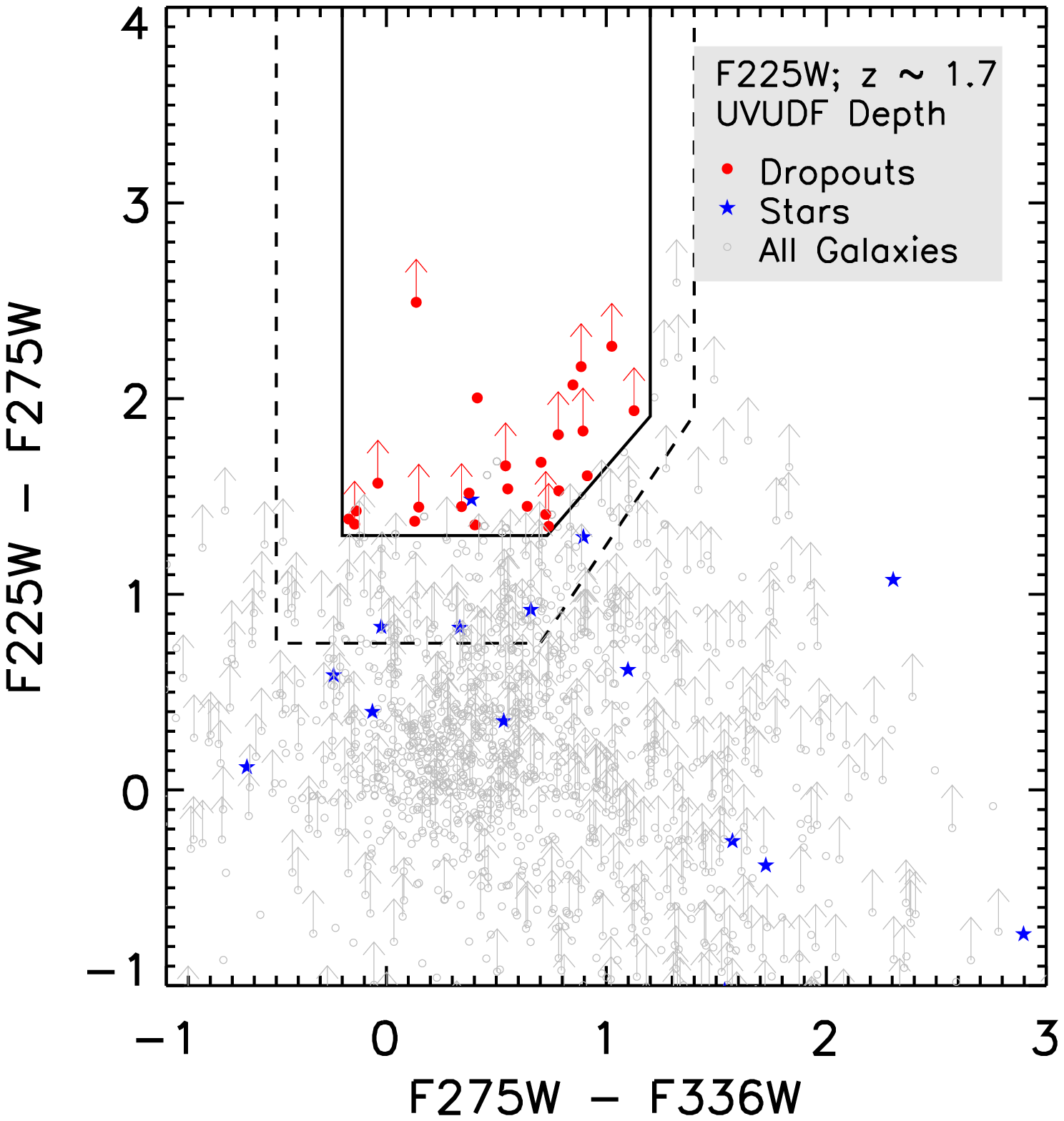}

\vspace{-5mm}

\end{center} \caption{(Top panel) Color-color diagram illustrating
  F225W dropout candidates in UVUDF to ERS depth using the method of
  Hathi et al 2010 (H10). Colors are computed from subtracting
  magnitudes in the appropriate HST wavebands. Sources are illustrated
  to ERS depth (F275W = 26.5 AB) and have S/N degraded to match ERS
  observations (as discussed in the text). Colors and symbols are the
  same as in Figure \ref{FIGURE:F336WDropoutDiagrams}. (Bottom panel)
  Color-color diagram illustrating F225W dropout candidates to the
  full depth of UVUDF. Colors and symbols are the same as in the top
  panel. There are 25 F225W dropouts. }

\label{FIGURE:F225WDropoutDiagrams}
\end{figure}

Figures \ref{FIGURE:F336WDropoutDiagrams}, \ref{FIGURE:F275WDropoutDiagrams},
and \ref{FIGURE:F225WDropoutDiagrams} illustrate the dropout
candidates selected according to the H10 criteria in color-color
diagrams. Sources with S/N $<1$ in the dropout band have their fluxes
replaced with $1\sigma$ upper limits and are indicated as arrows in
the figures. Stars are indicated as blue asterisks. Dropout
candidates (defined as meeting all of the criteria of H10) are
indicated as red symbols.  The mean and 1$\sigma$ redshift distributions reported in Hathi et
al. 2010 were: (F225W; 1.51, 0.13: F275W; 2.09, 0.42: F336W;
2.28, 0.4).  Likewise, the mean and 1$\sigma$ redshift distributions
reported in Oesch et al. 2010 were: (F225W; 1.5, 0.2: F275W; 1.9,
0.2: F336W; 2.5, 0.2)

For a direct comparison with previous observations, we use the UVUDF
data to examine source counts at the sensitivity of the shallower ERS
data and reproduce the selection at the shallow sensitivity. Because
the H10 and O10 dropout selection include a cut on source
significance, we scale the S/N ratio of the UVUDF measurements to what we would
expect from the ERS, using the HST Exposure Time Calculator (ETC). The
S/N ratios change by factors of 0.19, 0.33, 0.33 for F336W, F275W,
F225W respectively.  The dropout selection at the shallower limit is also
shown in the figures.

Table \ref{TABLE:DropoutLargeSummary} indicates the number of dropout
sources found using the methods of H10 and O10 in UVUDF data and, for
comparison, those reported previously in ERS data. We find 37, 22, 4
H10 dropouts in F336W, F275W, F225W bands, respectively, to ERS depth
in UVUDF, and we find 211, 88, 25 H10 dropouts in F336W, F275W, F225W
bands, respectively, to the full depth in UVUDF. The raw number of
dropouts in the narrow, deep UVUDF data is comparable to the numbers
found in the wider, shallower ERS data. Table
\ref{TABLE:DropoutLargeSummary} also compares the sky densities of
dropout candidates reported in H10 and O10 to those detected in UVUDF.
We find dropout sources with comparable sky densities as H10 and O10
at the same depth and S/N limits.

Table \ref{TABLE:DropoutLargeSummary} also compares the dropout
selection methods of H10 and O10 applied in the UVUDF to each other;
overall, the method of H10 is more conservative than the
method of O10. For example, the number of O10 F336W dropouts and their
resulting sky density exceeds the number of H10 F336W dropouts by a
factor of 1.8 (67 O10 dropouts vs. 37 H10 dropouts). This disparity
arises for several reasons. First, H10 uses a F435W-selected catalog, whereas
O10 uses a F606W-selected catalog; therefore, O10 dropout selection
begins with a larger sample of sources (531 F606W detected sources for
O10 versus 391 F435W detected sources for H10). Second, H10 has a
more stringent requirement for S/N in the bands blue-ward of the
dropout band than O10 (S/N(F275W,225W) $<1$ for H10 vs. S/N(F275W,F225W)
$<2$ for O10). Additional differences between the two methods include
the upper S/N limit for sources in the dropout band adopted by H10
(S/N(F336W) $<3$ for H10 vs. no S/N(F336W) requirement for O10), the
higher S/N requirement in the band redward of the dropout band in O10
(S/N(F435W) $>5$ for O10 vs. S/N(F435W) $>3$ for H10), and the
differences in color selection regions between the two methods, as
illustrated in Figure \ref{FIGURE:F336WDropoutDiagrams}. Stars are
found and rejected in each sample with approximately the same
percentage (15\% and 20\% for H10 and O10 respectively).

\vspace{5mm}

\subsubsection{Source Count Prediction}

We use the published luminosity functions of H10 and O10 dropout
sources to predict an expectation for the number of sources to be
found in the UVUDF.  Each luminosity function, expressed as a space
density of galaxies, $\phi$, in units of Mpc$^{-3}$ mag$^{-1}$ as a
function of absolute magnitude, $M$, at rest-frame 1500 \Angstrom ~is
described by a Schechter function \citep{Schechter:1976}.
We use the fitted parameters $\phi_*$, $M_*$ and $\alpha$ reported in
H10 and O10 for each dropout filter (F336W, F275W, F225W) to predict
the space densities of galaxies in each redshift range.  

The differential number of galaxies per unit redshift, dN/dz, for each
dropout filter, is given by integrals over the Schechter luminosity
function, $\Phi(M)$, expressed as a function of absolute magnitude,
$M$, multiplied by the (published in H10 and O10) gaussian galaxy
redshift distribution, g(z), and by the comoving volume element, $dV
/dz d\Omega$, and finally integrated over the survey solid
angle, $\Omega$.

\begin{equation}
\frac{dN}{dz} = \int d\Omega  \int^{M_{lim}}_{-27} \frac{dV}{dz d\Omega} \Phi(M) g(z) dM
\end{equation}

The lower limit of integration is chosen to include the brightest
observable galaxies.
The number of sources 
down to the magnitude limit, $M_{lim}$, is found
by integration over the mean, $z_m$, of the galaxy redshift
distribution within $\pm 1 \sigma$ limits

\begin{equation}
N(<M_{lim}) = \int^{z_m + \sigma}_{z_m - \sigma} \frac{dN}{dz} dz
\end{equation}

No correction is made for completeness or
selection efficiency effects (i.e. an effective volume correction),
since these corrections are specific to each H10 and O10 data set, and
are not transferrable to the UVUDF data.

The H10 and O10 luminosity functions were computed at rest-frame 1500 \Angstrom,
which for F225W, F275W, F336W dropout galaxies at redshifts $z\sim1.7,
2.1, 2.7$, correspond to observed-frame 4050, 4650, 5550 \Angstrom
~respectively. However, H10, O10 and the UVUDF catalogs are selected
from different wavebands (e.g. UVUDF uses a $F435W$-selected
catalog). To calculate the number density of LBGs,
the upper magnitude limit for ERS and UVUDF were modified by 
color-correction terms. These terms account for the difference between
rest-frame 1500 \Angstrom ~and the catalog detection band. An estimate
of the upper magnitude limit at rest-frame 1500 \Angstrom ~is found by
interpolating between the limits for the two closest photometric bands.
These correction factors, $dm$, were added to
the catalog detection limits, $m_{limit}$, to determine upper limits
of integration for the luminosity functions, $m_{limit} = m_{detect}
+ dm$. 
For UVUDF data, we used a limit of $m_{detect} = 28$\ to avoid the
magnitude range in which sources can be lost to CTI. Correction
factors were found to be $dm = +0.130, -0.066, -0.236$ for F225W,
F275W, F336W dropouts respectively.  

For verification, we use the reported LF to predict the number counts
that were observed in the ERS data themselves, the expected number
counts in the UVUDF at a depth comparable to the shallower ERS data
(at which the UVUDF is highly complete), and expected number counts in
the UVUDF to its full depth. We note that the O10 and H10 LFs included substantial
corrections for incompleteness, so we do not expect the number of
galaxies predicted in the ERS to match the observations.

Predicted source counts for each selection method and dropout filter
are presented in Table \ref{TABLE:DropoutLargeSummary}. In comparing
the predictions to the observations, it is important to consider
cosmic variance. For LBGs at these redshifts in fields the size of ERS
and UDF, cosmic variance could be a large effect 
\citep[$\sim$20-30\%, bias=1.5,][]{Somerville:2004, Rafelski:2009, Moster:2011}. 
However, in practice these fields are so
close together on the sky (see Figure~\ref{fig:acs_par_footprint}) that
they are not independent in terms of large scale structure. In
addition, we do not correct for incompleteness effects at the faint
end of the UVUDF data. With these caveats in mind, we conclude that
the predictions are generally consistent with the LBGs that we observe.

\subsection{Resolving Galaxy Structure}

\begin{figure}[t!]
\begin{center}
  \includegraphics[scale=0.35]{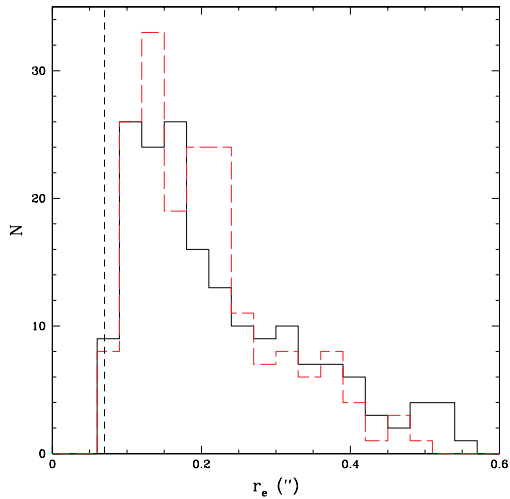} \end{center}
  \vspace{-3mm}
\caption{ \label{fig:sizes} Distribution of effective radii for
  galaxies in the UVUDF with $m_{275}<27.5$ and $0.5 < z_{phot} <
  1.5$. We plot both the F275W (solid) and I-band (long-dashed)
  distributions, with the resolution limit marked with a vertical
  dashed line. Despite visible differences in the morphology between
  the rest-frame UV and rest-frame optical, the sizes remain
  approximately constant.}
\end{figure}

The deep WFC3 UVIS data provide the depth and resolution that allow us
to study star-forming regions at $ z \sim 1$\ in the rest-frame UV. We
identified a sample of 179 galaxies with $m_{275}<27.5$ and
$0.5<z_{phot}<1.5$, where photometric redshifts are taken from
\cite{Rafelski:2009}. We find that galaxies frequently exhibit UV
irregular morphologies and compact sizes (Figure \ref{fig:sizes}), with
a median effective radius of $0\farcs19 \pm 0\farcs01$ ($1.5$~kpc at
$z=1$) in the F275W filter. The F275W sizes are broadly consistent
with those measured at rest-frame $\sim 4000$\AA\, which is probed by
ACS I-band for $0.5\leq z<1$ and z-band for $1\leq z<1.5$. At these
wavelengths, we find a median size of $0\farcs18 \pm 0\farcs01$, suggesting that
the recent star formation is occurring on the same spatial scale as
previous generations of stars. However, when we deconstruct the
galaxies clump-by-clump, clear morphological differences begin to
emerge.

\begin{figure}[t!]
\begin{center}
\includegraphics[scale=0.7]{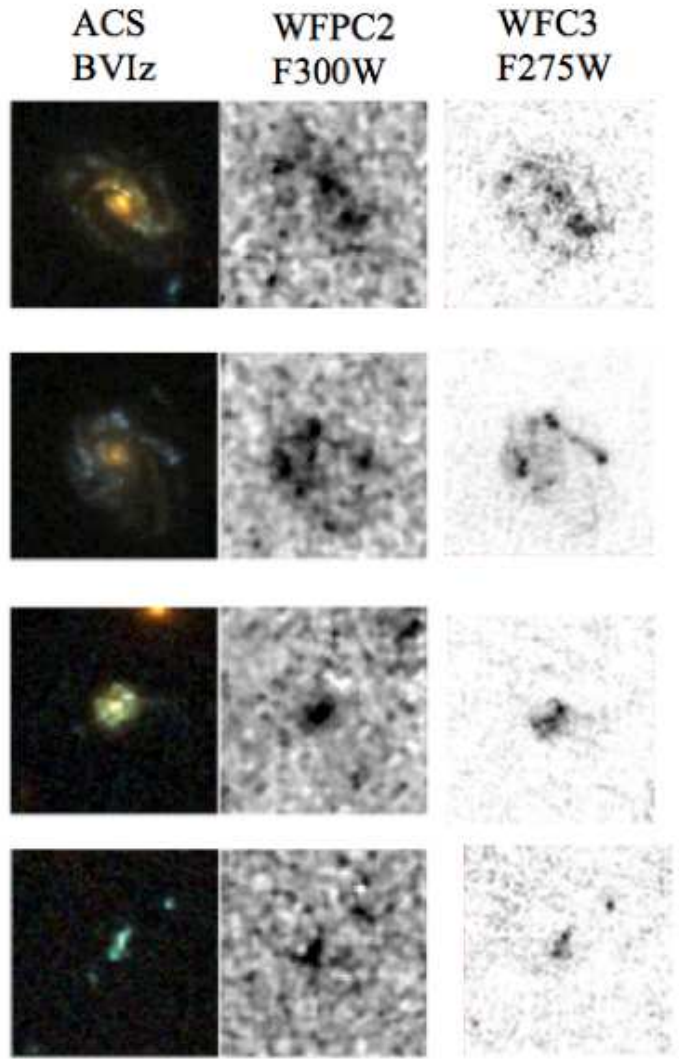}
\end{center}
\caption{ \label{fig:morph1} HST gallery of clumpy spiral and
  irregular galaxies in the HUDF.  (From left to right) Each panel
  shows (a) ACS BViz color combined image, UV WFPC2 F300W images from
  Voyer et al. (2009), and UV WFC3 F275W image of the combined Epoch 1
  and 2.  All images are 5\arcsec$\times$5\arcsec. Photometric
  redshifts of each galaxy from top to bottom are 0.63, 0.57, 0.77,
  1.18.  }
\end{figure}

In Figure \ref{fig:morph1}\ we show some examples of $z\sim1$ galaxies.
The object in the second row is at $z_{\rm phot}=0.67$, where the
F275W probes the light from short-lived O and B stars. In the UV, most
of the light is concentrated in a few bright clumps. However, images
at longer wavelengths (left column of Figure \ref{fig:morph1}) reveal
that these clumps are within the disk of a well defined spiral galaxy,
with a clear bulge component at the center. If seen only in the UV,
this object may resemble the clump-cluster galaxies observed at $z>2$
\citep[e.g.][]{Elmegreen:2005a}, and predicted to form by
fragmentation within gas rich disks \citep{Ceverino:2012}. Clumps are
predicted to migrate toward the center of the disk and coaelesce to
form a bulge, which eventually should stabilize the disk \citep{Dekel:2009b}. 

UV-bright clumps were previously seen in the same object using
WFPC2/F300W images (Voyer et al 2009), but the significantly higher
resolution of the WFC3 UVIS data (WFPC2/F300W FWHM=0\farcs27 compared
to the WFC3/F275W FWHM=0\farcs11), enables us to measure star-forming
regions as small as $\sim$0.8~kpc (at $z=0.67$), reaching 8$\sigma$
above the background level. One of the clumps that is unresolved in
the WFPC2 images is clearly resolved into two clumps with diameters of
1.0~kpc and 1.5~kpc in the WFC3 image. We have also identified clumps
that do not appear to reside in a larger optical disk (bottom row Figure
\ref{fig:morph1}). This object is at $z_{\rm phot}=1.18$ and contains
clumps with sizes ranging from $0.7 - 1.6$~kpc.

\section{Summary}
\label{summary}

The UVUDF project obtained WFC3/UVIS observations of the Hubble
Ultra Deep Field in three NUV filters, F225W, F275W, and F336W (Figure
\ref{fig:filter}).  The UDF was observed with each filter for a total
of thirty orbits. The data were taken in three observing epochs with
three orientation angles (Figure \ref{fig:uvis_footprint}).  The data
in the first two epochs were taken with $2\times2$\ binning of the CCD
readout in order to reduce the read noise that limited the
sensitivity. For Epoch 3, as described in this paper (Section
\ref{cteintro}), the observing strategy was changed to use the WFC3
post-flash capability to add additional background to the observations
in order to mitigate the degradation of the CCD charge transfer
efficiency. The post-flashed data were taken without binning the CCD
readout, because the additional background noise dominated the read noise.
Coordinated parallel observations were obtained with ACS/WFC3 in order
to provide very deep B-band fields, and the Epoch 3 parallels fall on
top of one of the HUDF09 deep optical parallel fields (Figure~\ref{fig:acs_par_footprint}).

The UVUDF observations present several data reduction challenges.  The
team has produced new calibration files for the binned data using
binned and unbinned calibration observations obtained by STScI.
In addition, we have reprocessed darks for all the data, modeling each 
dark's gradient and using a more aggressive approach to flag hot and warm pixels
(see Section \ref{calpipe}). 

The UVUDF data provide a stark demonstration of the effects of charge
transfer inefficiency.  In this paper, we provide evidence that CTI
causes increased scatter in the photometry of sources far from the CCD
readout (Figures \ref{fig:magepoch12}, \ref{fig:magepoch2ab}, and
\ref{fig:readepoch12}).  The application of a statistical correction
to the source flux densities based upon distance from readout is shown
to reduce the scatter but not down to the level predicted from sky and
read noise.  We also find evidence that some faint sources far from
the readout are lost to traps on the CCD in the data that were taken
without post-flash (Figures \ref{fig:ctemaghist}, \ref{fig:ctelost},
and \ref{fig:ctelost2}).   We agree with the STScI recommendation that future
WFC3 UVIS observations that require very sensitive measurements use
the post-flash. The CTI also has demonstrated effects on the
observed shapes of sources in the UVIS images, elongating them in the
direction of the readout (Figure~\ref{fig:ctemorph}).  This effect is
problematic for both astrometric alignment and morphological analysis.
STScI has released a preliminary version of software to apply a pixel-based
correction for the CTI, but it will need significant testing and
verification before it is stable enough to justify its use in
producing enhanced science products for the archive.

The UVUDF data complete HST's panchromatic coverage of the Hubble
Ultra Deep Field. These data are applicable to a wide range of
science topics. The measurement of the UV luminosity function,
together with the mass function measured at longer wavelengths, will
provide a statistical picture of the history of star formation
during its peak epoch. The superb spatial resolution of UVIS will
allow detailed analysis of star-forming ``clumps'' in galaxies,
extending results obtained from optical images of $z\sim 2$\ sources
to later times and exploring the build up of normal galaxies.  The UV
sensitivity and spatial resolution will provide vital tests of the
escape fraction of Lyman continuum photons from sources at $z<3$, and
of the star formation rate efficiency of neutral atomic-dominated hydrogen
gas at $z\sim1-3$.  Finally, the new UV measurements enable
significant improvements in the estimation of photometric redshifts.
These several science investigations will be presented by the UVUDF
team in later papers.

In the current paper, we have presented a preliminary analysis of the
galaxies observed in the UVUDF. We used the UVUDF data to select Lyman break
galaxies at redshifts 1.7, 2.1, and 2.7.  We find that the number
density of dropouts is largely consistent with the number predicted by the
published luminosity functions based on measurements in the ERS.  In
addition, we confirm that UVUDF images of clumpy galaxies
at $z\sim 1$\ have sufficient sensitivity and spatial resolution to
support the planned analysis of the evolution of star-forming clumps.

There are many science uses for UV images of the UDF beyond those
outlined above. This Treasury project will support a broad range of
archival research. At the moment, we are limited by the need to
continue characterizing and correcting the CTI effects. We expect
that the CTI correction software will become stable in the coming
year. When it has been robustly verified, we will produce enhanced
science products to be distributed by the Mikulski Archive for Space
Telescopes (MAST).

The UVUDF observations are currently the most sensitive component of a
new generation of surveys that exploit the unique capabilities of
WFC3/UVIS.  The surveys will leave a legacy of UV imaging for use in a
wide range of research.  The next logical step in expanding HST's UV
legacy will be deep observations over a wider area than the UDF, in
order to sample the variety of galaxy populations and their
environments.  These vital observations will greatly augment studies
with the next generation of telescopes such as ALMA and JWST.

\acknowledgments

 We would like to thank the WFC3 team at the Space Telescope Science
 Institute for their help with solving new calibration and CTE challenges in
 the binned data.  We also thank our Program Coordinator, Anthony
 Roman, and our Contact Scientist, John Mackenty.
 Support for HST Program GO-12534 was provided by NASA
 through grants from the Space Telescope Science Institute, which is
 operated by the Association of Universities for Research in
 Astronomy, Inc., under NASA contract NAS5-26555.

 \facility{} Facilities:  HST (ACS/WF3, WFC3/UVIS)

\bibliography{uvudf_overview,manual}

\begin{thebibliography}{121}
\expandafter\ifx\csname natexlab\endcsname\relax\def\natexlab#1{#1}\fi

\bibitem[{Adelberger \& Steidel(2000)}]{Adelberger:2000}
Adelberger, K.~L. \& Steidel, C.~C. 2000, \apj, 544, 218

\bibitem[{Adelberger {et~al.}(2004)Adelberger, Steidel, Shapley, Hunt, Erb,
  Reddy, \& Pettini}]{Adelberger:2004}
Adelberger, K.~L., Steidel, C.~C., Shapley, A.~E., Hunt, M.~P., Erb, D.~K.,
  Reddy, N.~A., \& Pettini, M. 2004, \apj, 607, 226

\bibitem[{Anderson(2013)}]{Anderson:2013a}
Anderson, J. 2013, Instructions for Using the Alpha-Release of the WFC3/UVIS
  Pixel-based CTE Correction, Tech. rep., STScI

\bibitem[{Anderson \& Bedin(2010)}]{Anderson:2010}
Anderson, J. \& Bedin, L.~R. 2010, \pasp, 122, 1035

\bibitem[{Anderson {et~al.}(2012)Anderson, MacKenty, Baggett, \&
  Noeske}]{Anderson:2012c}
Anderson, J., MacKenty, J.~W., Baggett, S., \& Noeske, K. 2012, The Efficacy of
  Post-Flashing for Mitigating CTE-Losses in WFC3/UVIS Images, Tech. rep.,
  STScI

\bibitem[{Balestra {et~al.}(2010)Balestra, Mainieri, Popesso, Dickinson,
  Nonino, Rosati, Teimoorinia, Vanzella, {et~al.}}]{Balestra:2010}
Balestra, I., Mainieri, V., Popesso, P., Dickinson, M., Nonino, M., Rosati, P.,
  Teimoorinia, H., Vanzella, E., {et~al.} 2010, \aap, 512, 12

\bibitem[{Becker {et~al.}(2007)Becker, Silvestri, Owen, Ivezi{\'c}, \&
  Lupton}]{Becker:2007}
Becker, A.~C., Silvestri, N.~M., Owen, R.~E., Ivezi{\'c}, {\v Z}., \& Lupton,
  R.~H. 2007, \pasp, 119, 1462

\bibitem[{Beckwith {et~al.}(2006)Beckwith, Stiavelli, Koekemoer, Caldwell,
  Ferguson, Hook, Lucas, Bergeron, {et~al.}}]{Beckwith:2006}
Beckwith, S. V.~W., Stiavelli, M., Koekemoer, A.~M., Caldwell, J. A.~R.,
  Ferguson, H.~C., Hook, R., Lucas, R.~A., Bergeron, L.~E., {et~al.} 2006, \aj,
  132, 1729

\bibitem[{Bertin \& Arnouts(1996)}]{Bertin:1996}
Bertin, E. \& Arnouts, S. 1996, \aaps, 117, 393

\bibitem[{Boissier {et~al.}(2007)Boissier, Gil~de Paz, Boselli, Madore, Buat,
  Cortese, Burgarella, Mu{\~n}oz-Mateos, {et~al.}}]{Boissier:2007}
Boissier, S., Gil~de Paz, A., Boselli, A., Madore, B.~F., Buat, V., Cortese,
  L., Burgarella, D., Mu{\~n}oz-Mateos, J.~C., {et~al.} 2007, \apjs, 173, 524

\bibitem[{Borders \& Baggett(2009)}]{Borders:2009}
Borders, T. \& Baggett, S. 2009, WFC3 Instrument Science Report, 2009-16

\bibitem[{Bouwens {et~al.}(2004)Bouwens, Illingworth, Blakeslee, Broadhurst, \&
  Franx}]{Bouwens:2004}
Bouwens, R.~J., Illingworth, G.~D., Blakeslee, J.~P., Broadhurst, T.~J., \&
  Franx, M. 2004, arXiv

\bibitem[{Bouwens {et~al.}(2006)Bouwens, Illingworth, Blakeslee, \&
  Franx}]{Bouwens:2006}
Bouwens, R.~J., Illingworth, G.~D., Blakeslee, J.~P., \& Franx, M. 2006, \apj,
  653, 53

\bibitem[{Bouwens {et~al.}(2012)Bouwens, Illingworth, Oesch, Franx, Labb{\'e},
  Trenti, van Dokkum, Carollo, {et~al.}}]{Bouwens:2012b}
Bouwens, R.~J., Illingworth, G.~D., Oesch, P.~A., Franx, M., Labb{\'e}, I.,
  Trenti, M., van Dokkum, P., Carollo, C.~M., {et~al.} 2012, \apj, 754, 83

\bibitem[{Bouwens {et~al.}(2011)Bouwens, Illingworth, Oesch, Labb{\'e}, Trenti,
  van Dokkum, Franx, Stiavelli, {et~al.}}]{Bouwens:2011b}
Bouwens, R.~J., Illingworth, G.~D., Oesch, P.~A., Labb{\'e}, I., Trenti, M.,
  van Dokkum, P., Franx, M., Stiavelli, M., {et~al.} 2011, \apj, 737, 90

\bibitem[{Bouwens {et~al.}(2010)Bouwens, Illingworth, Oesch, Stiavelli, Dokkum,
  Trenti, Magee, Labb{\'e}, {et~al.}}]{Bouwens:2010}
Bouwens, R.~J., Illingworth, G.~D., Oesch, P.~A., Stiavelli, M., Dokkum, P.~v.,
  Trenti, M., Magee, D., Labb{\'e}, I., {et~al.} 2010, \apjl, 709, L133

\bibitem[{Bridge {et~al.}(2010)Bridge, Teplitz, Siana, Scarlata, Conselice,
  Ferguson, Brown, Salvato, {et~al.}}]{Bridge:2010}
Bridge, C.~R., Teplitz, H.~I., Siana, B., Scarlata, C., Conselice, C.~J.,
  Ferguson, H.~C., Brown, T.~M., Salvato, M., {et~al.} 2010, \apj, 720, 465

\bibitem[{Buat {et~al.}(2010)Buat, Giovannoli, Burgarella, Altieri, Amblard,
  Arumugam, Aussel, Babbedge, {et~al.}}]{Buat:2010}
Buat, V., Giovannoli, E., Burgarella, D., Altieri, B., Amblard, A., Arumugam,
  V., Aussel, H., Babbedge, T., {et~al.} 2010, \mnras, 409, L1

\bibitem[{Cawley {et~al.}(2001)Cawley, Goudfrooij, Whitmore, \&
  Stiavelli}]{Cawley:2001}
Cawley, L., Goudfrooij, P., Whitmore, B., \& Stiavelli, M. 2001, Instrument
  Science Report WFC3, 2001-05

\bibitem[{Ceverino {et~al.}(2012)Ceverino, Dekel, Mandelker, Bournaud, Burkert,
  Genzel, \& Primack}]{Ceverino:2012}
Ceverino, D., Dekel, A., Mandelker, N., Bournaud, F., Burkert, A., Genzel, R.,
  \& Primack, J. 2012, \mnras, 420, 3490

\bibitem[{Chen {et~al.}(2002)Chen, McCarthy, Marzke, Wilson, Carlberg, Firth,
  Persson, Sabbey, {et~al.}}]{Chen:2002}
Chen, H.-W., McCarthy, P.~J., Marzke, R.~O., Wilson, J., Carlberg, R.~G.,
  Firth, A.~E., Persson, S.~E., Sabbey, C.~N., {et~al.} 2002, \apj, 570, 54

\bibitem[{Cimatti {et~al.}(2008)Cimatti, Cassata, Pozzetti, Kurk, Mignoli,
  Renzini, Daddi, Bolzonella, {et~al.}}]{Cimatti:2008}
Cimatti, A., Cassata, P., Pozzetti, L., Kurk, J., Mignoli, M., Renzini, A.,
  Daddi, E., Bolzonella, M., {et~al.} 2008, \aap, 482, 21

\bibitem[{Coe {et~al.}(2006)Coe, Ben{\'\i}tez, S{\'a}nchez, Jee, Bouwens, \&
  Ford}]{Coe:2006}
Coe, D., Ben{\'\i}tez, N., S{\'a}nchez, S.~F., Jee, M., Bouwens, R., \& Ford,
  H. 2006, \aj, 132, 926

\bibitem[{Coe {et~al.}(2013)Coe, Zitrin, Carrasco, Shu, Zheng, Postman,
  Bradley, Koekemoer, {et~al.}}]{Coe:2013}
Coe, D., Zitrin, A., Carrasco, M., Shu, X., Zheng, W., Postman, M., Bradley,
  L., Koekemoer, A., {et~al.} 2013, \apj, 762, 32

\bibitem[{Daddi {et~al.}(2010)Daddi, Bournaud, Walter, Dannerbauer, Carilli,
  Dickinson, Elbaz, Morrison, {et~al.}}]{Daddi:2010a}
Daddi, E., Bournaud, F., Walter, F., Dannerbauer, H., Carilli, C.~L.,
  Dickinson, M., Elbaz, D., Morrison, G.~E., {et~al.} 2010, \apj, 713, 686

\bibitem[{Daddi {et~al.}(2007)Daddi, Dickinson, Morrison, Chary, Cimatti,
  Elbaz, Frayer, Renzini, {et~al.}}]{Daddi:2007}
Daddi, E., Dickinson, M., Morrison, G., Chary, R., Cimatti, A., Elbaz, D.,
  Frayer, D., Renzini, A., {et~al.} 2007, \apj, 670, 156

\bibitem[{Dekel {et~al.}(2009)Dekel, Sari, \& Ceverino}]{Dekel:2009b}
Dekel, A., Sari, R., \& Ceverino, D. 2009, \apj, 703, 785

\bibitem[{Dressler {et~al.}(2012)Dressler, Spergel, Mountain, Postman, Elliott,
  Bendek, Bennett, Dalcanton, {et~al.}}]{Dressler:2012a}
Dressler, A., Spergel, D., Mountain, M., Postman, M., Elliott, E., Bendek, E.,
  Bennett, D., Dalcanton, J., {et~al.} 2012, arXiv, 7809

\bibitem[{Elbaz {et~al.}(2011)Elbaz, Dickinson, Hwang, D{\'\i}az-Santos,
  Magdis, Magnelli, Le~Borgne, Galliano, {et~al.}}]{Elbaz:2011}
Elbaz, D., Dickinson, M., Hwang, H.~S., D{\'\i}az-Santos, T., Magdis, G.,
  Magnelli, B., Le~Borgne, D., Galliano, F., {et~al.} 2011, \aap, 533, 119

\bibitem[{Ellis {et~al.}(2013)Ellis, McLure, Dunlop, Robertson, Ono, Schenker,
  Koekemoer, Bowler, {et~al.}}]{Ellis:2013}
Ellis, R.~S., McLure, R.~J., Dunlop, J.~S., Robertson, B.~E., Ono, Y.,
  Schenker, M.~A., Koekemoer, A., Bowler, R. A.~A., {et~al.} 2013, \apjl, 763,
  L7

\bibitem[{Elmegreen {et~al.}(2008)Elmegreen, Bournaud, \&
  Elmegreen}]{Elmegreen:2008}
Elmegreen, B.~G., Bournaud, F., \& Elmegreen, D.~M. 2008, \apj, 688, 67

\bibitem[{Elmegreen \& Elmegreen(2005)}]{Elmegreen:2005b}
Elmegreen, B.~G. \& Elmegreen, D.~M. 2005, \apj, 627, 632

\bibitem[{Elmegreen {et~al.}(2005)Elmegreen, Elmegreen, \&
  Ferguson}]{Elmegreen:2005a}
Elmegreen, D.~M., Elmegreen, B.~G., \& Ferguson, T.~E. 2005, \apj, 623, L71

\bibitem[{Feldmeier {et~al.}(2002)Feldmeier, Mihos, Morrison, Rodney, \&
  Harding}]{Feldmeier:2002}
Feldmeier, J.~J., Mihos, J.~C., Morrison, H.~L., Rodney, S.~A., \& Harding, P.
  2002, \apj, 575, 779

\bibitem[{Finkelstein {et~al.}(2012)Finkelstein, Papovich, Salmon, Finlator,
  Dickinson, Ferguson, Giavalisco, Koekemoer, {et~al.}}]{Finkelstein:2012a}
Finkelstein, S.~L., Papovich, C., Salmon, B., Finlator, K., Dickinson, M.,
  Ferguson, H.~C., Giavalisco, M., Koekemoer, A.~M., {et~al.} 2012, \apj, 756,
  164

\bibitem[{{Ford} {et~al.}(2002){Ford}, {Illingworth}, {Clampin},
  {et~al.}}]{Ford:2002}
{Ford}, H.~C., {Illingworth}, G.~D., {Clampin}, M., {et~al.} 2002, in Bulletin
  of the American Astronomical Society, Vol.~34, American Astronomical Society
  Meeting Abstracts \#200, 675

\bibitem[{Fruchter \& Hook(2002)}]{Fruchter:2002}
Fruchter, A.~S. \& Hook, R.~N. 2002, \pasp, 114, 144

\bibitem[{Gardner {et~al.}(2006)Gardner, Mather, Clampin, Doyon, Greenhouse,
  Hammel, Hutchings, Jakobsen, {et~al.}}]{Gardner:2006}
Gardner, J.~P., Mather, J.~C., Clampin, M., Doyon, R., Greenhouse, M.~A.,
  Hammel, H.~B., Hutchings, J.~B., Jakobsen, P., {et~al.} 2006, \ssr, 123, 485

\bibitem[{Gawiser {et~al.}(2006)Gawiser, van Dokkum, Herrera, Maza, Castander,
  Infante, Lira, Quadri, {et~al.}}]{Gawiser:2006}
Gawiser, E., van Dokkum, P.~G., Herrera, D., Maza, J., Castander, F.~J.,
  Infante, L., Lira, P., Quadri, R., {et~al.} 2006, \apjs, 162, 1

\bibitem[{Genzel {et~al.}(2008)Genzel, Burkert, Bouch{\'e}, Cresci,
  F{\"o}rster~Schreiber, Shapley, Shapiro, Tacconi, {et~al.}}]{Genzel:2008}
Genzel, R., Burkert, A., Bouch{\'e}, N., Cresci, G., F{\"o}rster~Schreiber,
  N.~M., Shapley, A., Shapiro, K., Tacconi, L.~J., {et~al.} 2008, \apj, 687, 59

\bibitem[{Giavalisco {et~al.}(2004)Giavalisco, Ferguson, Koekemoer, Dickinson,
  Alexander, Bauer, Bergeron, Biagetti, {et~al.}}]{Giavalisco:2004}
Giavalisco, M., Ferguson, H.~C., Koekemoer, A.~M., Dickinson, M., Alexander,
  D.~M., Bauer, F.~E., Bergeron, J., Biagetti, C., {et~al.} 2004, \apj, 600,
  L93

\bibitem[{Gnedin \& Kravtsov(2010)}]{Gnedin:2010}
Gnedin, N.~Y. \& Kravtsov, A.~V. 2010, \apj, 714, 287

\bibitem[{Grogin {et~al.}(2011)Grogin, Kocevski, Faber, Ferguson, Koekemoer,
  Riess, Acquaviva, Alexander, {et~al.}}]{Grogin:2011}
Grogin, N.~A., Kocevski, D.~D., Faber, S.~M., Ferguson, H.~C., Koekemoer,
  A.~M., Riess, A.~G., Acquaviva, V., Alexander, D.~M., {et~al.} 2011, \apjs,
  197, 35

\bibitem[{Hathi {et~al.}(2010)Hathi, Ryan, Cohen, Yan, Windhorst, McCarthy,
  O'Connell, Koekemoer, {et~al.}}]{Hathi:2010}
Hathi, N.~P., Ryan, R. E.~J., Cohen, S.~H., Yan, H., Windhorst, R.~A.,
  McCarthy, P.~J., O'Connell, R.~W., Koekemoer, A.~M., {et~al.} 2010, \apj,
  720, 1708

\bibitem[{Iwata {et~al.}(2009)Iwata, Inoue, Matsuda, Furusawa, Hayashino,
  Kousai, Akiyama, Yamada, {et~al.}}]{Iwata:2009}
Iwata, I., Inoue, A.~K., Matsuda, Y., Furusawa, H., Hayashino, T., Kousai, K.,
  Akiyama, M., Yamada, T., {et~al.} 2009, \apj, 692, 1287

\bibitem[{Kajisawa {et~al.}(2006)Kajisawa, Konishi, Suzuki, Tokoku, Uchimoto,
  {Katsuno}, Yoshikawa, Akiyama, {et~al.}}]{Kajisawa:2006}
Kajisawa, M., Konishi, M., Suzuki, R., Tokoku, C., Uchimoto, Y., {Katsuno},
  Yoshikawa, T., Akiyama, M., {et~al.} 2006, Publications of the Astronomical
  Society of Japan, 58, 951

\bibitem[{Kennicutt(1998{\natexlab{a}})}]{Kennicutt:1998b}
Kennicutt, R.~C. 1998{\natexlab{a}}, \araa, 36, 189

\bibitem[{Kennicutt(1998{\natexlab{b}})}]{Kennicutt:1998a}
---. 1998{\natexlab{b}}, \apj, 498, 541

\bibitem[{Kere{\v s} {et~al.}(2009)Kere{\v s}, Katz, Fardal, Dav{\'e}, \&
  Weinberg}]{Keres:2009a}
Kere{\v s}, D., Katz, N., Fardal, M., Dav{\'e}, R., \& Weinberg, D.~H. 2009,
  \mnras, 395, 160

\bibitem[{Koekemoer {et~al.}(2012)Koekemoer, Ellis, McLure, Dunlop, Robertson,
  Ono, Schenker, Ouchi, {et~al.}}]{Koekemoer:2013}
Koekemoer, A.~M., Ellis, R.~S., McLure, R.~J., Dunlop, J.~S., Robertson, B.~E.,
  Ono, Y., Schenker, M.~A., Ouchi, M., {et~al.} 2012, arXiv

\bibitem[{Koekemoer {et~al.}(2011)Koekemoer, Faber, Ferguson, Grogin, Kocevski,
  Koo, Lai, Lotz, {et~al.}}]{Koekemoer:2011}
Koekemoer, A.~M., Faber, S.~M., Ferguson, H.~C., Grogin, N.~A., Kocevski,
  D.~D., Koo, D.~C., Lai, K., Lotz, J.~M., {et~al.} 2011, \apjs, 197, 36

\bibitem[{Kron(1980)}]{Kron:1980}
Kron, R.~G. 1980, \apjs, 43, 305

\bibitem[{Kurczynski {et~al.}(2012)Kurczynski, Gawiser, Huynh, Ivison,
  Treister, Smail, Blanc, Cardamone, {et~al.}}]{Kurczynski:2012}
Kurczynski, P., Gawiser, E., Huynh, M., Ivison, R.~J., Treister, E., Smail, I.,
  Blanc, G.~A., Cardamone, C.~N., {et~al.} 2012, \apj, 750, 117

\bibitem[{Kurk {et~al.}(2013)Kurk, Cimatti, Daddi, Mignoli, Pozzetti,
  Dickinson, Bolzonella, Zamorani, {et~al.}}]{Kurk:2013}
Kurk, J., Cimatti, A., Daddi, E., Mignoli, M., Pozzetti, L., Dickinson, M.,
  Bolzonella, M., Zamorani, G., {et~al.} 2013, \aap, 549, 63

\bibitem[{Labb{\'e} {et~al.}(2006)Labb{\'e}, Bouwens, Illingworth, \&
  Franx}]{Labbe:2006}
Labb{\'e}, I., Bouwens, R., Illingworth, G.~D., \& Franx, M. 2006, \apj, 649,
  L67

\bibitem[{Labb{\'e} {et~al.}(2003)Labb{\'e}, Franx, Rudnick, Schreiber, Rix,
  Moorwood, van Dokkum, van~der Werf, {et~al.}}]{Labbe:2003}
Labb{\'e}, I., Franx, M., Rudnick, G., Schreiber, N. M.~F., Rix, H.-W.,
  Moorwood, A., van Dokkum, P.~G., van~der Werf, P., {et~al.} 2003, \aj, 125,
  1107

\bibitem[{Laureijs {et~al.}(2012)Laureijs, Gondoin, Duvet, Saavedra~Criado,
  Hoar, Amiaux, Augu{\`e}res, Cole, {et~al.}}]{Laureijs:2012}
Laureijs, R., Gondoin, P., Duvet, L., Saavedra~Criado, G., Hoar, J., Amiaux,
  J., Augu{\`e}res, J.~L., Cole, R., {et~al.} 2012, Space Telescopes and
  Instrumentation 2012: Optical, 8442

\bibitem[{Le~F{\`e}vre {et~al.}(2005)Le~F{\`e}vre, Vettolani, Garilli, Tresse,
  Bottini, Le~Brun, Maccagni, Picat, {et~al.}}]{LeFevre:2005}
Le~F{\`e}vre, O., Vettolani, G., Garilli, B., Tresse, L., Bottini, D., Le~Brun,
  V., Maccagni, D., Picat, J.~P., {et~al.} 2005, \aap, 439, 845

\bibitem[{Lee {et~al.}(2012{\natexlab{a}})Lee, Alberts, Atlee, Dey, Pope,
  Jannuzi, Reddy, \& Brown}]{Lee:2012b}
Lee, K.-S., Alberts, S., Atlee, D., Dey, A., Pope, A., Jannuzi, B.~T., Reddy,
  N., \& Brown, M. J.~I. 2012{\natexlab{a}}, \apjl, 758, L31

\bibitem[{Lee {et~al.}(2012{\natexlab{b}})Lee, Ferguson, Wiklind, Dahlen,
  Dickinson, Giavalisco, Grogin, Papovich, {et~al.}}]{Lee:2012}
Lee, K.-S., Ferguson, H.~C., Wiklind, T., Dahlen, T., Dickinson, M.~E.,
  Giavalisco, M., Grogin, N., Papovich, C., {et~al.} 2012{\natexlab{b}}, \apj,
  752, 66

\bibitem[{MacKenty \& Smith(2012)}]{MacKenty:2012}
MacKenty, J.~W. \& Smith, L. 2012, CTE White Paper, Tech. rep., STScI

\bibitem[{Martel {et~al.}(2008)Martel, Baggett, Bushouse, \&
  Sabbi}]{Martel:2008}
Martel, A., Baggett, S., Bushouse, H., \& Sabbi, E. 2008, Technical Instrument
  Report WFC3, 2008-01

\bibitem[{Massey(2010)}]{Massey:2010b}
Massey, R. 2010, \mnras, 409, L109

\bibitem[{Massey {et~al.}(2010)Massey, Stoughton, Leauthaud, Rhodes, Koekemoer,
  Ellis, \& Shaghoulian}]{Massey:2010a}
Massey, R., Stoughton, C., Leauthaud, A., Rhodes, J., Koekemoer, A., Ellis, R.,
  \& Shaghoulian, E. 2010, \mnras, 401, 371

\bibitem[{Meurer {et~al.}(1999)Meurer, Heckman, \& Calzetti}]{Meurer:1999}
Meurer, G.~R., Heckman, T.~M., \& Calzetti, D. 1999, \apj, 521, 64

\bibitem[{Moster {et~al.}(2011)Moster, Somerville, Newman, \&
  Rix}]{Moster:2011}
Moster, B.~P., Somerville, R.~S., Newman, J.~A., \& Rix, H.-W. 2011, \apj, 731,
  113

\bibitem[{Nestor {et~al.}(2013)Nestor, Shapley, Kornei, Steidel, \&
  Siana}]{Nestor:2013}
Nestor, D.~B., Shapley, A.~E., Kornei, K.~A., Steidel, C.~C., \& Siana, B.
  2013, \apj, 765, 47

\bibitem[{Noeske {et~al.}(2012)Noeske, Baggett, Bushouse, Petro, Gilliland, \&
  Khozurina-Platais}]{Noeske:2012}
Noeske, K., Baggett, S., Bushouse, H., Petro, L., Gilliland, R., \&
  Khozurina-Platais, V. 2012, Instrument Science Report WFC3, 2012-09

\bibitem[{Nonino {et~al.}(2009)Nonino, Dickinson, Rosati, Grazian, Reddy,
  Cristiani, Giavalisco, Kuntschner, {et~al.}}]{Nonino:2009}
Nonino, M., Dickinson, M., Rosati, P., Grazian, A., Reddy, N., Cristiani, S.,
  Giavalisco, M., Kuntschner, H., {et~al.} 2009, \apjs, 183, 244

\bibitem[{Oesch {et~al.}(2010{\natexlab{a}})Oesch, Bouwens, Carollo,
  Illingworth, Magee, Trenti, Stiavelli, Franx, {et~al.}}]{Oesch:2010eb}
Oesch, P.~A., Bouwens, R.~J., Carollo, C.~M., Illingworth, G.~D., Magee, D.,
  Trenti, M., Stiavelli, M., Franx, M., {et~al.} 2010{\natexlab{a}}, \apjl,
  725, L150

\bibitem[{Oesch {et~al.}(2010{\natexlab{b}})Oesch, Bouwens, Carollo,
  Illingworth, Trenti, Stiavelli, Magee, Labb{\'e}, {et~al.}}]{Oesch:2010b}
Oesch, P.~A., Bouwens, R.~J., Carollo, C.~M., Illingworth, G.~D., Trenti, M.,
  Stiavelli, M., Magee, D., Labb{\'e}, I., {et~al.} 2010{\natexlab{b}}, \apjl,
  709, L21

\bibitem[{Oesch {et~al.}(2010{\natexlab{c}})Oesch, Bouwens, Illingworth,
  Carollo, Franx, Labb{\'e}, Magee, Stiavelli, {et~al.}}]{Oesch:2010a}
Oesch, P.~A., Bouwens, R.~J., Illingworth, G.~D., Carollo, C.~M., Franx, M.,
  Labb{\'e}, I., Magee, D., Stiavelli, M., {et~al.} 2010{\natexlab{c}}, \apjl,
  709, L16

\bibitem[{Oesch {et~al.}(2007)Oesch, Stiavelli, Carollo, Bergeron, Koekemoer,
  Lucas, Pavlovsky, Trenti, {et~al.}}]{Oesch:2007}
Oesch, P.~A., Stiavelli, M., Carollo, C.~M., Bergeron, L.~E., Koekemoer, A.~M.,
  Lucas, R.~A., Pavlovsky, C.~M., Trenti, M., {et~al.} 2007, \apj, 671, 1212

\bibitem[{Overzier {et~al.}(2011)Overzier, Heckman, Wang, Armus, Buat, Howell,
  Meurer, Seibert, {et~al.}}]{Overzier:2011}
Overzier, R.~A., Heckman, T.~M., Wang, J., Armus, L., Buat, V., Howell, J.,
  Meurer, G., Seibert, M., {et~al.} 2011, \apjl, 726, L7

\bibitem[{Pirzkal {et~al.}(2005)Pirzkal, Sahu, Burgasser, Moustakas, Xu,
  Malhotra, Rhoads, Koekemoer, {et~al.}}]{Pirzkal:2005}
Pirzkal, N., Sahu, K.~C., Burgasser, A., Moustakas, L.~A., Xu, C., Malhotra,
  S., Rhoads, J.~E., Koekemoer, A.~M., {et~al.} 2005, \apj, 622, 319

\bibitem[{Pirzkal {et~al.}(2004)Pirzkal, Xu, Malhotra, Rhoads, Koekemoer,
  Moustakas, Walsh, Windhorst, {et~al.}}]{Pirzkal:2004}
Pirzkal, N., Xu, C., Malhotra, S., Rhoads, J.~E., Koekemoer, A.~M., Moustakas,
  L.~A., Walsh, J.~R., Windhorst, R.~A., {et~al.} 2004, \apjs, 154, 501

\bibitem[{Popesso {et~al.}(2009)Popesso, Dickinson, Nonino, Vanzella, Daddi,
  Fosbury, Kuntschner, Mainieri, {et~al.}}]{Popesso:2009}
Popesso, P., Dickinson, M., Nonino, M., Vanzella, E., Daddi, E., Fosbury, R.
  A.~E., Kuntschner, H., Mainieri, V., {et~al.} 2009, \aap, 494, 443

\bibitem[{Rafelski {et~al.}(2011)Rafelski, Wolfe, \& Chen}]{Rafelski:2011}
Rafelski, M., Wolfe, A.~M., \& Chen, H.-W. 2011, \apj, 736, 48

\bibitem[{Rafelski {et~al.}(2009)Rafelski, Wolfe, Cooke, Chen, Armandroff, \&
  Wirth}]{Rafelski:2009}
Rafelski, M., Wolfe, A.~M., Cooke, J., Chen, H.-W., Armandroff, T.~E., \&
  Wirth, G.~D. 2009, \apj, 703, 2033

\bibitem[{Rafelski {et~al.}(2012)Rafelski, Wolfe, Prochaska, Neeleman, \&
  Mendez}]{Rafelski:2012}
Rafelski, M., Wolfe, A.~M., Prochaska, J.~X., Neeleman, M., \& Mendez, A.~J.
  2012, \apj, 755, 89

\bibitem[{Ravindranath {et~al.}(2006)Ravindranath, Giavalisco, Ferguson,
  Conselice, Katz, Weinberg, Lotz, Dickinson, {et~al.}}]{Ravindranath:2006}
Ravindranath, S., Giavalisco, M., Ferguson, H.~C., Conselice, C., Katz, N.,
  Weinberg, M., Lotz, J., Dickinson, M., {et~al.} 2006, \apj, 652, 963

\bibitem[{Reddy {et~al.}(2012{\natexlab{a}})Reddy, Dickinson, Elbaz, Morrison,
  Giavalisco, Ivison, Papovich, Scott, {et~al.}}]{Reddy:2012a}
Reddy, N., Dickinson, M., Elbaz, D., Morrison, G., Giavalisco, M., Ivison, R.,
  Papovich, C., Scott, D., {et~al.} 2012{\natexlab{a}}, \apj, 744, 154

\bibitem[{Reddy {et~al.}(2010)Reddy, Erb, Pettini, Steidel, \&
  Shapley}]{Reddy:2010}
Reddy, N.~A., Erb, D.~K., Pettini, M., Steidel, C.~C., \& Shapley, A.~E. 2010,
  \apj, 712, 1070

\bibitem[{Reddy {et~al.}(2012{\natexlab{b}})Reddy, Pettini, Steidel, Shapley,
  Erb, \& Law}]{Reddy:2012b}
Reddy, N.~A., Pettini, M., Steidel, C.~C., Shapley, A.~E., Erb, D.~K., \& Law,
  D.~R. 2012{\natexlab{b}}, \apj, 754, 25

\bibitem[{Reddy \& Steidel(2009)}]{Reddy:2009}
Reddy, N.~A. \& Steidel, C.~C. 2009, \apj, 692, 778

\bibitem[{Reddy {et~al.}(2008)Reddy, Steidel, Pettini, Adelberger, Shapley,
  Erb, \& Dickinson}]{Reddy:2008}
Reddy, N.~A., Steidel, C.~C., Pettini, M., Adelberger, K.~L., Shapley, A.~E.,
  Erb, D.~K., \& Dickinson, M. 2008, \apjs, 175, 48

\bibitem[{Retzlaff {et~al.}(2010)Retzlaff, Rosati, Dickinson, Vandame,
  Rit{\'e}, Nonino, Cesarsky, \& Team}]{Retzlaff:2010}
Retzlaff, J., Rosati, P., Dickinson, M., Vandame, B., Rit{\'e}, C., Nonino, M.,
  Cesarsky, C., \& Team, T.~G. 2010, \aap, 511, A50

\bibitem[{Rhodes {et~al.}(2010)Rhodes, Leauthaud, Stoughton, Massey, Dawson,
  Kolbe, \& Roe}]{Rhodes:2010}
Rhodes, J., Leauthaud, A., Stoughton, C., Massey, R., Dawson, K., Kolbe, W., \&
  Roe, N. 2010, \pasp, 122, 439

\bibitem[{Rhodes {et~al.}(2007)Rhodes, Massey, Albert, Collins, Ellis, Heymans,
  Gardner, Kneib, {et~al.}}]{Rhodes:2007}
Rhodes, J.~D., Massey, R.~J., Albert, J., Collins, N., Ellis, R.~S., Heymans,
  C., Gardner, J.~P., Kneib, J.-P., {et~al.} 2007, \apjs, 172, 203

\bibitem[{Riess \& Mack(2004)}]{Riess:2004}
Riess, A. \& Mack, J. 2004, Instrument Science Report ACS, 2004-006

\bibitem[{Rix {et~al.}(2004)Rix, Barden, Beckwith, Bell, Borch, Caldwell,
  Haussler, Jahnke, {et~al.}}]{Rix:2004}
Rix, H.-W., Barden, M., Beckwith, S. V.~W., Bell, E.~F., Borch, A., Caldwell,
  J. A.~R., Haussler, B., Jahnke, K., {et~al.} 2004, \apjs, 152, 163

\bibitem[{Rosati {et~al.}(2002)Rosati, Tozzi, Giacconi, Gilli, Hasinger,
  Kewley, Mainieri, Nonino, {et~al.}}]{Rosati:2002}
Rosati, P., Tozzi, P., Giacconi, R., Gilli, R., Hasinger, G., Kewley, L.,
  Mainieri, V., Nonino, M., {et~al.} 2002, \apj, 566, 667

\bibitem[{Sawicki \& Thompson(2005)}]{Sawicki:2005}
Sawicki, M. \& Thompson, D. 2005, \apj, 635, 100

\bibitem[{Scarlata {et~al.}(2007)Scarlata, Carollo, Lilly, Feldmann, Kampczyk,
  Renzini, Cimatti, Halliday, {et~al.}}]{Scarlata:2007}
Scarlata, C., Carollo, C.~M., Lilly, S.~J., Feldmann, R., Kampczyk, P.,
  Renzini, A., Cimatti, A., Halliday, C., {et~al.} 2007, \apjs, 172, 494

\bibitem[{Schechter(1976)}]{Schechter:1976}
Schechter, P. 1976, \apj, 203, 297

\bibitem[{Schmidt(1959)}]{Schmidt:1959}
Schmidt, M. 1959, \apj, 129, 243

\bibitem[{Shapley {et~al.}(2006)Shapley, Steidel, Pettini, Adelberger, \&
  Erb}]{Shapley:2006}
Shapley, A.~E., Steidel, C.~C., Pettini, M., Adelberger, K.~L., \& Erb, D.~K.
  2006, \apj, 651, 688

\bibitem[{Siana {et~al.}(2009)Siana, Smail, Swinbank, Richard, Teplitz, Coppin,
  Ellis, Stark, {et~al.}}]{Siana:2009}
Siana, B., Smail, I., Swinbank, A.~M., Richard, J., Teplitz, H.~I., Coppin, K.
  E.~K., Ellis, R.~S., Stark, D.~P., {et~al.} 2009, \apj, 698, 1273

\bibitem[{Siana {et~al.}(2007)Siana, Teplitz, Colbert, Ferguson, Dickinson,
  Brown, Conselice, de~Mello, {et~al.}}]{Siana:2007}
Siana, B., Teplitz, H.~I., Colbert, J., Ferguson, H.~C., Dickinson, M., Brown,
  T.~M., Conselice, C.~J., de~Mello, D.~F., {et~al.} 2007, \apj, 668, 62

\bibitem[{Siana {et~al.}(2010)Siana, Teplitz, Ferguson, Brown, Giavalisco,
  Dickinson, Chary, de~Mello, {et~al.}}]{Siana:2010}
Siana, B., Teplitz, H.~I., Ferguson, H.~C., Brown, T.~M., Giavalisco, M.,
  Dickinson, M., Chary, R.-R., de~Mello, D.~F., {et~al.} 2010, \apj, 723, 241

\bibitem[{Somerville {et~al.}(2004)Somerville, Lee, Ferguson, Gardner,
  Moustakas, \& Giavalisco}]{Somerville:2004}
Somerville, R.~S., Lee, K., Ferguson, H.~C., Gardner, J.~P., Moustakas, L.~A.,
  \& Giavalisco, M. 2004, \apj, 600, L171

\bibitem[{Steidel {et~al.}(1999)Steidel, Adelberger, Giavalisco, Dickinson, \&
  Pettini}]{Steidel:1999}
Steidel, C.~C., Adelberger, K.~L., Giavalisco, M., Dickinson, M., \& Pettini,
  M. 1999, \apj, 519, 1

\bibitem[{Steidel {et~al.}(2003)Steidel, Adelberger, Shapley, Pettini,
  Dickinson, \& Giavalisco}]{Steidel:2003}
Steidel, C.~C., Adelberger, K.~L., Shapley, A.~E., Pettini, M., Dickinson, M.,
  \& Giavalisco, M. 2003, \apj, 592, 728

\bibitem[{Steidel {et~al.}(1996{\natexlab{a}})Steidel, Giavalisco, Dickinson,
  \& Adelberger}]{Steidel:1996b}
Steidel, C.~C., Giavalisco, M., Dickinson, M., \& Adelberger, K.~L.
  1996{\natexlab{a}}, \aj, 112, 352

\bibitem[{Steidel {et~al.}(1996{\natexlab{b}})Steidel, Giavalisco, Pettini,
  Dickinson, \& Adelberger}]{Steidel:1996a}
Steidel, C.~C., Giavalisco, M., Pettini, M., Dickinson, M., \& Adelberger,
  K.~L. 1996{\natexlab{b}}, \apjl, 462, L17

\bibitem[{Swinbank {et~al.}(2009)Swinbank, Webb, Richard, Bower, Ellis,
  Illingworth, Jones, Kriek, {et~al.}}]{Swinbank:2009}
Swinbank, A.~M., Webb, T.~M., Richard, J., Bower, R.~G., Ellis, R.~S.,
  Illingworth, G., Jones, T., Kriek, M., {et~al.} 2009, \mnras, 400, 1121

\bibitem[{Szokoly {et~al.}(2004)Szokoly, Bergeron, Hasinger, Lehmann, Kewley,
  Mainieri, Nonino, Rosati, {et~al.}}]{Szokoly:2004}
Szokoly, G.~P., Bergeron, J., Hasinger, G., Lehmann, I., Kewley, L., Mainieri,
  V., Nonino, M., Rosati, P., {et~al.} 2004, \apjs, 155, 271

\bibitem[{Teplitz {et~al.}(2011)Teplitz, Chary, Elbaz, Dickinson, Bridge,
  Colbert, Le~Floc'h, Frayer, {et~al.}}]{Teplitz:2011}
Teplitz, H.~I., Chary, R., Elbaz, D., Dickinson, M., Bridge, C., Colbert, J.,
  Le~Floc'h, E., Frayer, D.~T., {et~al.} 2011, \aj, 141, 1

\bibitem[{Thilker {et~al.}(2005)Thilker, Bianchi, Boissier, Gil~de Paz, Madore,
  Martin, Meurer, Neff, {et~al.}}]{Thilker:2005}
Thilker, D.~A., Bianchi, L., Boissier, S., Gil~de Paz, A., Madore, B.~F.,
  Martin, D.~C., Meurer, G.~R., Neff, S.~G., {et~al.} 2005, \apj, 619, L79

\bibitem[{Thompson {et~al.}(2005)Thompson, Illingworth, Bouwens, Dickinson,
  Eisenstein, Fan, Franx, Riess, {et~al.}}]{Thompson:2005}
Thompson, R.~I., Illingworth, G., Bouwens, R., Dickinson, M., Eisenstein, D.,
  Fan, X., Franx, M., Riess, A., {et~al.} 2005, \aj, 130, 1

\bibitem[{van Dokkum {et~al.}(2008)van Dokkum, Franx, Kriek, Holden,
  Illingworth, Magee, Bouwens, Marchesini, {et~al.}}]{vanDokkum:2008}
van Dokkum, P.~G., Franx, M., Kriek, M., Holden, B., Illingworth, G.~D., Magee,
  D., Bouwens, R., Marchesini, D., {et~al.} 2008, \apj, 677, L5

\bibitem[{Vanzella {et~al.}(2008)Vanzella, Cristiani, Dickinson, Giavalisco,
  Kuntschner, Haase, Nonino, Rosati, {et~al.}}]{Vanzella:2008}
Vanzella, E., Cristiani, S., Dickinson, M., Giavalisco, M., Kuntschner, H.,
  Haase, J., Nonino, M., Rosati, P., {et~al.} 2008, \aap, 478, 83

\bibitem[{Vanzella {et~al.}(2005)Vanzella, Cristiani, Dickinson, Kuntschner,
  Moustakas, Nonino, Rosati, Stern, {et~al.}}]{Vanzella:2005}
Vanzella, E., Cristiani, S., Dickinson, M., Kuntschner, H., Moustakas, L.~A.,
  Nonino, M., Rosati, P., Stern, D., {et~al.} 2005, \aap, 434, 53

\bibitem[{Vanzella {et~al.}(2006)Vanzella, Cristiani, Dickinson, Kuntschner,
  Nonino, Rettura, Rosati, Vernet, {et~al.}}]{Vanzella:2006}
Vanzella, E., Cristiani, S., Dickinson, M., Kuntschner, H., Nonino, M.,
  Rettura, A., Rosati, P., Vernet, J., {et~al.} 2006, \aap, 454, 423

\bibitem[{Vanzella {et~al.}(2009)Vanzella, Giavalisco, Dickinson, Cristiani,
  Nonino, Kuntschner, Popesso, Rosati, {et~al.}}]{Vanzella:2009}
Vanzella, E., Giavalisco, M., Dickinson, M., Cristiani, S., Nonino, M.,
  Kuntschner, H., Popesso, P., Rosati, P., {et~al.} 2009, \apj, 695, 1163

\bibitem[{Vanzella {et~al.}(2012)Vanzella, Guo, Giavalisco, Grazian,
  Castellano, Cristiani, Dickinson, Fontana, {et~al.}}]{Vanzella:2012a}
Vanzella, E., Guo, Y., Giavalisco, M., Grazian, A., Castellano, M., Cristiani,
  S., Dickinson, M., Fontana, A., {et~al.} 2012, \apj, 751, 70

\bibitem[{Voyer {et~al.}(2009)Voyer, de~Mello, Siana, Gardner, Quirk, \&
  Teplitz}]{Voyer:2009}
Voyer, E.~N., de~Mello, D.~F., Siana, B., Gardner, J.~P., Quirk, C., \&
  Teplitz, H.~I. 2009, \aj, 138, 598

\bibitem[{Windhorst {et~al.}(2011)Windhorst, Cohen, Hathi, McCarthy, Ryan, Yan,
  Baldry, Driver, {et~al.}}]{Windhorst:2011}
Windhorst, R.~A., Cohen, S.~H., Hathi, N.~P., McCarthy, P.~J., Ryan, R.~E.,
  Yan, H., Baldry, I.~K., Driver, S.~P., {et~al.} 2011, \apjs, 193, 27

\bibitem[{Wolfe \& Chen(2006)}]{Wolfe:2006}
Wolfe, A.~M. \& Chen, H.-W. 2006, \apj, 652, 981

\bibitem[{Wolfe {et~al.}(2005)Wolfe, Gawiser, \& Prochaska}]{Wolfe:2005}
Wolfe, A.~M., Gawiser, E., \& Prochaska, J.~X. 2005, \araa, 43, 861

\bibitem[{Yan {et~al.}(2004)Yan, Dickinson, Eisenhardt, Ferguson, Grogin,
  Paolillo, Chary, Casertano, {et~al.}}]{Yan:2004}
Yan, H., Dickinson, M., Eisenhardt, P. R.~M., Ferguson, H.~C., Grogin, N.~A.,
  Paolillo, M., Chary, R.-R., Casertano, S., {et~al.} 2004, \apj, 616, 63

\end{thebibliography}

\end{document}